\documentclass[twocolumn,showpacs,aps,prd,floatfix]{revtex4}
\usepackage{graphicx}
\usepackage{dcolumn}
\usepackage{amsmath}
\usepackage{epsfig}
\usepackage{multirow}
\usepackage{enumerate}

\graphicspath{{figures/}}

\newcommand{\BABARPubYear}{11}
\newcommand{\BABARPubNumber}{020}
\newcommand{\SLACPubNumber}{14730}

\input{babarsym}
\newcommand{\gevmone}{\ensuremath{\mathrm{\,Ge\kern -0.1em V^{-1}}}\xspace}

\newcommand{\pippim}                {\mbox{$\Bz \to 3\KS(\pip\pim)$}}
\newcommand{\pizpiz}                {\mbox{$\Bz \to 2\KS(\pip\pim)\KS(\piz\piz)$}}
\newcommand{\bei}{\begin{itemize}}
\newcommand{\eei}{\end{itemize}}
\newcommand{\beq}{\begin{equation}}
\newcommand{\eeq}{\end{equation}}
\newcommand{\beqn}{\begin{eqnarray}}
\newcommand{\eeqn}{\end{eqnarray}}
\newcommand{\beqns}{\begin{eqnarray*}}
\newcommand{\eeqns}{\end{eqnarray*}}

\def\lnL{\ln{\cal L}}

\def\KS{\ensuremath{K^0_S}}
\def\de{\Delta E}
\def\dt{\deltat}

\def\fscfave{\kern 0.18em\overline{\kern -0.18em f}_{\rm SCF}}

\def\Abar{\kern 0.18em\overline{\kern -0.18em A}{}}
\def\Amptp{{\cal A}}

\def\Amptpbar{\kern 0.18em\overline{\kern -0.18em {\cal A}}}

\def\detJ{|\det J|}

\def\rar{\rightarrow}

\def\pipi   {\ensuremath{\pip\pim}\xspace}

\newcommand{\fI}                 {\mbox{$f_0(980)$}}

\newcommand{\fV}                {\mbox{$f_0(1710)$}}

\newcommand{\fVII}                {\mbox{$f_2(2010)$}}

\newcommand{\fXI}                 {\mbox{$f_X(1500)$}}
\newcommand{\smin}         {\mbox{$s_{\min}$}}
\newcommand{\smax}         {\mbox{$s_{\max}$}}
\newcommand{\smed}         {\mbox{$s_{\rm med}$}}

\newcommand{\hmin}         {\mbox{$h_{\min}$}}
\newcommand{\hmax}         {\mbox{$h_{\max}$}}

\newcommand{\half}{\mbox{$\frac{1}{2}$}}

\def\eg                 {e.g.}

\def\ie                 {i.e.}

\def\mBz{m_{\Bz}}
\def\mKS{m_{\KS}}

\long\def\inst#1{\par\nobreak\kern 4pt\nobreak
    {\it #1}\par\vskip 10pt plus 3pt minus 3pt}

\begin{document}

\begin{flushleft}
%arXiv:\LANLNumber\ [hep-ex] \\
SLAC-PUB-\SLACPubNumber \\
\babar-PUB-\BABARPubYear/\BABARPubNumber
%\rm \babar$\;$Analysis Document \#2359, Version 13\\
\end{flushleft}

% Title of the paper
\title{
\large \bf
\boldmath
Amplitude analysis and measurement of the time-dependent \CP asymmetry of $\Bz\to\KS\KS\KS$ decays
} % end title

% Input author list file
%\input ../pubboard/authors_jul2011
%% author list as of 01-Jul-2011 (387 authors)
%
\author{J.~P.~Lees}
\author{V.~Poireau}
\author{V.~Tisserand}
\affiliation{Laboratoire d'Annecy-le-Vieux de Physique des Particules (LAPP), Universit\'e de Savoie, CNRS/IN2P3,  F-74941 Annecy-Le-Vieux, France}
\author{J.~Garra~Tico}
\author{E.~Grauges}
\affiliation{Universitat de Barcelona, Facultat de Fisica, Departament ECM, E-08028 Barcelona, Spain }
\author{M.~Martinelli$^{ab}$}
\author{D.~A.~Milanes$^{a}$}
\author{A.~Palano$^{ab}$ }
\author{M.~Pappagallo$^{ab}$ }
\affiliation{INFN Sezione di Bari$^{a}$; Dipartimento di Fisica, Universit\`a di Bari$^{b}$, I-70126 Bari, Italy }
\author{G.~Eigen}
\author{B.~Stugu}
\affiliation{University of Bergen, Institute of Physics, N-5007 Bergen, Norway }
% ...
\collaboration{The \babar\ Collaboration}
\noaffiliation

%\date{\today}

% Abstract
\begin{abstract}
   \noindent
   We present the first results on the Dalitz-plot structure and improved 
measurements of the time-dependent \CP-violation parameters of the 
process $\Bz \rightarrow \KS\KS\KS$ obtained using $468\times10^{6}$ 
\BB\ decays collected with the \babar\ detector at the \pep2\ 
asymmetric-energy \B\ factory at SLAC.
The Dalitz-plot structure is probed by a time-integrated amplitude analysis that 
does not distinguish between \Bz and \Bzb decays.
We measure
the total inclusive branching fraction ${\cal B}(\Bz \rightarrow \KS\KS\KS)=\rm (6.19 \pm 0.48 \pm 0.15 \pm 0.12)\times 10^{-6}$, where
the first uncertainty is statistical, the second is systematic, and the third represents the Dalitz-plot signal model dependence.
We also observe evidence for the intermediate resonant states $\fI$, $f_0(1710)$, and $f_2(2010)$. Their respective product branching fractions are measured to be $\rm(2.70\,^{+1.25}_{-1.19} \pm 0.36 \pm 1.17)\times 10^{-6}$, $\rm(0.50\,^{+0.46}_{-0.24} \pm 0.04 \pm 0.10)\times 10^{-6}$, and $\rm(0.54\,^{+0.21}_{-0.20} \pm 0.03 \pm 0.52)\times 10^{-6}$. Additionally, we determine the mixing-induced \CP-violation parameters to be ${\cal S} = -0.94\,^{+0.24}_{-0.21} \pm 0.06 $ and ${\cal C} = -0.17 \pm 0.18 \pm  0.04 $, where
the first uncertainty is statistical and the second is systematic. These values are in agreement with the standard model expectation.
For the first time, we report evidence of \CP\ violation in $\Bz\to\KS\KS\KS$ decays; \CP\ conservation is excluded at $3.8$ standard deviations including systematic uncertainties.

\end{abstract}

% PACS

\maketitle

% The body of the paper starts here

\section{INTRODUCTION}
\label{sec:introduction}

Over the past ten years, the $B$ factories have shown that the
Cabibbo-Kobayashi-Maskawa paradigm in the standard model (SM),
with a single weak phase in the quark mixing matrix, accounts for the
observed \CP-symmetry violation in the quark sector.
However, there may be other \CP-violating sources beyond the SM.  
Charmless hadronic $B$ decays, like $\Bz\to\KS\KS\KS$,  are of great interest because they are  dominated by loop diagrams and are thus sensitive to new physics effects at large energy scales~\cite{Grossman:1996ke}. 
In the SM, the mixing-induced \CP-violation parameters in this decay are expected to be the same, up to $\sim 1\%$~\cite{Cheng:2005ug}, as in the tree-diagram-dominated modes such as $\Bz\to \jpsi\KS$.  
Both \babar~\cite{KsKsKsBabar} and Belle~\cite{belleCP} have previously performed time-dependent \CP-violation measurements of the inclusive mode $\Bz\to\KS\KS\KS$, which is permissible because the final state is \CP\ definite~\cite{Gershon:2004tk}. 

The structure of the Dalitz plot (DP), however, is of interest; although the
time-dependent \CP-violation parameters ${\cal S}$ and ${\cal C}$ [see Eq.~\eqref{eq:dtmeas}] can be measured inclusively without taking into account the phase space, different resonant contributions may have different values of these parameters in the presence of new physics. The statistical precision is not sufficient to perform a time-dependent amplitude analysis, but as we show below, it is possible to extract branching fractions from resonant contributions to the decay using a time-integrated amplitude analysis.
Additionally, the amplitude analysis could shed light on the controversial \fXI\ resonance: recent measurements of $\Bz\to \Kp\Km\KS$ and $\Bpm \to \Kp\Km\Kpm$ from \babar\ \cite{babarKKK, Aubert:2007sd, babarKKKs} and Belle~\cite{belleKKK, belleKKKs} have shown evidence of a wide structure in the $m_{\Kp\Km}$ spectrum around $1.5\gev$.
In these measurements, it was assumed that this structure is a single scalar resonance; however a vector hypothesis could not be ruled out. The \babar\ measurement of $\Bp \rightarrow \Kp\Km \pip$ \cite{babarKKpi} appears to show an enhancement  around $1.5\gev$, while the \babar\ analysis of $\Bpm \rightarrow \KS \KS \pipm$~\cite{KsKspi} finds no evidence of a possible \fXI, suggesting that the structure is either a vector meson or something exotic. An amplitude analysis of $\Bz\to\KS\KS\KS$ will provide further insight into the nature of this structure, as only intermediate states of even spin are permitted due to Bose-Einstein statistics; an observation of the \fXI\ decaying to \KS\KS\ would require an even-spin state.
Finally, the amplitude analyses of $\B \to K\pi\pi$ and $\B \to KKK$ modes may be used to extract the Cabibbo-Kobayashi-Maskawa angle $\gamma$~\cite{London_Lorier:2011}.

This paper presents the first amplitude analysis and the final \babar\ update of the time-dependent \CP-asymmetry measurement of $\Bz\to\KS\KS\KS$ using the full \FourS\ dataset.
The amplitude analysis is time-integrated \CP-averaged (i.e., it does not use flavor-tagging information to  distinguish between \Bz\ and \Bzb mesons). It takes advantage of the interference pattern in the DP to measure relative magnitudes and phases for the different resonant modes using $\Bz\to\KS\KS\KS$ decays with $\KS\to\pip\pim$, denoted by \pippim.
The magnitudes and phases are then translated into individual branching fractions for the resonant modes.
The time-dependent analysis extracts the ${\cal S}$ and ${\cal C}$ parameters by modeling the proper-time distribution. This part of the analysis uses both \pippim\ events and events where one of the \KS\ mesons decays to $\piz\piz$, denoted by \pizpiz.

In Sec.~\ref{sec:babar} we briefly describe the \babar\ detector
and the dataset. The amplitude analysis is described in Sec.~\ref{sec:DP_analysis} and the time-dependent analysis in Sec.~\ref{sec:TD_analysis}. Finally we summarize the results in Sec.~\ref{sec:Summary}.

\section{THE \babar\ DETECTOR AND DATASET}
\label{sec:babar}

The data used in this analysis were collected with the \babar\ 
detector at the \pep2\ asymmetric-energy $e^+e^-$ storage ring at 
SLAC.
%between 1999 and 2007.
The sample consists of
an integrated luminosity of
$426.0\;\mathrm{fb}^{-1}$, corresponding to $(467.8\pm 5.1)\times10^{6}$ 
$B\Bbar$ pairs collected at the \FourS resonance (``on-resonance''),
and $44.5$~\invfb collected about $40$~\mev 
below the~\FourS (``off-resonance'').

A detailed description of the \babar\ detector is presented in 
Ref.~\cite{babar}. The tracking system used for track and vertex 
reconstruction has two  components: a silicon vertex tracker 
and a drift chamber, both operating within a 1.5~T 
magnetic field generated by a superconducting solenoidal magnet. 
A detector of internally reflected Cherenkov light 
associates Cherenkov photons with tracks for particle 
identification.
The energies of photons and electrons
are determined from the measured light produced
in electromagnetic showers inside a CsI crystal electromagnetic calorimeter.
Muon
candidates are identified with the
use of the instrumented flux return of the solenoid.

\section{Amplitude analysis}
\label{sec:DP_analysis}

In Secs.~\ref{sec:kinematics} and \ref{sec:SquareDP} we describe the DP formalism and introduce
the signal parameters that are extracted from
data. In Sec.~\ref{sec:dp_selection} we describe the requirements used to select the signal candidates and
suppress backgrounds. In Sec.~\ref{sec:dp_likelihood} we describe the fit method
and the approach used to account for experimental effects such
as resolution. In Sec.~\ref{sec:fitresults} we present the results of the fit,
and finally, in Sec.~\ref{sec:dp_systematics} we discuss systematic
uncertainties in the results.

\subsection{Decay amplitudes}
\label{sec:kinematics}
The $\Bz\to\KS\KS\KS$ decay contains three identical particles in the final state and therefore the amplitude needs to be symmetrized. 
We consider the decay of a spin-zero $\Bz$
%with four-momentum $p_B$
into three daughters, $\KS(1)$, $\KS(2)$, and $\KS(3)$, with four-momenta $p_1$, $p_2$, and $p_3$. The decay amplitude is given by~\cite{Cheng:2005ug}
\begin{eqnarray}
\label{eq:paper_soni_2}
{\cal A}[\Bz&\to&\KS(1)\KS(2)\KS(3)] \\
 \nonumber &=& \left(\frac{1}{2}\right)^{3/2}\Big\{{\cal A}_1[\Bz\to\Kzb(1)\Kz(2)\Kz(3)] \\
 \nonumber &\ &        \ \ \ \ \ \ \ \ \ + \ {\cal A}_2[\Bz\to\Kzb(2)\Kz(3)\Kz(1)] \\
 \nonumber &\ &  \ \ \ \ \ \ \ \ \ + \ {\cal A}_3[\Bz\to\Kzb(3)\Kz(1)\Kz(2)]\Big\}, 
 \end{eqnarray}
which takes into account the three permitted paths from the initial state to the final state. For instance for the \Bz\ decay this consists of an intermediate state $K^0 K^0 \Kbar^0$.
Since the labeling of the three identical particles is arbitrary, we classify the final-state particles according to the square of the invariant mass, $s_{ij}$, defined as
\begin{equation}
\label{eq:minmaxVariables} 
       s_{ij} \;=\;s_{ji} \;=\; m_{\KS(i)\KS(j)}^2 \;=\; (p_i + p_j)^2,
\end{equation}
where $i$ and $j$ are the \KS\ indices.
We use as independent (Mandelstam) variables the minimum and the maximum of the squared masses $\smin$ and $\smax$:
\begin{eqnarray}
\label{eq:dalitzVariables} 
       \smin \;&=&\; \min(s_{12},s_{23},s_{13}),\\ \nonumber
       \smax \;&=&\; \max(s_{12},s_{23},s_{13}).
\end{eqnarray}
The third (median) invariant squared mass $\smed$ can be obtained from energy and momentum conservation:
\beq
\label{eq:magicSum}
        \smed \;=\; \mBz^2 + 3m_{\KS}^2
                   - \smin - \smax.
\eeq
The differential $\B$ meson decay width with respect to the 
variables defined in Eq.~\eqref{eq:dalitzVariables} (\ie,~the 
DP variables) reads
\beq
\label{eq:partialWidth}
        d\Gamma(\B\to\KS\KS\KS) \;=\; 
        \frac{1}{(2\pi)^3}\frac{|\Amptp|^2}{32 \mBz^3}\,d\smin d\smax,
\eeq
where ${\cal A}$ is the Lorentz-invariant amplitude
of the three-body decay. 
This amplitude analysis does not take into account any flavor tagging or time dependence, thus it is \CP\ averaged and time integrated. The term $\left|{\cal A}\right|^2$ is therefore simply the average of squares of the contributions ${\cal A}[\Bz\to\KS\KS\KS]$ and ${\cal A}[\Bzb\to\KS\KS\KS$].

The choice of the variables $\smin$ and $\smax$ gives a uniquely defined coordinate in the symmetrized DP. Therefore only one sixth of the DP is populated, i.e., the event density is six times larger compared to an amplitude analysis involving three distinct particles.

We describe the distribution of signal events 
in the DP using an isobar approximation,
which models the total amplitude as 
resulting from a coherent sum of amplitudes from the $N$ individual decay channels
of the $B$ meson, either into an intermediate resonance and a bachelor particle or in a nonresonant manner:
\begin{eqnarray}
  \label{eq:isobar}
  {\cal A}(\smin,\smax) 
  & = & \sum_{j=1}^{N} c_j F_j(\smin,\smax).
\end{eqnarray}
Here $F_j$ (described in detail below) are DP-dependent amplitudes containing the decay dynamics
and $c_j$ are complex coefficients describing the relative
magnitudes and phases of the different decay channels.
This description, which contains a single complex number $c_j$ for each decay channel regardless of the \B flavor (\Bz\ or \Bzb), reflects the assumptions of no direct \CP\ violation and of
a common weak phase for all the decay channels.
With this description we cannot extract any weak phase information; this would require using per-\B\-flavor complex amplitudes.
The factor $F_j$ contains strong dynamics only, and thus does not change under \CP\ conjugation.

Intermediate resonances decay to $\Kz\!\Kzb$. In terms of the isobar approximation, the amplitude in Eq.~\eqref{eq:paper_soni_2} for a resonant state $j$ becomes
\begin{eqnarray}
\label{eq:ResKzKzb}
{\cal A}[\Bz&\to&\KS(1)\KS(2)\KS(3)] \\
\nonumber &\propto& c_j\left[F_j(s_{12},s_{13})+F_j(s_{12},s_{23})+F_j(s_{13},s_{23}) \right].
\end{eqnarray}
This reflects the fact that it is impossible to associate a given \KS\ to a flavor eigenstate \Kz\ or \Kzb. In practice, this sum of three $F_j$ terms, corresponding to an even-spin resonance, is implicitly taken into account by the description in terms of $\smin$ and $\smax$.

The $F_j$ terms are represented
by the product of the invariant mass and angular distributions; \ie,
\begin{equation}
\label{eq:ResDynEqn}
F_j(\smin,\smax, L) = R_j(m) X_L(|\vec{p}\,^{\star}|\,r') X_L(|\vec{q}\,|\,r) T_j(L,\vec{p},\vec{q}\,),
\end{equation}
where
\begin{enumerate}[(i)]
\item $m$ is the invariant mass of the decay products of the resonance;
\item $R_j(m)$ is the resonance mass term or ``line shape'', \eg,~relativistic
Breit--Wigner (RBW);
\item $L$ is the orbital angular momentum between the resonance and the
      bachelor particle;
\item $\vec{p}\,^{\star}$ is the momentum of the bachelor particle
      evaluated in the rest frame of the $B$;
\item $\vec{p}$ and $\vec{q}$ are the momenta of the bachelor particle and
      one of the resonance daughters, respectively, both evaluated in the
      rest frame of the resonance;
\item $X_L(|\vec{p}\,^{\star}|\,r')$ and $X_L(|\vec{q}\,|\,r)$ are Blatt--Weisskopf barrier factors~\cite{blatt-weisskopf}
with barrier radii of $r$ and $r'$, and
\item $T_j(L,\vec{p},\vec{q})$ is the angular distribution:
\begin{eqnarray}
L=0 &:& T_j = 1,\\
L=2 &:& T_j = \frac{8}{3} \left[3(\vec{p}\cdot\vec{q}\,)^2 - (|\vec{p}\,||\vec{q}\,|)^2\right].
\end{eqnarray}
\end{enumerate}
The Blatt--Weisskopf barrier factor is unity for all the zero-spin resonances. In our analysis it is relevant only for the $\fVII$.
Since for this resonance $r$ and $r'$ are not measured, we take them both to be $1.5\gevmone$ and vary by $\pm 0.5\gevmone$ to estimate the systematic uncertainty.

The helicity angle of a resonance is defined as the angle between $\vec{p}$ and $\vec{q}$.
Explicitly, the helicity angle $\theta$
for a given resonance
is defined between the momenta of the bachelor particle and one of the daughters of the resonance in the resonance rest frame.
Because of the identical final-state particles this definition is ambiguous,
but the ambiguity disappears because of the description of the DP in terms of \smin\ and \smax.
There are three possible invariant-mass combinations: $\smin$, $\smed$, and $\smax$. We denote the corresponding helicity angles as $\theta_{\min}$, $\theta_{\rm med}$, and $\theta_{\max}$.
The three angles are defined between $0$ and $\pi/2$.

As the present study is the first amplitude analysis of this decay, we use the method outlined in Sec.~\ref{sec:signal_model} to determine the contributing intermediate states. The components of the nominal signal model are summarized in Table~\ref{tab:model}.

For most resonances in this analysis the $R_j$ are taken to be RBW~\cite{Amsler:2008zz} line shapes:
\begin{equation}
R_j(m) = \frac{1}{(m^2_0 - m^2) - i m_0 \Gamma(m)},
\label{eqn:BreitWigner}
\end{equation}
where $m_0$ is the nominal mass of the resonance and $\Gamma(m)$ is the
mass-dependent width.
In the general case of a spin-$J$ resonance, the latter can be expressed as
\begin{equation}
\Gamma(m) = \Gamma_0 \left( \frac{q}{q_0}\right)^{2J+1} 
\left(\frac{m_0}{m}\right) \frac{X^2_J(|\vec{q}\,|\,r)}{X^2_J(|\vec{q_0}\,|\,r)}.
\label{eqn:resWidth}
\end{equation}
The symbol $\Gamma_0$ denotes the nominal width of the resonance.
The values of $m_0$ and $\Gamma_0$ are listed in Table~\ref{tab:model}.
The symbol $q_0$ denotes the value of $q$ when $m = m_0$.

For the \fI\ line shape the Flatt\'e form~\cite{Flatte} is used.
In this case the mass-dependent width is given by the sum 
of the widths in the $\pi\pi$ and $KK$ systems:
\begin{equation}
\Gamma(m) = \Gamma_{\pi\pi}(m) + \Gamma_{KK}(m),
\label{eqn:FlatteW1}
\end{equation}
where
\begin{eqnarray}
\Gamma_{\pi\pi}(m) &=&
g_{\pi} \Bigg(\frac{1}{3}\sqrt{1 - 4m_{\piz}^2/m^2} + \\ \nonumber
&& \phantom{g_{\pi} \Bigg(\frac{1}{3}} \frac{2}{3}\sqrt{1 - 4m_{\pipm}^2/m^2}\Bigg)\,,\\
\Gamma_{KK}(m) &=&
g_{K} \Bigg(\frac{1}{2}\sqrt{1 - 4m_{\Kpm}^2/m^2} + \\ \nonumber
&& \phantom{g_{K} \Bigg(\frac{1}{2}} \frac{1}{2}\sqrt{1 - 4m_{\Kz}^2/m^2}\Bigg)\,.
\label{eqn:FlatteW2}
\end{eqnarray}
The fractional coefficients arise from isospin conservation and $g_{\pi}$
and $g_{K}$ are coupling constants for which the values are given in Table~\ref{tab:model}.
The nonresonant (NR) component is modeled using an exponential function:
\begin{equation}
R_{\rm NR}(m) = e^{\alpha m^2}.
\label{eqn:NREXP}
\end{equation}
As in the resonant case, here $m$ is the invariant mass of the relevant \KS\KS\ pair.
The parameter $\alpha$ is taken from the \babar\ $\Bp \to \Kp\Km\Kp$ analysis~\cite{Aubert:2007sd,babarKKKs} and is given in Table~\ref{tab:model}. This value was found to be compatible with the one resulting from varying $\alpha$ in the maximum-likelihood fit in the present analysis.
There is no satisfactory theoretical description of the NR component; it has to be determined empirically. The exponential function of Eq.~\eqref{eqn:NREXP} was used by other amplitude analyses of \B-meson decays to three kaons~\cite{babarKKK, Aubert:2007sd, babarKKKs, belleKKK, belleKKKs}. Adopting the same parametrization for the NR term allows the comparison of results for other components. 

\begin{table*}[htbp]
\begin{center}
\caption{Parameters of the DP model used in the fit.
The Blatt--Weisskopf barrier parameters ($r$ and $r'$) of the $\fVII$, which have not been measured, are varied by $\pm 0.5\gevmone$  for the model uncertainty.
\label{tab:model}}
\begin{tabular}{cccc}
\hline\hline
Resonance      & Parameters                      & Line shape                  & Reference        \\ \hline\\[-9pt]
$\fI$          & $m_0=(965 \pm 10)\mevcc$        & Flatt\'e                    & \cite{valFlatte} \\
               & $g_{\pi}=(165 \pm 18)\mevcc$    & Eq.~\eqref{eqn:FlatteW1}    &                  \\
               & $g_{K}=(695 \pm 93)\mevcc$      &                             &                  \\ \hline\\[-9pt]
$\fV$          & $m_0=(1724 \pm 7)\mevcc$        & RBW                         & \cite{Amsler:2008zz}  \\
               & $\Gamma_0=(137 \pm 8)\mevcc$    & Eq.~\eqref{eqn:BreitWigner} &                  \\ \hline\\[-9pt]
$\fVII$        & $m_0=(2011\,^{+60}_{-80})\mevcc$& RBW                         & \cite{Amsler:2008zz}  \\
               & $\Gamma_0=(202 \pm 60)\mevcc$   & Eq.~\eqref{eqn:BreitWigner} &                  \\ \\[-9pt]
               & $r=r'=1.5 \gev^{-1}$            &                             &                  \\ \hline\\[-9pt]
NR decays \hspace*{5 mm}& $\alpha=(-0.14 \pm 0.02)\gev^{-2} c^4$ \hspace*{5 mm}& Exponential NR & \cite{babarKKKs} \\
               &                                 & Eq.~\eqref{eqn:NREXP}       &  \\ \hline\\[-9pt]
$\chiczero$    & $m_0=(3414.75 \pm 0.31)\mevcc$  & RBW                         & \cite{Amsler:2008zz}    \\
               & $\Gamma_0=(10.2 \pm 0.7)\mevcc$ & Eq.~\eqref{eqn:BreitWigner} &                  \\
\hline \hline
\end{tabular}
\end{center}
\end{table*}

\subsection{The square Dalitz plot}
\label{sec:SquareDP}
\begin{figure*}[htbp]
\begin{center}
\includegraphics[width=8.0cm,keepaspectratio]{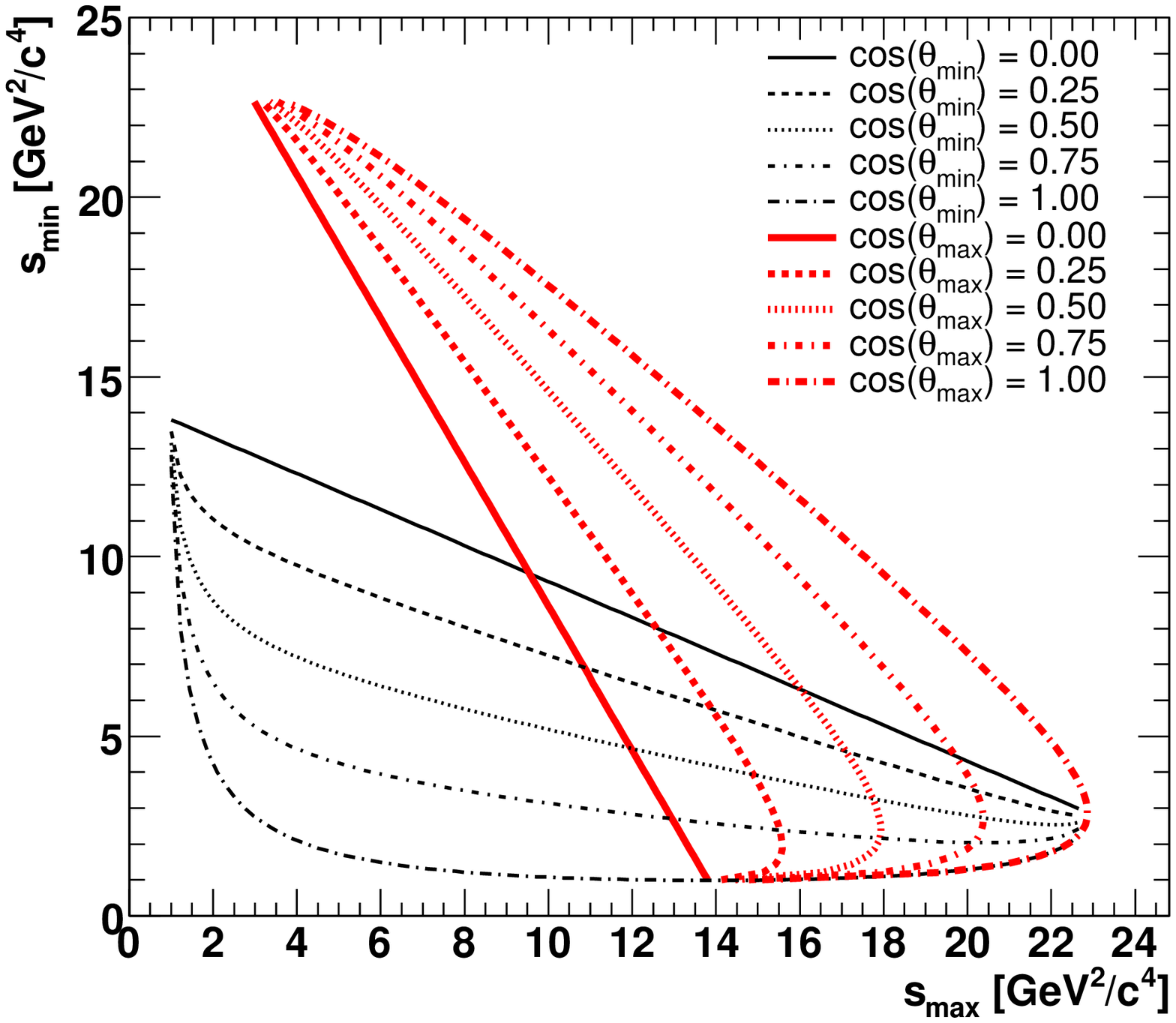}
\includegraphics[width=8.0cm,keepaspectratio]{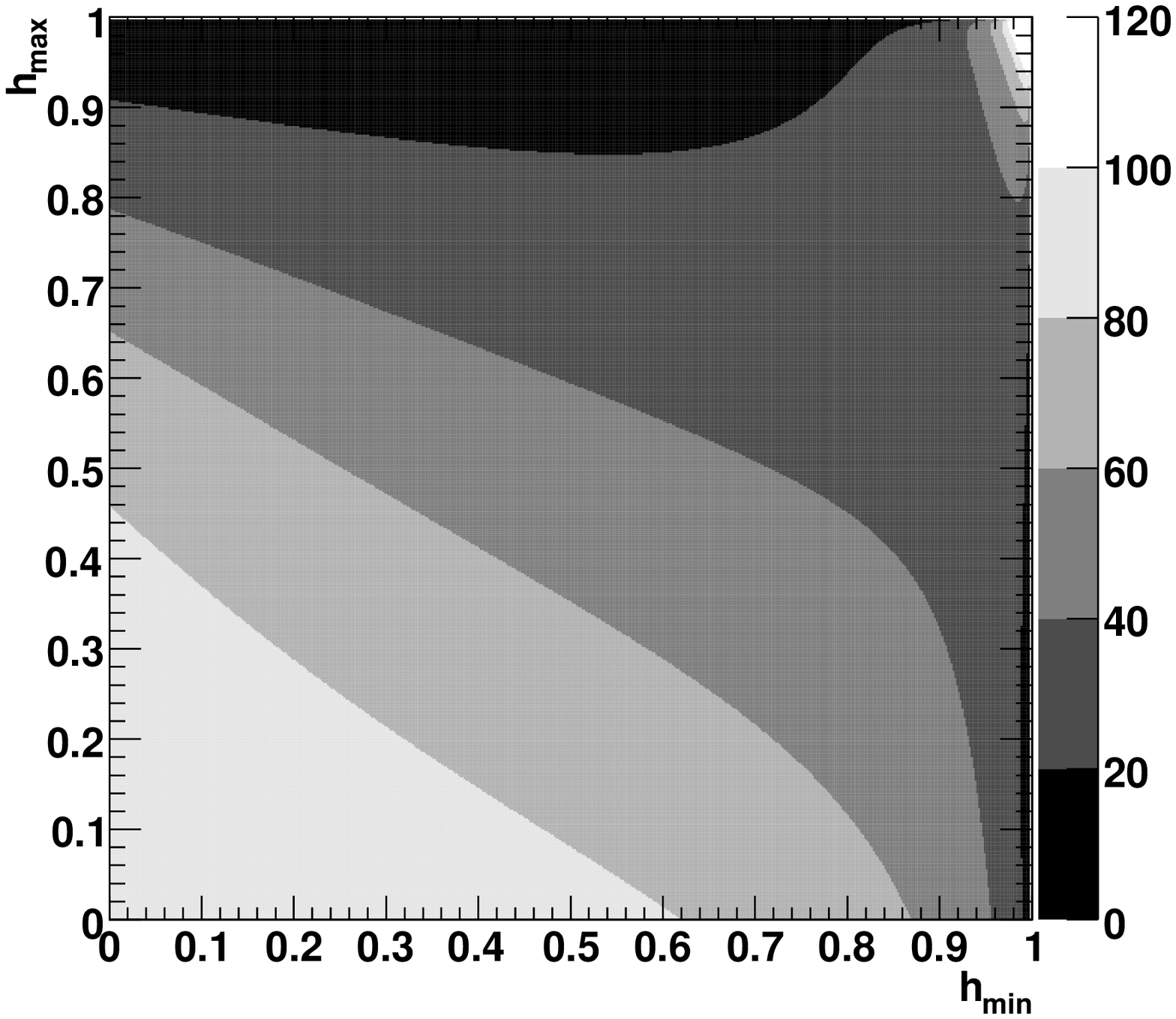}
\caption{(color online). 
Lines of constant helicity angle in the Dalitz plot of \smin\ versus \smax\ (left), and 
the magnitude of the Jacobian (gray scale on the right) mapping (\smin, \smax) to (\hmin, \hmax). For the latter see Eq.~\eqref{eq:SqDalitzTrans2}.}
\label{fig:SquareDalitz}
\end{center}
\end{figure*}

We use two-dimensional histograms to describe the phase-space dependent reconstruction efficiency and to model the background over the DP. When the phase-space boundaries of the DP do not coincide with the histogram bin boundaries this may introduce biases. We therefore define $\hmin$ and $\hmax$ as
$\cos \theta_{\min}$ and $\cos \theta_{\max}$, respectively, and apply the transformation
\beq
\label{eq:SqDalitzTrans1}
        \left(\smin, \smax\right) \longrightarrow \left(\hmin, \hmax\right).
\eeq
The $\left(\hmin, \hmax\right)$ plane is referred to as the square Dalitz plot (SDP), where both $\hmin$ and $\hmax$ range between $0$ and $1$ 
due to the convention adopted for the helicity angles (see Fig.~\ref{fig:SquareDalitz}).
Explicitly, the transformation is
\begin{eqnarray}
\hmin &=&
\frac{\smin(\smax-\smed)}{\sqrt{\smin^2-4\mKS^2\smin}}\times \\ \nonumber
&& \frac{1}{\sqrt{(\mBz^2-\mKS^2-\smin)^2-4\mKS^2\smin}}, \\
\hmax &=&
\frac{\smax(\smed-\smin)}{\sqrt{\smax^2-4\mKS^2\smax}}\times \\ \nonumber
&& \frac{1}{\sqrt{(\mBz^2-\mKS^2-\smax)^2-4\mKS^2\smax}}, \nonumber
\label{eq:SqDalitzTransExplicit}
\end{eqnarray}
where the numerators may easily be expressed in terms of \smin\ and \smax\ using Eq.~\eqref{eq:magicSum}.
The differential surface elements of the DP and the SDP are related by
\beq
\label{eq:SqDalitzTrans2}
        d \smin \,d \smax = \detJ\, d \hmin \, d \hmax,
\eeq
where $J=J\left(\hmin, \hmax\right)$ is the appropriate Jacobian matrix.
The backward transformations $\smin\left(\hmin, \hmax\right)$ and $\smax\left(\hmin, \hmax\right)$, and therefore the Jacobian $\detJ$, cannot be found analytically; they are obtained numerically.
The variables \hmin\ and \hmax\ as a function of the invariant masses are shown in Fig.~\ref{fig:SquareDalitz} together with the Jacobian.

\subsection{Event selection and backgrounds}
\label{sec:dp_selection}

We reconstruct $\Bz\to\KS\KS\KS$ candidates from three $\KS\to\pip\pim$ candidates that form a good quality vertex; i.e., the fit of the \Bz\ vertex is required to converge 
and the $\chi^2$ probability of each \KS\ vertex fit has to be greater than $10^{-6}$.
Each $\KS$ candidate must have $\pip\pim$ invariant mass within $12.1\mevcc$ of the nominal $\Kz$ mass~\cite{Amsler:2008zz}, and decay length with respect to the $B$ vertex between $0.22$ and $45\cm$. The last criterion ensures that the decay vertices of the \Bz and the \KS\ are well separated.
In addition, combinatorial background is suppressed by selecting events for which the angle between
the momentum vector of each \KS\ candidate
and the vector connecting the beamspot and the $\KS$ vertex is smaller than
$0.0185$~radians.
We ensure a good $B$ vertex fit quality by requiring that the charged pions of at least one of the $\KS$ candidates have hits in the two inner layers of the vertex tracker.

A $B$ meson candidate is characterized kinematically by the energy-substituted mass $\mes\equiv\sqrt{(s/2+\vec {p}_i\cdot\vec{p}_B)^2/E_i^2-p_B^2}$ and the energy difference $\de \equiv E_B^*-\half\sqrt{s}$, where $(E_B,\vec{p}_B)$ and $(E_i,\vec {p}_i)$ are the four-vectors in the laboratory frame of the $B$-candidate and the initial electron-positron system,
respectively, and $p_B$ is the magnitude of $\vec{p}_B$. The asterisk denotes the \FourS\  frame, and $s$ is the square of the invariant mass of the electron-positron system. We require $5.27 < \mes <5.29\gevcc$ and $|\de|<0.1\gev$. Following the calculation of these kinematic variables, each of the $B$ candidates is refitted with its mass constrained to the world average value of the $B$ meson mass~\cite{Amsler:2008zz} in order to improve the DP position resolution, and ensure that Eq.~\eqref{eq:magicSum} holds. The sideband used for background studies is in the range $5.20 < \mes <5.27\gevcc$ and $|\de|<0.1\gev$.

Backgrounds arise primarily from random combinations in continuum $e^+e^- \to q \bar{q}$ events ($q=d,u,s,c$). To enhance discrimination between signal and continuum background, we  use a neural network (NN)~\cite{Gay:1995sm} to combine four discriminating variables: the angles with respect to the beam axis of the $B$ momentum and $B$ thrust axis in the \FourS\ frame, and the zeroth- and second-order monomials $L_{0,2}$ of the energy flow about the $B$ thrust axis.  The monomials are defined by $ L_n = \sum_i p_i\times\left|\cos\theta_i\right|^n$, where $\theta_i$ is the angle with respect to the $B$ thrust axis of track or neutral cluster $i$ and $p_i$ is the magnitude of its momentum. The sum excludes the $B$ candidate and all quantities are calculated in the \FourS\ frame. The NN is trained with off-resonance data, sideband data and simulated signal events that pass the selection criteria.
Approximately $0.5\%$ of events passing the full selection have more than one candidate. When this occurs, we select the candidate for which the error-weighted average of the masses of the \KS\ candidates is closest to the world average \KS\ mass~\cite{Amsler:2008zz}.
With the above selection criteria, we  obtain a signal reconstruction efficiency of $6.6\%$ that has been determined 
from a signal Monte Carlo (MC) sample generated using
the same DP model and parameters as obtained from the data fit results.
We estimate from this MC that $1.4\%$ of the selected signal events are misreconstructed, and assign a systematic uncertainty (see Sec.~\ref{sec:dp_systematics}).
We use MC events to study the background from other $B$ decays ($B$ background). We expect fewer than 6 such events in our data sample. As these events are wrongly reconstructed, the $\mes$ and $\de$ distributions are continuumlike and as a result the events are mostly absorbed in the continuum background category. We assign a systematic uncertainty for $B$ background contamination in the signal.

\subsection{The maximum-likelihood fit}
\label{sec:dp_likelihood}
We perform an unbinned extended maximum-likelihood fit to extract the $\Bz\to\KS\KS\KS$ event yield, as well as the resonant and nonresonant amplitudes. 
The fit for the amplitude analysis uses the variables $\mes$ and $\de$, the NN output, and the SDP variables to discriminate signal from background. The selected on-resonance data sample is assumed to consist of signal and continuum background.
The feed-through from \B decays other than the signal is found to be negligible.
Misreconstructed signal events are not considered as a separate event species, but are taken into account as a part of the signal.
The likelihood function ${\cal L}_i$ for event $i$ is the sum
\begin{equation}
\label{eq:PropDenSingle}
 {\cal L}_i=\sum_j{N_j{\cal P}^i_j(\mes,\Delta E,{\rm NN},h_{\min},h_{\max})},
\end{equation}
where $j$ stands for the species (signal, continuum background) and $N_j$ is the corresponding yield. Each probability density function (PDF) ${\cal P}^i_j$ is the product of four individual PDFs:
\begin{equation}
\label{eq:PropProd}
 {\cal P}^i_j={\cal P}^i_j(\mes) \; {\cal P}^i_j(\Delta E) \; {\cal P}^i_j({\rm NN}) \; {\cal P}^i_j(h_{\min},h_{\max}).
\end{equation}
A study with fully reconstructed MC samples shows that correlations between the PDF variables are small and therefore we neglect them. However, possible small discrepancies in the fit results due to these correlations are accounted for in the systematic uncertainty (see Sec.~\ref{sec:dp_systematics}).
The total likelihood is given by
\begin{equation}
\label{eq:dp_Likelihood}
 {\cal L}=\exp(-\sum_j N_j)\prod_i {\cal L}_i.
\end{equation}

\subsubsection{The \mes, \DeltaE, and NN PDFs}
\label{sec:dp_likeDiscrim}

The \mes\  and \DeltaE\ distributions of signal events are parametrized by an asymmetric Gaussian with power-law tails:
\begin{eqnarray}
\label{eq:cruijff}
& & {\rm Cr}(x;m_0,\sigma_l,\sigma_r,\alpha_l,\alpha_r)=\\
\nonumber & & \exp\left(-\frac{(x-m_0)^2}{2\sigma_i^2+\alpha_i(x-m_0)^2}\right)
\left\{
\begin{array}{ll}
x-m_0<0 :    & i=l\\
x-m_0\geq0 : & i=r
\end{array}.
\right.
\end{eqnarray}%
The $m_0$ parameters for both \mes\ and \DeltaE\ 
are free in the fit to data, while the other parameters are fixed to values determined from a fit to MC simulation. For the NN distributions of signal we use a histogram PDF from MC simulation.

For continuum events the \mes and \DeltaE\ PDFs are parametrized by an ARGUS shape function~\cite{argus} and a straight line, respectively. The NN PDF is described by a sum of power functions:
\begin{eqnarray}
\label{eq:spexp}
& & E(x;c_1,a,b_0,b_1,b_2,b_3,c_2,c_3)=\\
\nonumber & & \cos^2(c_1) \: [\cos^2(a)\: {\cal N}(b_0,b_1)\: x^{b_0}\: (1-x)^{b_1}  \\
\nonumber & & + \sin^2(a)\:{\cal N}(b_2,b_3)\: x^{b_2}\: (1-x)^{b_3}]  \\
\nonumber & & + \sin^2(c_1)\: {\cal N}(c_2,c_3)\: x^{c_2}\: (1-x)^{c_3},
\end{eqnarray}
where
$x=\left({\rm NN}-{\rm NN}_{\rm min}\right)/\left({\rm NN}_{\rm max}-{\rm NN}_{\rm min}\right)$
and the ${\cal N}$ are normalization factors, computed analytically using the standard $\Gamma$ function:
\begin{equation}
{\cal N}(\alpha,\beta) = \frac{\Gamma(\beta+2+\alpha)}{\Gamma(\alpha+1)\Gamma(\beta+1)}.
\end{equation}
The parameters for all the continuum PDFs are determined by a fit to sideband data and then fixed for the fit in the signal region.

\subsubsection{Dalitz-plot PDFs}
\label{sec:dp_likeDalitz}

The SDP PDF for continuum background is a histogram obtained from \mes sideband on-resonance events.
The SDP signal PDFs require as input the DP-dependent selection efficiency, $\varepsilon=\varepsilon(\hmin,\hmax)$, that is described by a histogram and is taken from MC simulation.
For each event we define the SDP signal PDF:
\begin{equation}
\label{eq:sigSDPPDF}
 {\cal P}^i_{\rm sig} (\hmin, \hmax)\propto\varepsilon(\hmin, \hmax)\:\left|\Amptp(\hmin,\hmax)\right|^2.
\end{equation}
The normalization of the PDF is implemented by numerical integration.
%We describe the experimental resolution in the SDP variables by convolving the SDP signal PDFs with a resolution function taken from MC simulation.
To describe the experimental resolution in the SDP variables, we use an ensemble of two-dimensional histograms that represents the probability to reconstruct at the coordinate (\hmin', \hmax') an event that has the true coordinate (\hmin, \hmax). These histograms are taken from MC simulation and are convolved with the signal PDF.

\subsubsection{Determination of the signal Dalitz-plot model}
\label{sec:signal_model}

\begin{figure*}[htbp]
\begin{center}
\includegraphics[width=8.0cm,keepaspectratio]{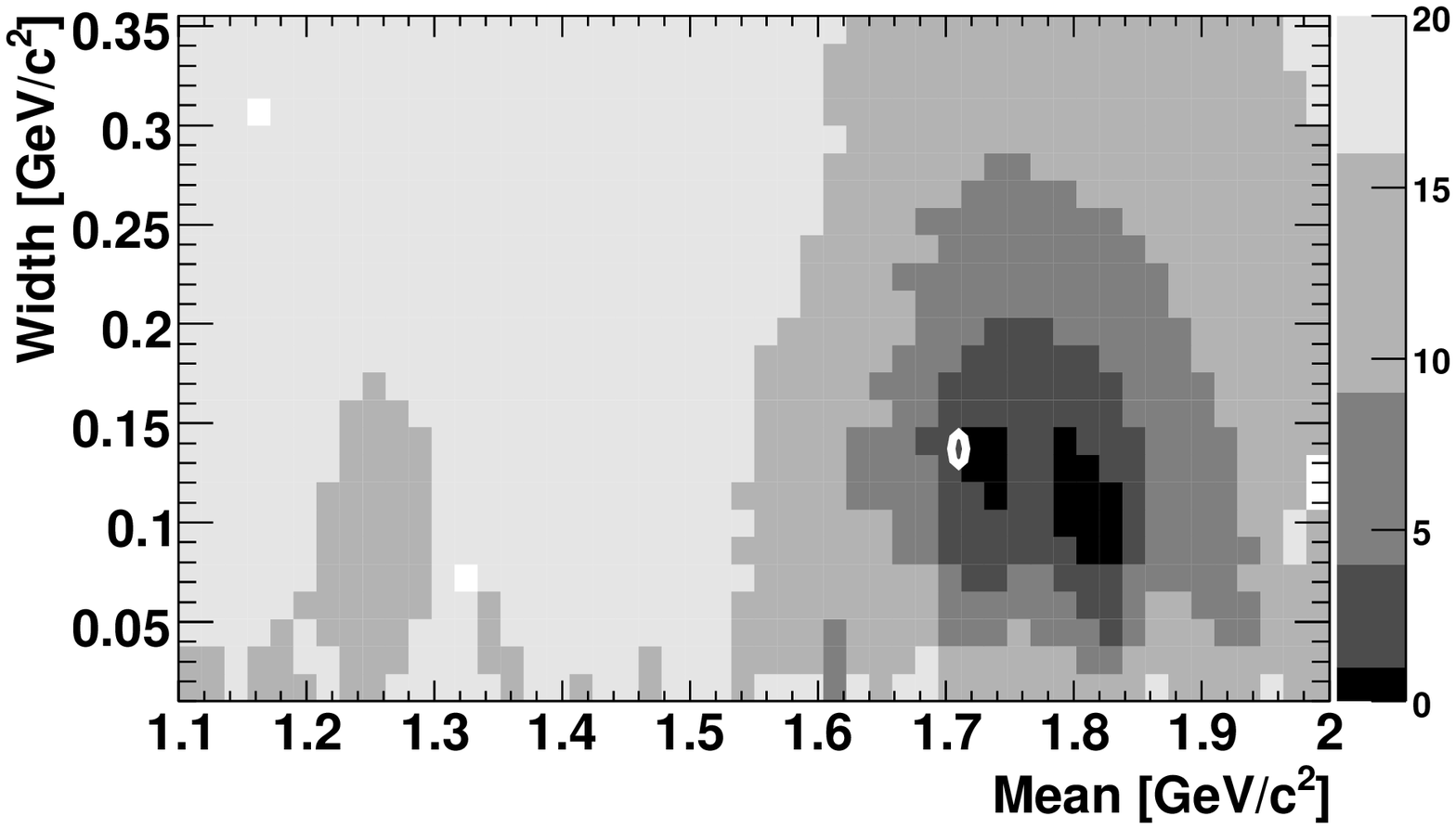}
\includegraphics[width=8.0cm,keepaspectratio]{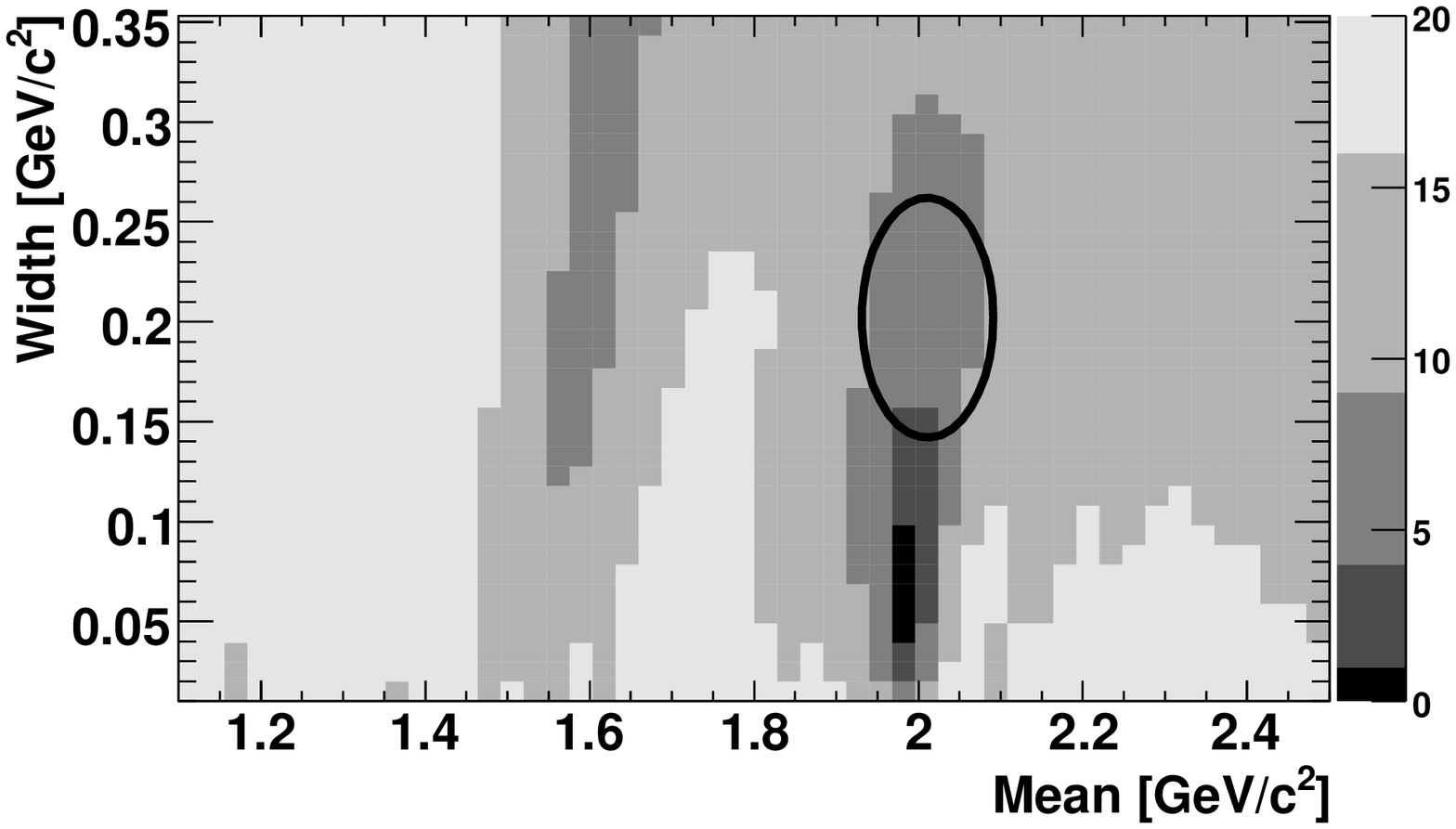}
\caption{
%(color online).
Two-dimensional scans of $-2 \Delta \lnL$
%(color scale)
(gray scale)
as a function of the mass and the width of an additional resonance. These scans were performed to look for an additional scalar resonance (left) and an additional tensor resonance (right). The baseline model of the scans for additional scalar resonances contains $\fI$, $\chiczero$, and NR intermediate states. The baseline model of the scans for additional tensor resonances contains $\fI$, $\chiczero$, NR, and $\fV$ intermediate states. The
ellipses indicate the world average parameters~\cite{Amsler:2008zz} for the $\fV$ and $\fVII$ resonances that are added to the model.}
\label{fig:ModelScans}
\end{center}
\end{figure*}

Using on-resonance data, we determine a nominal signal DP model by making likelihood scans with various combinations of isobars.
We start from a baseline model that includes $\fI$, $\chiczero$, and NR components. We then add another scalar resonance described by the RBW parametrization.
We scan the likelihood by fixing the width and mass of this additional resonance at several consecutive values, for each of which
the fit to the data is repeated.
All isobar magnitudes and phases are floating in these fits.
From the scans we observe a significant improvement of the fit around a width and mass that are compatible with the values of the $\fV$ resonance~\cite{Amsler:2008zz}.
After adding the $\fV$ to the nominal model we repeat the same procedure for an additional tensor particle. We find that the $\fVII$ has a significant contribution. The results of the likelihood scans are shown in Fig.~\ref{fig:ModelScans} in terms of $-2 \Delta {\ln}{\cal L}=-2\rm\lnL-(-2\lnL)_{min}$, where $(-2\lnL)_{\rm min}$ corresponds to the minimal value obtained in the particular scan. To conclude the search for possible resonant contributions we add all well established resonances~\cite{Amsler:2008zz} and check if the likelihood increases. We do not find any other significant resonant contribution but as we cannot exclude small contributions from the $f_{0}(1370)$, $f_{2}(1270)$, $f_{2}^{'}(1525)$, $a_{0}(1450)$, and $f_{0}(1500)$ resonances, we assign model uncertainties (see Sec.~\ref{sec:dp_systematics}) due to not taking these resonances into account.        

\subsection{Results}
\label{sec:fitresults}

\begin{figure*}[htbp]
\includegraphics[width=8.0cm,keepaspectratio]{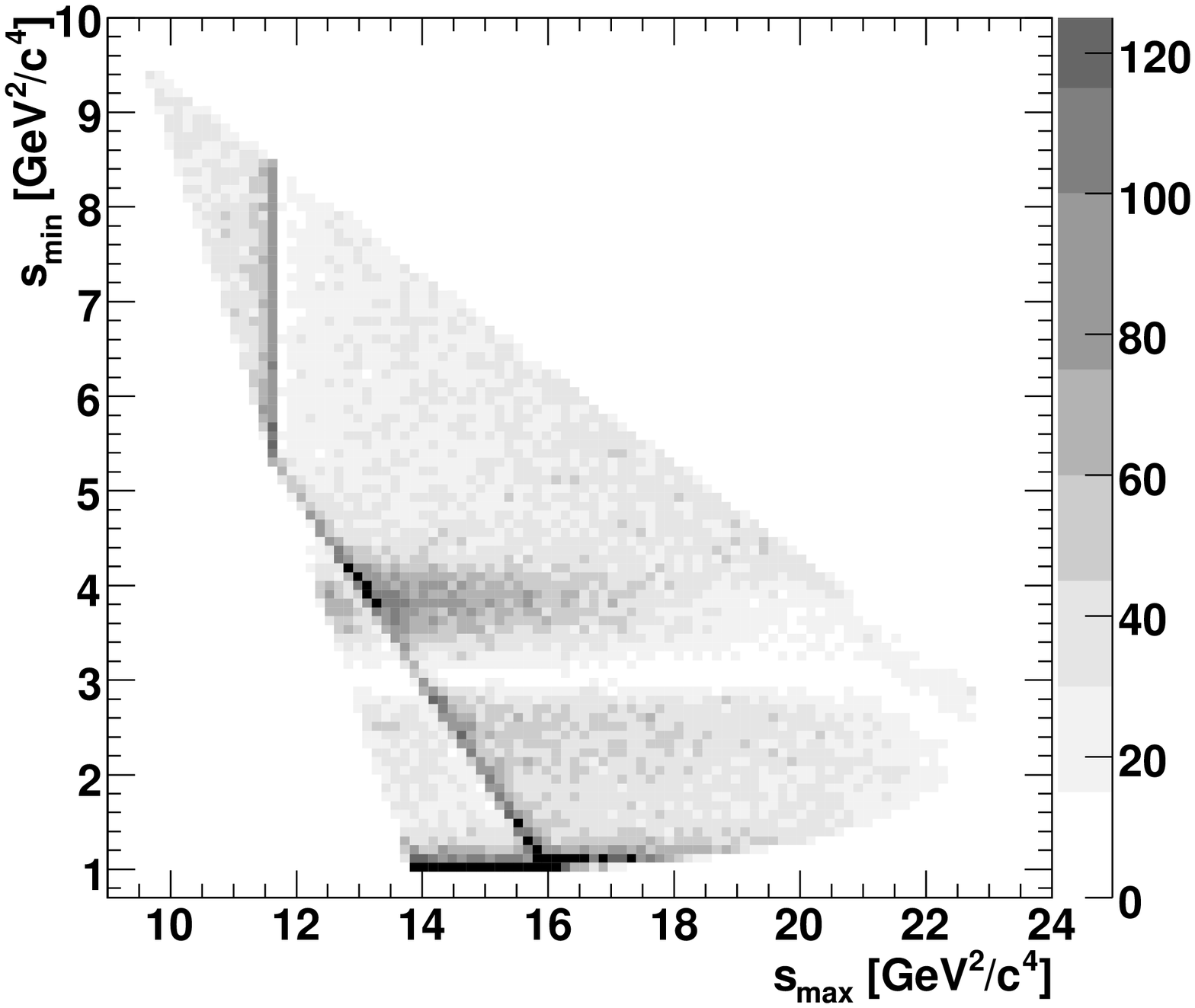}
\includegraphics[width=8.0cm,keepaspectratio]{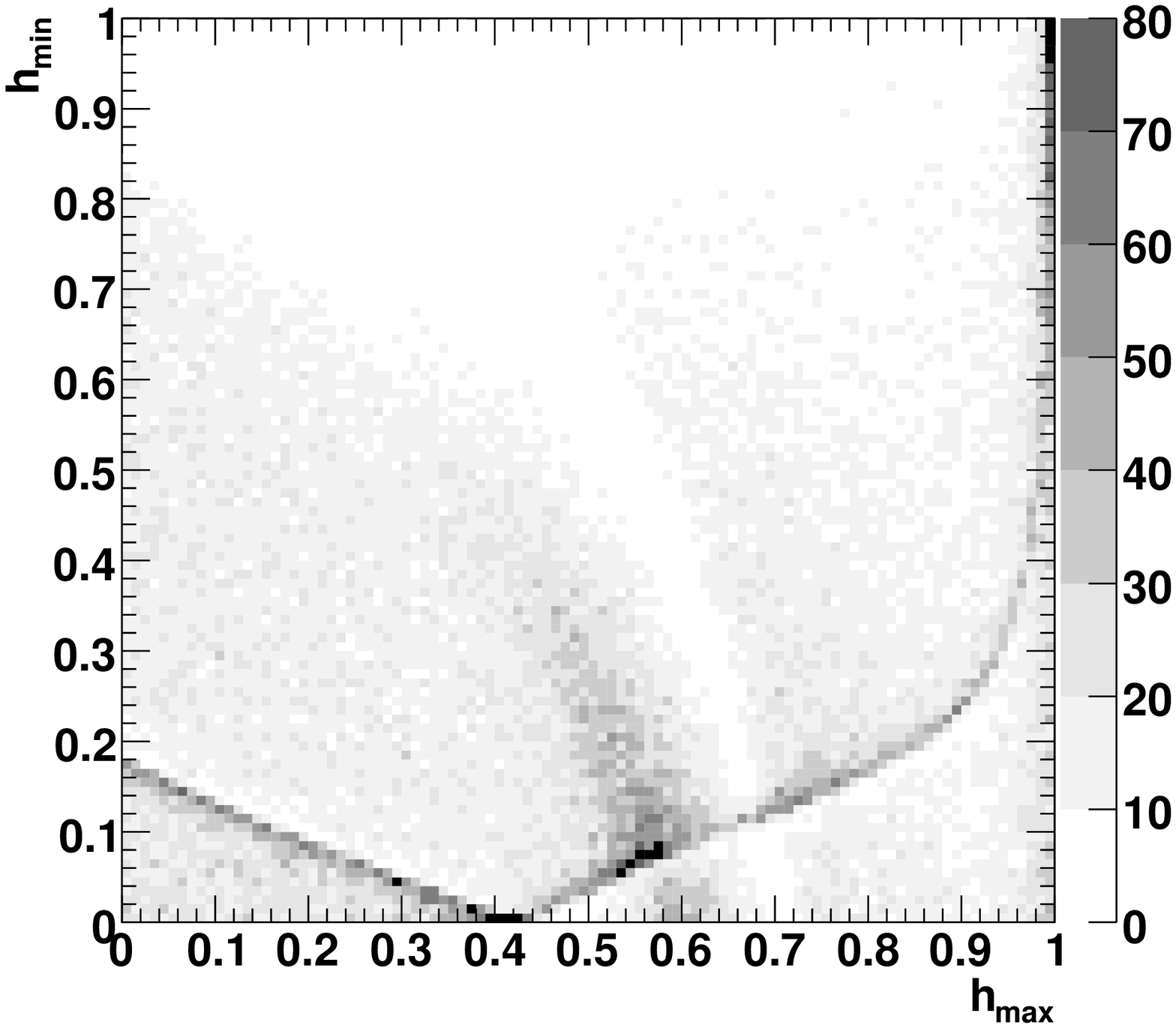}
\caption{
%(color online).
Symmetrized (left) and square (right) DP for MC simulated signal events using the amplitudes obtained from the fit to data. The low population in bins along the edge of the symmetrized DP is due to the fact that the phase space boundaries do not coincide with the histogram bin boundaries.}
\label{fig:MCsignal}
\end{figure*}

\begin{figure*}[htbp]
\includegraphics[width=8.0cm,keepaspectratio]{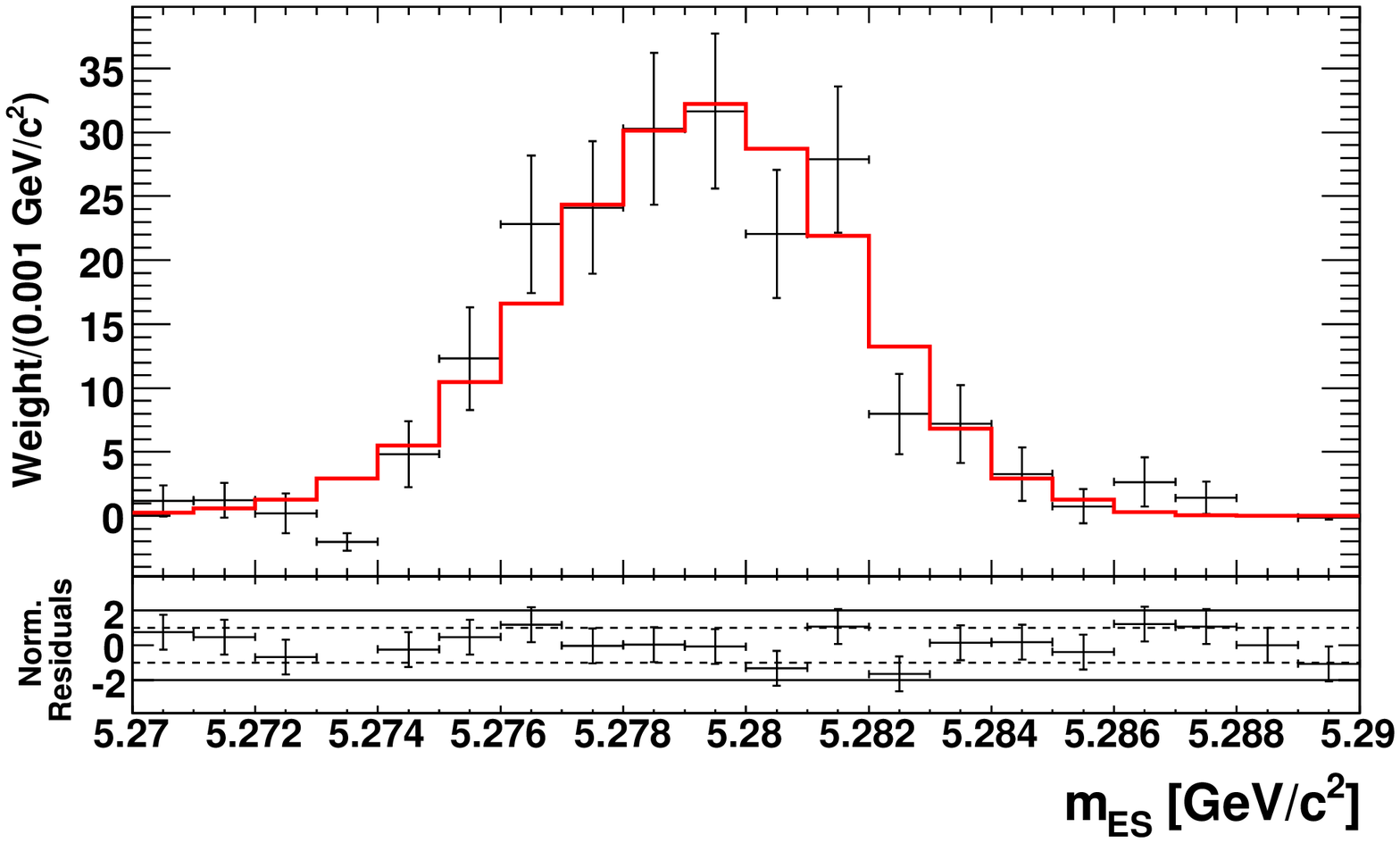}
\includegraphics[width=8.0cm,keepaspectratio]{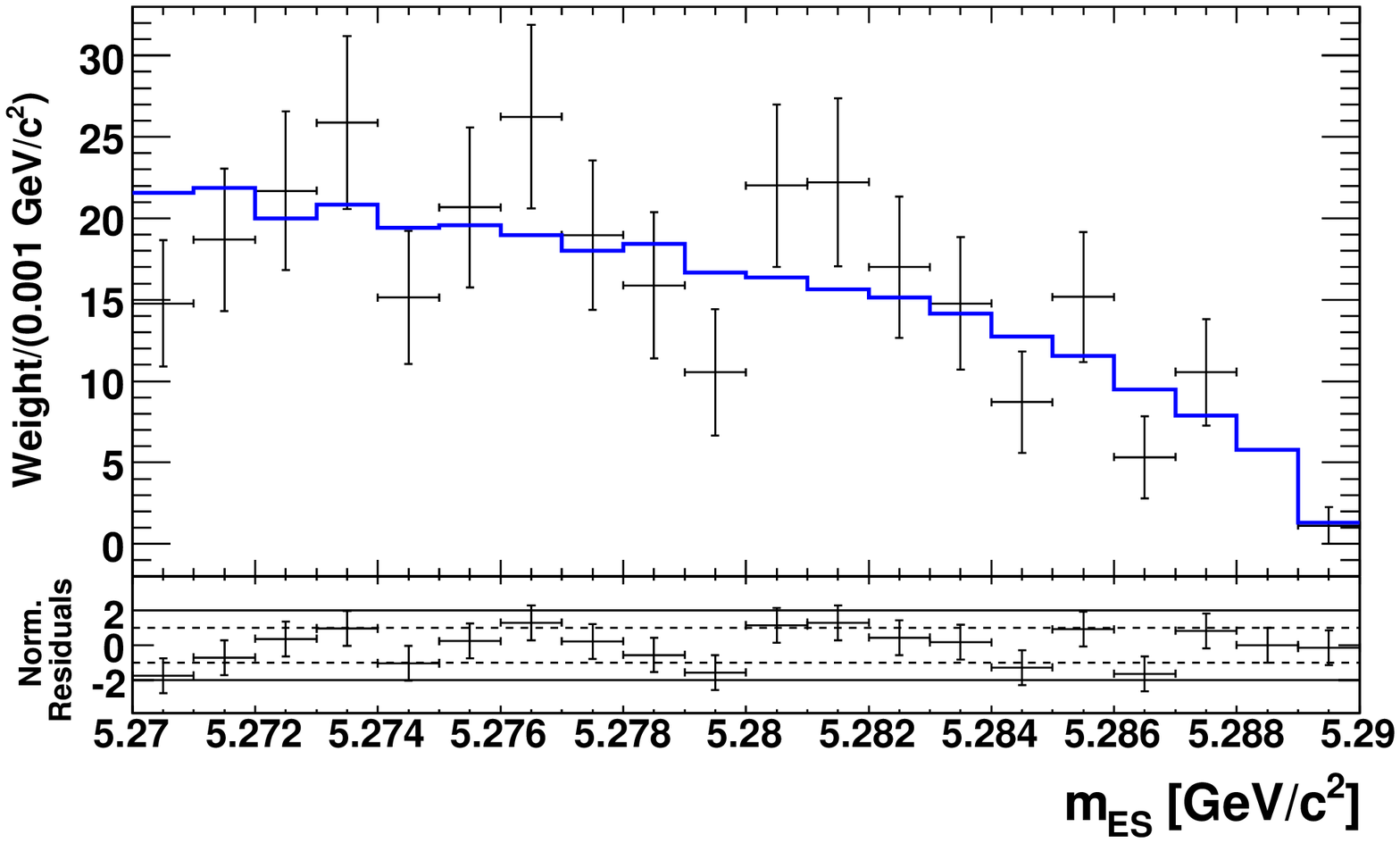}
\includegraphics[width=8.0cm,keepaspectratio]{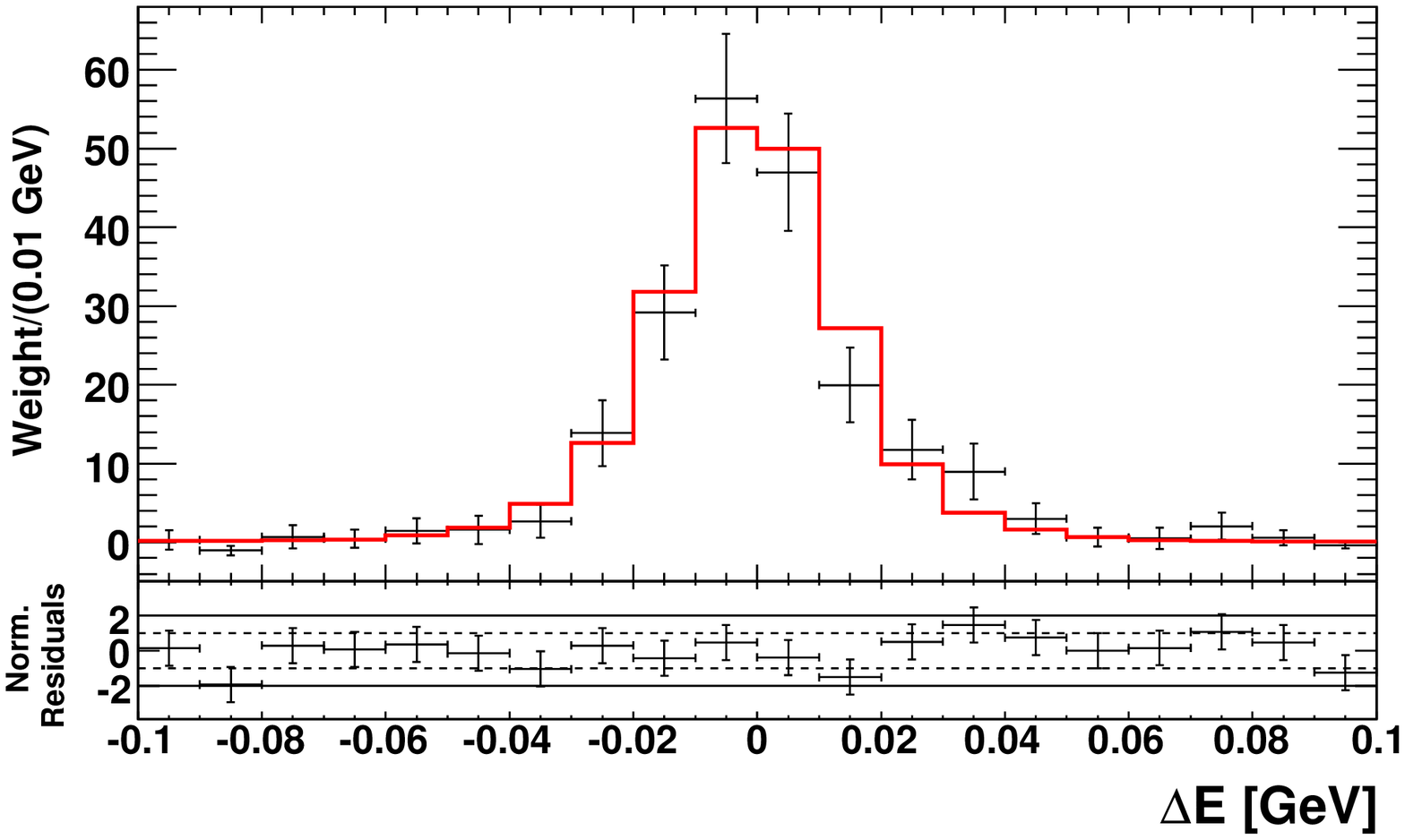}
\includegraphics[width=8.0cm,keepaspectratio]{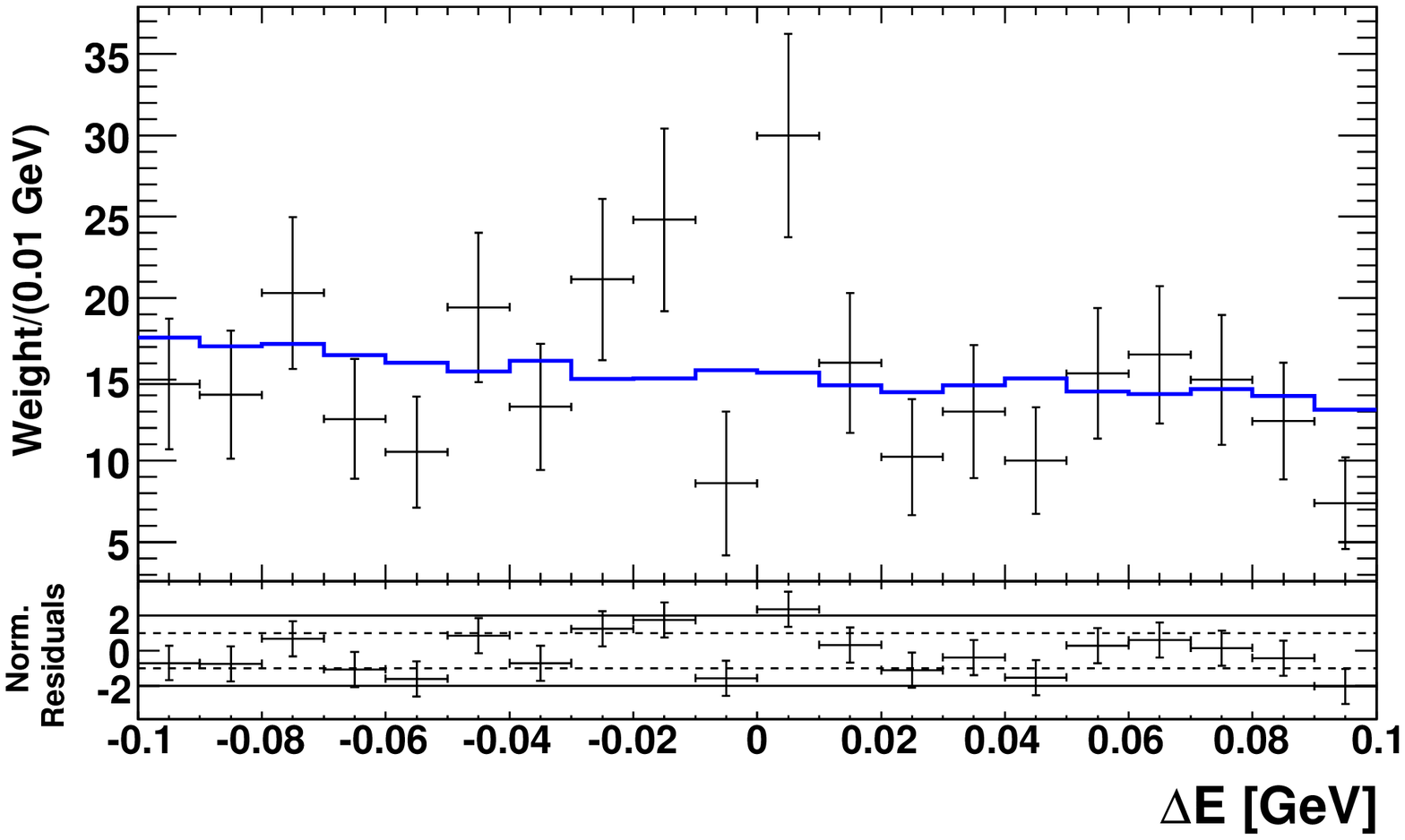}
\includegraphics[width=8.0cm,keepaspectratio]{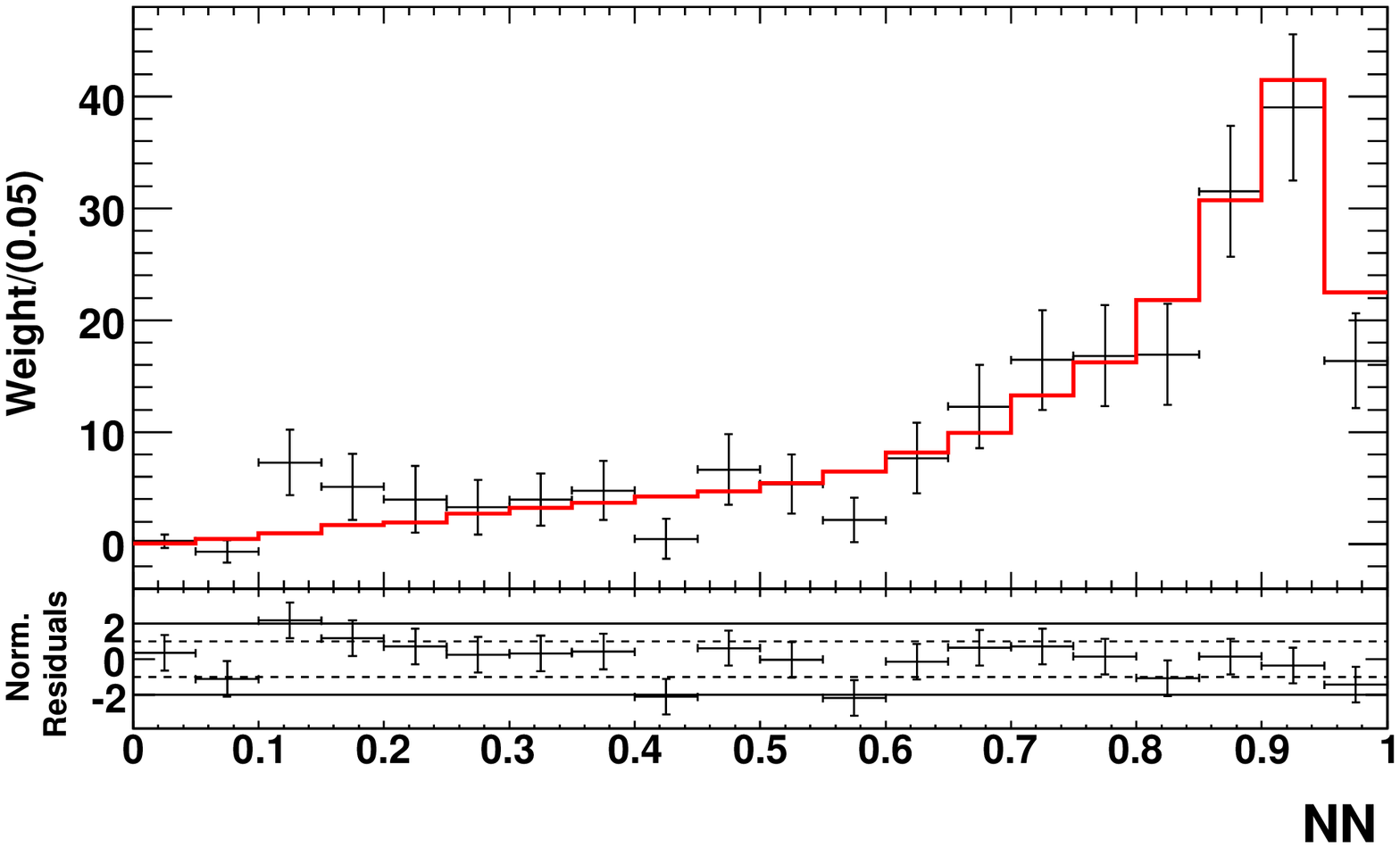}
\includegraphics[width=8.0cm,keepaspectratio]{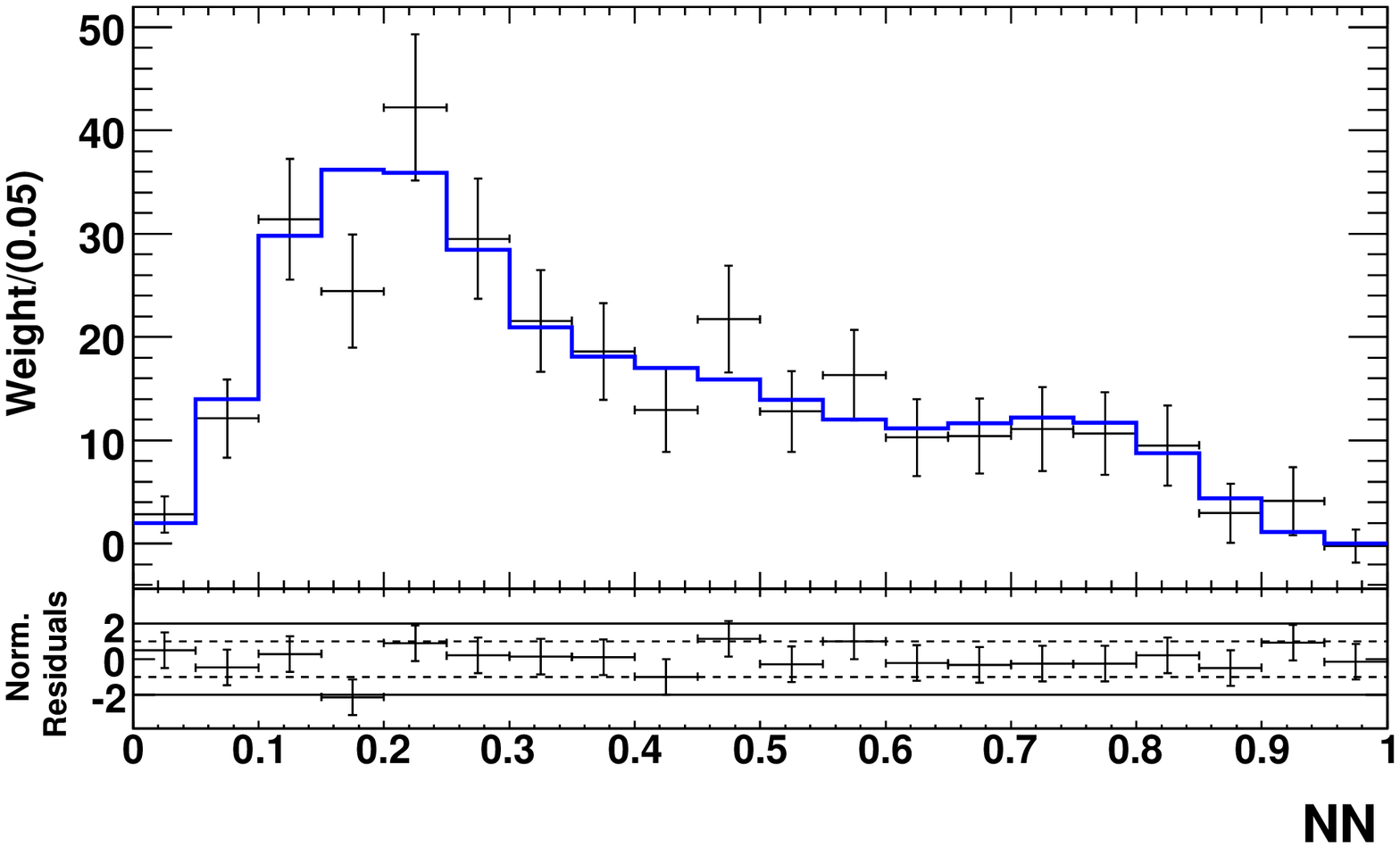}
\caption{(color online). {\ensuremath{\hbox{$_s$}{\cal P}lots}} (points with error bars) and PDFs (histograms) of the discriminating variables: $\mes$ (top), $\de$ (middle), and NN (bottom), for signal events (left) and  continuum events (right).
Below each bin are shown the residuals, normalized in error units. The horizontal dotted and full lines mark the one- and two-standard deviation levels, respectively.
}
\label{fig:sPlots}
\end{figure*}

\begin{figure*}[htbp]
\includegraphics[width=8.0cm,keepaspectratio]{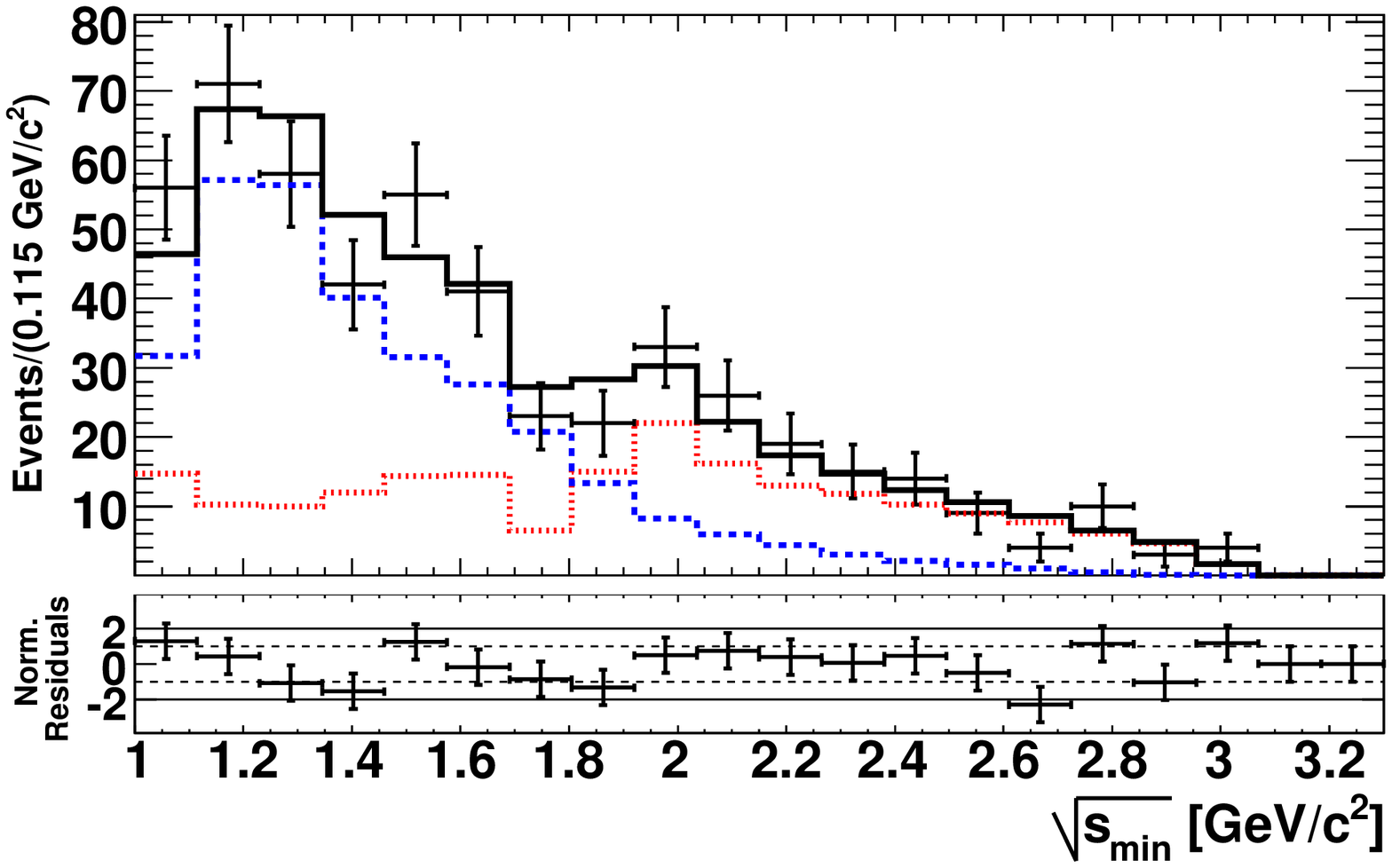}
\includegraphics[width=8.0cm,keepaspectratio]{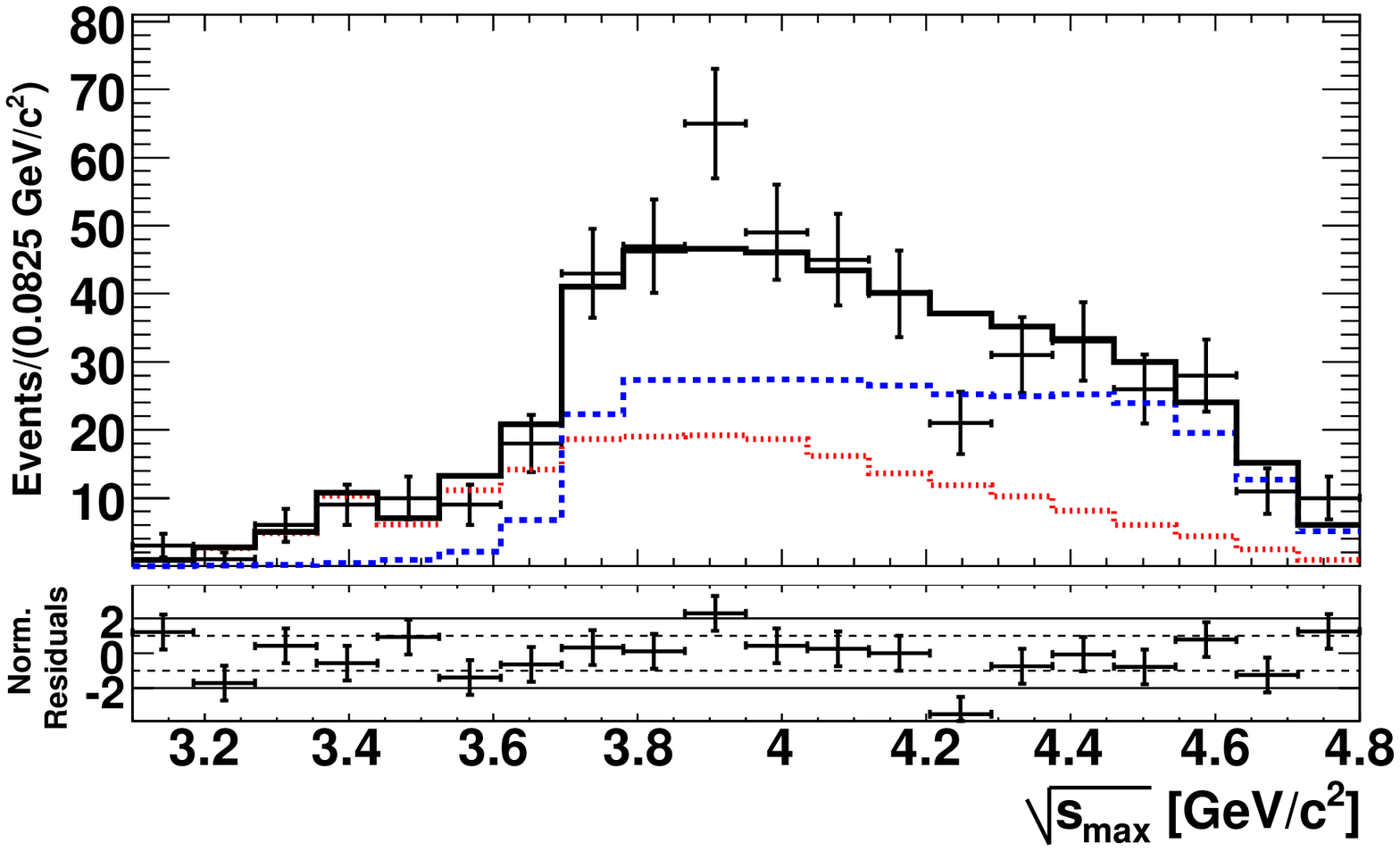}
\caption{(color online). Projections onto $\sqrt{\smin}$ (left) and $\sqrt{\smax}$ (right). On-resonance data are shown as points with error bars while the dashed (dotted) histogram represents the signal (continuum) component.  The solid-line histogram is the total PDF. Below each bin are shown the residuals, normalized in error units. The horizontal dotted and full lines mark the one- and two-standard deviation levels, respectively.}
\label{fig:projPlots}
\end{figure*}

The maximum-likelihood fit of $505$ candidates results in a $\Bz\to\KS\KS\KS$ event yield of $200\pm 15$ and a continuum yield of $305\pm 18$, where the uncertainties are statistical only.
The symmetrized and square Dalitz plots of a signal DP-model MC sample generated with the result of the fit to data are shown in Fig.~\ref{fig:MCsignal}.
Figure~\ref{fig:sPlots} shows plots of $\de$, $\mes$, and the NN for isolated signal and continuum background events obtained by the {\ensuremath{\hbox{$_s$}{\cal P}lots}}~\cite{Pivk:2004ty} technique.
Figure~\ref{fig:projPlots} shows projections of the data onto the invariant masses $\smin$ and $\smax$.

When the fit is repeated with initial parameter values randomly chosen within wide ranges above and below the nominal values for the magnitudes and within the $[-\pi,\,\pi]$ interval for the phases, we observe convergence towards two solutions with minimum values of the negative log likelihood function $-2\lnL$ separated by $3.25$ units. In the following, we refer to them as
Solution 1 (the global minimum) and Solution 2 (a local minimum). No other local minima were found.

\begin{figure*}[htbp]
\includegraphics[width=7.64cm,keepaspectratio]{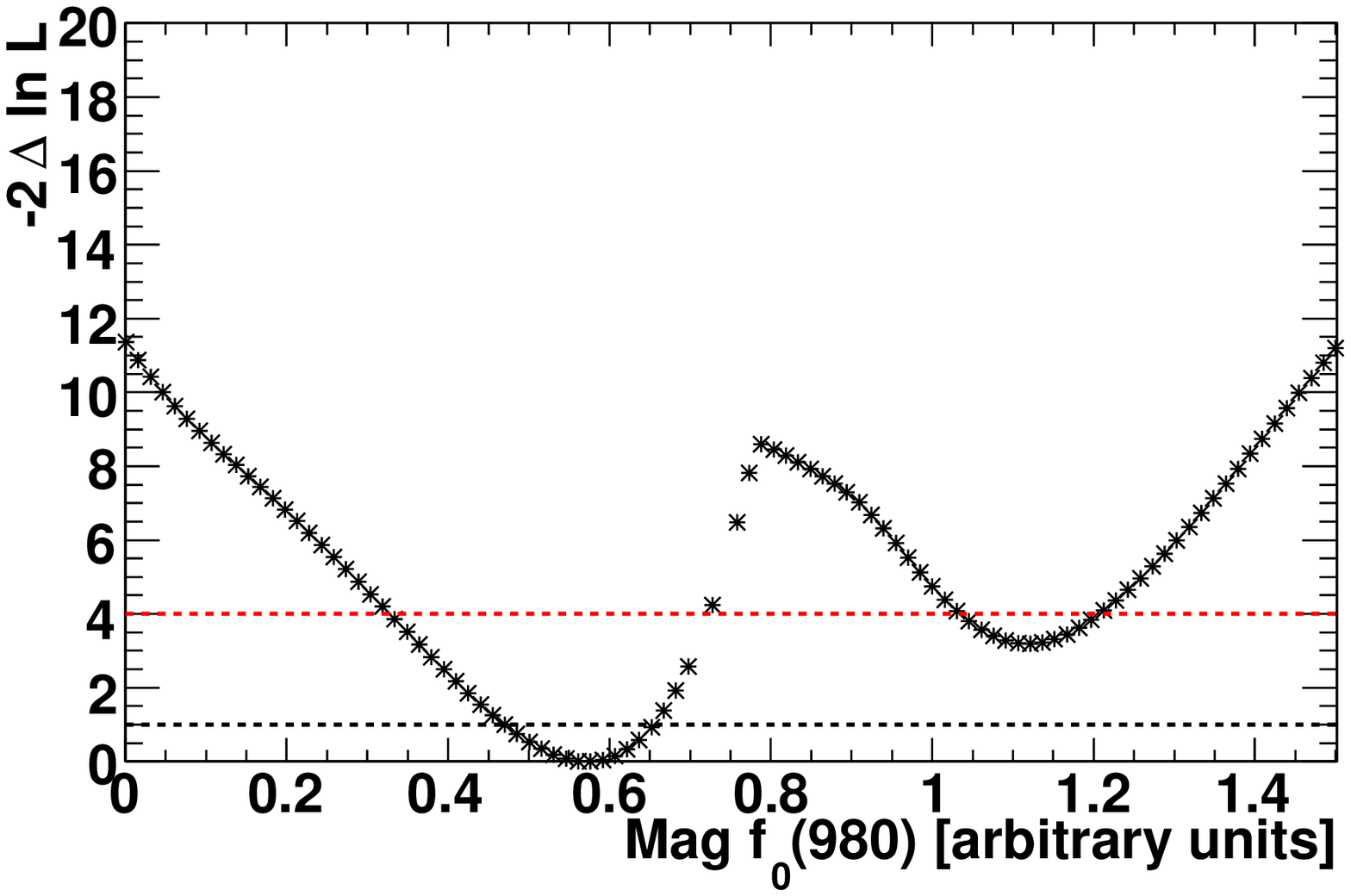}
\includegraphics[width=7.64cm,keepaspectratio]{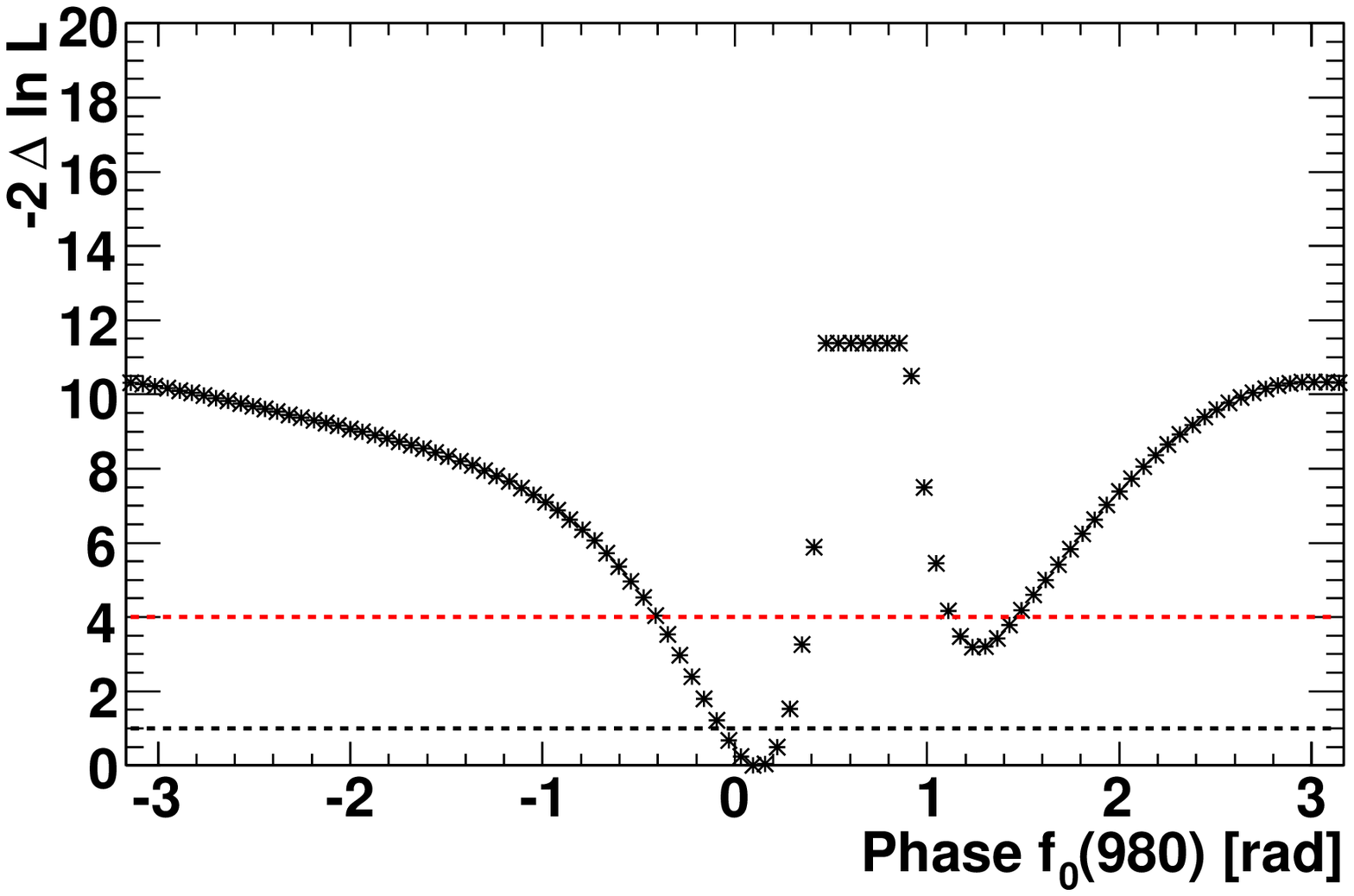}
\includegraphics[width=7.64cm,keepaspectratio]{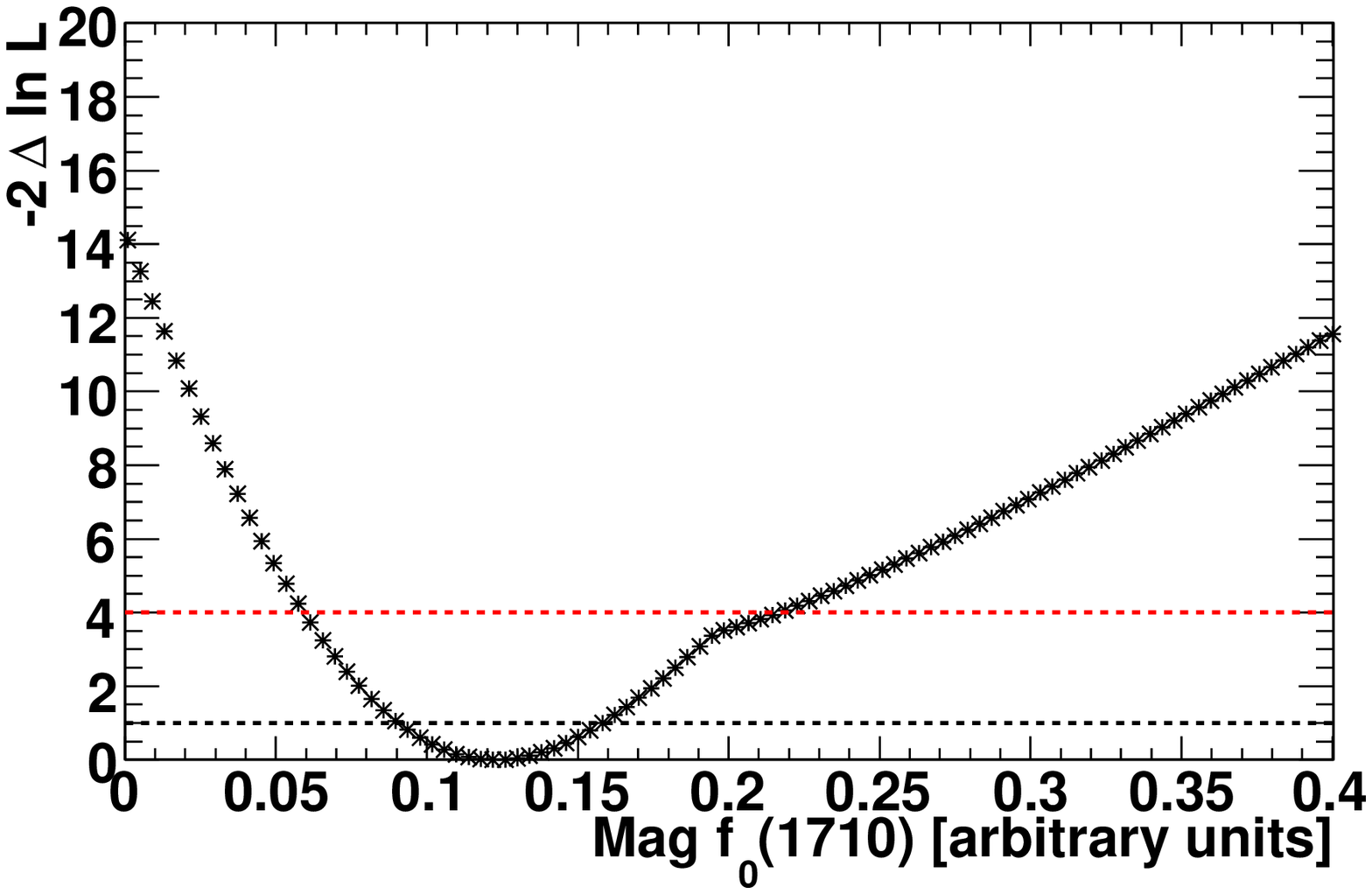}
\includegraphics[width=7.64cm,keepaspectratio]{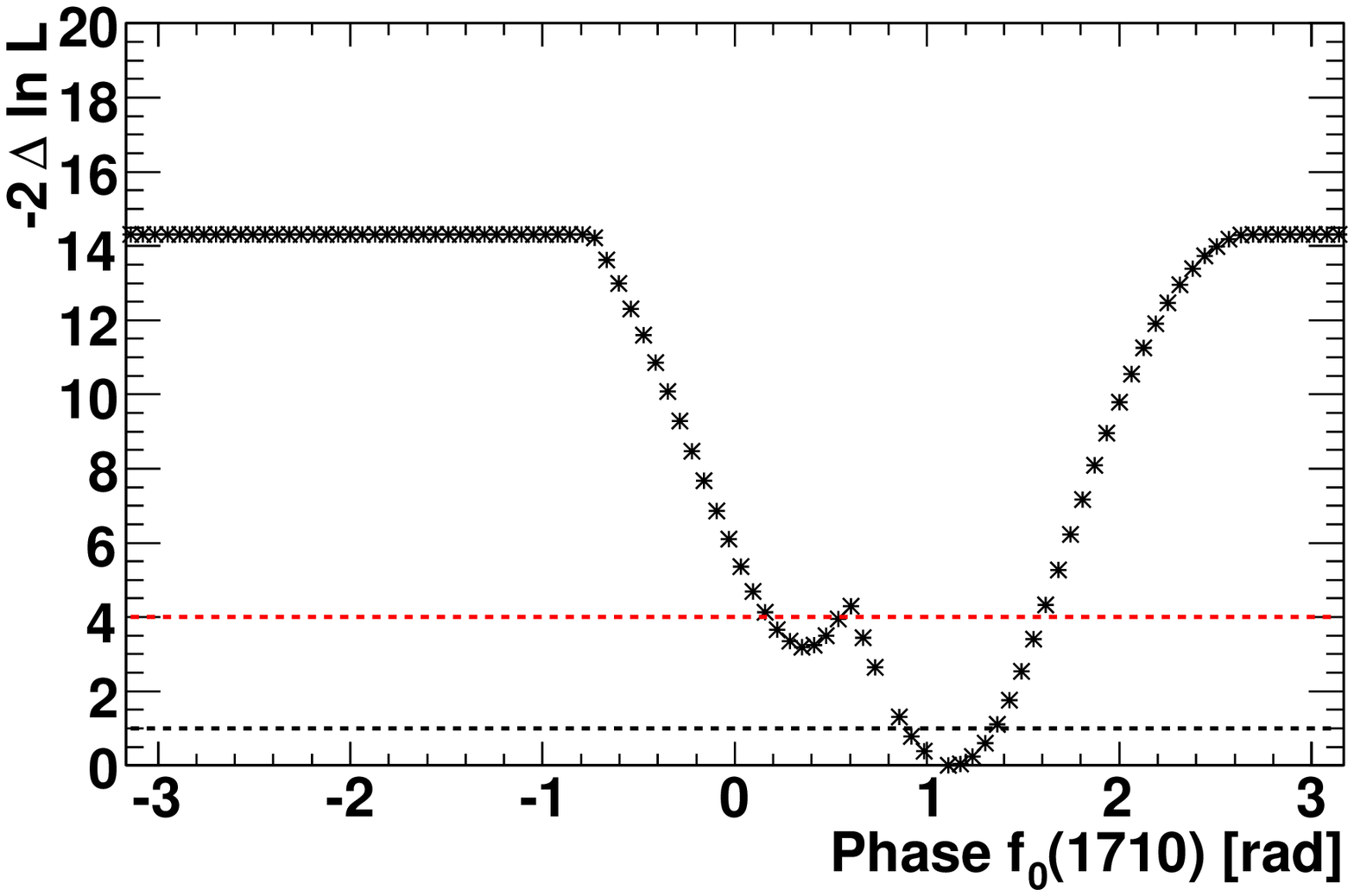}
\includegraphics[width=7.64cm,keepaspectratio]{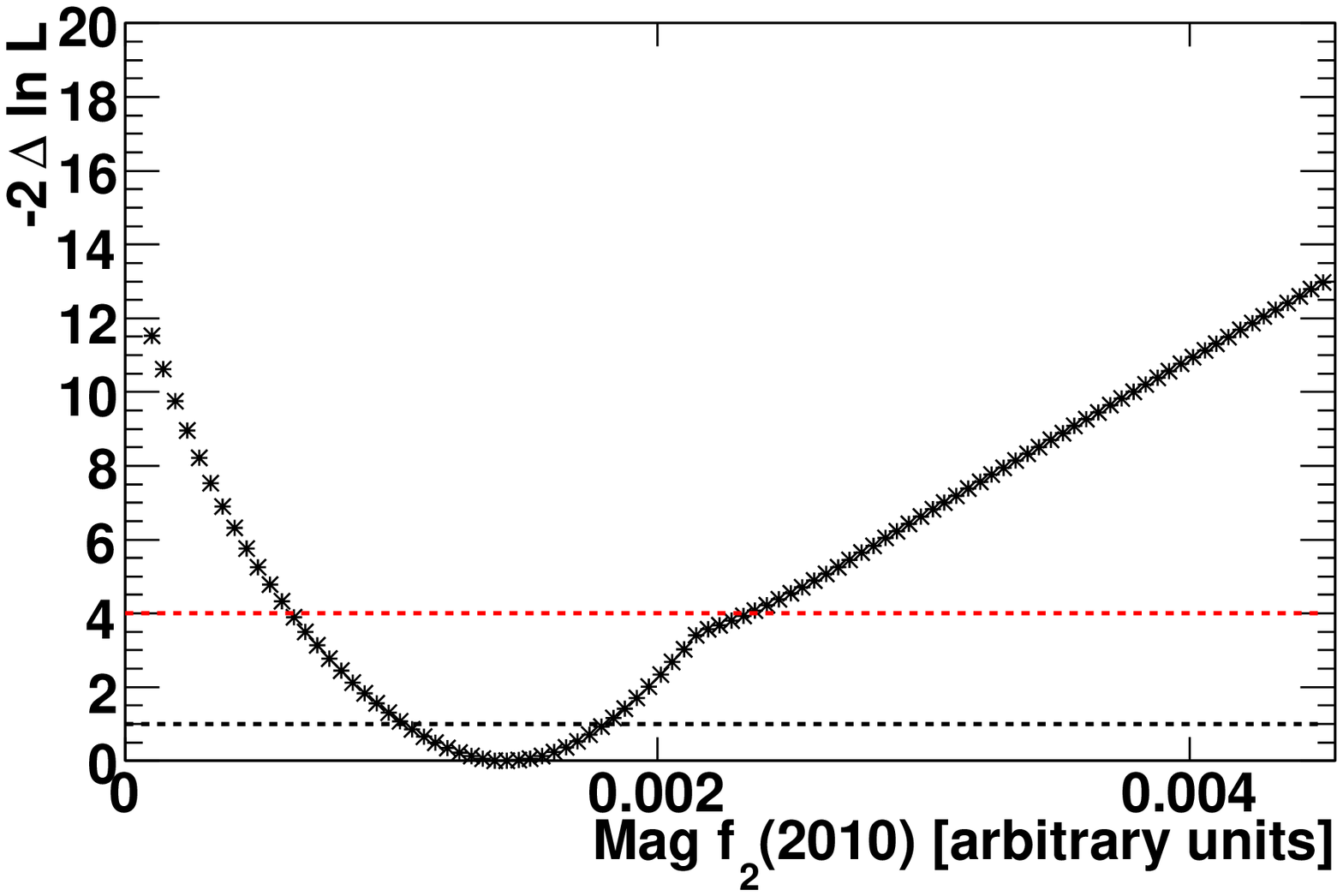}
\includegraphics[width=7.64cm,keepaspectratio]{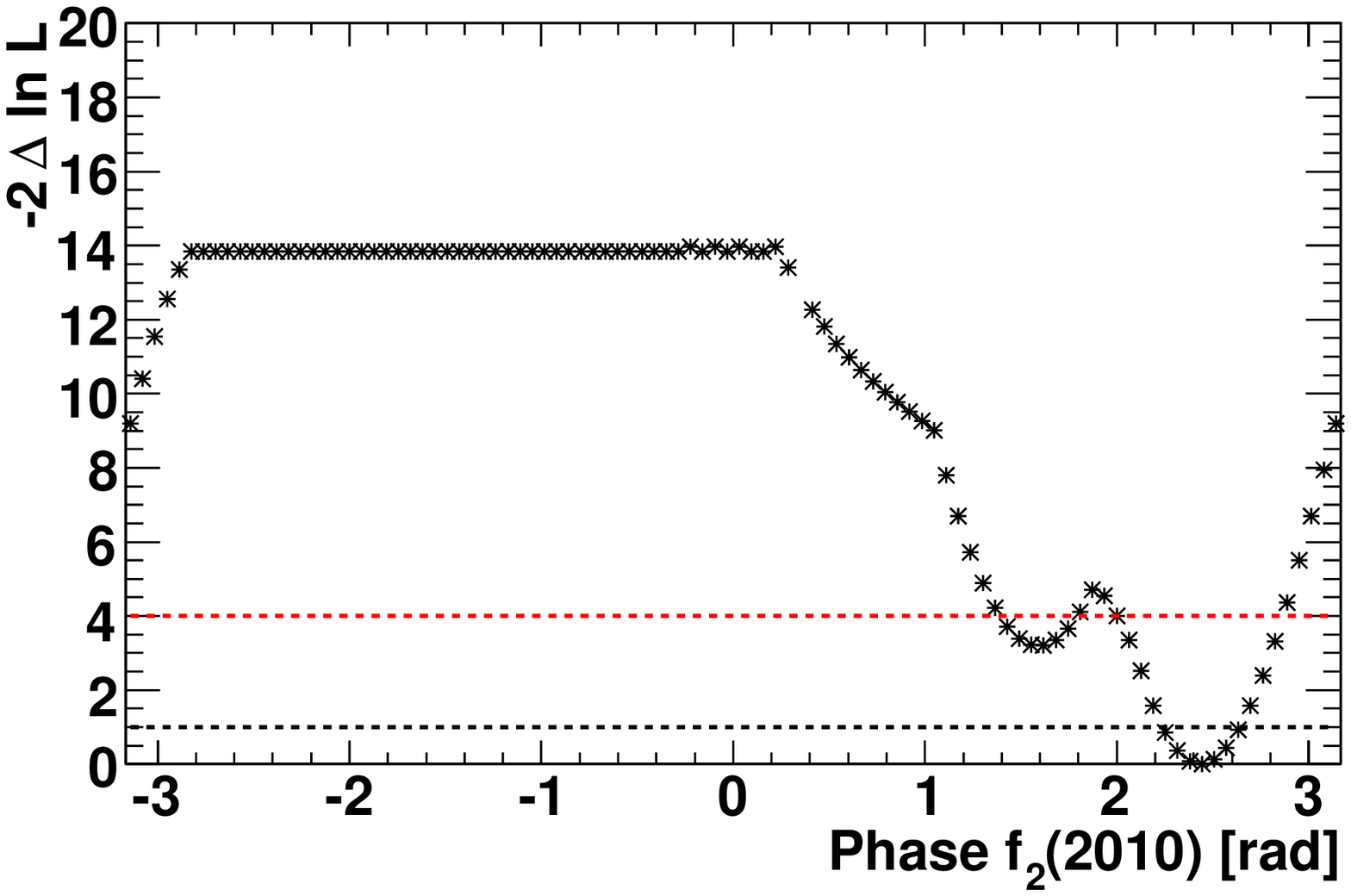}
\includegraphics[width=7.64cm,keepaspectratio]{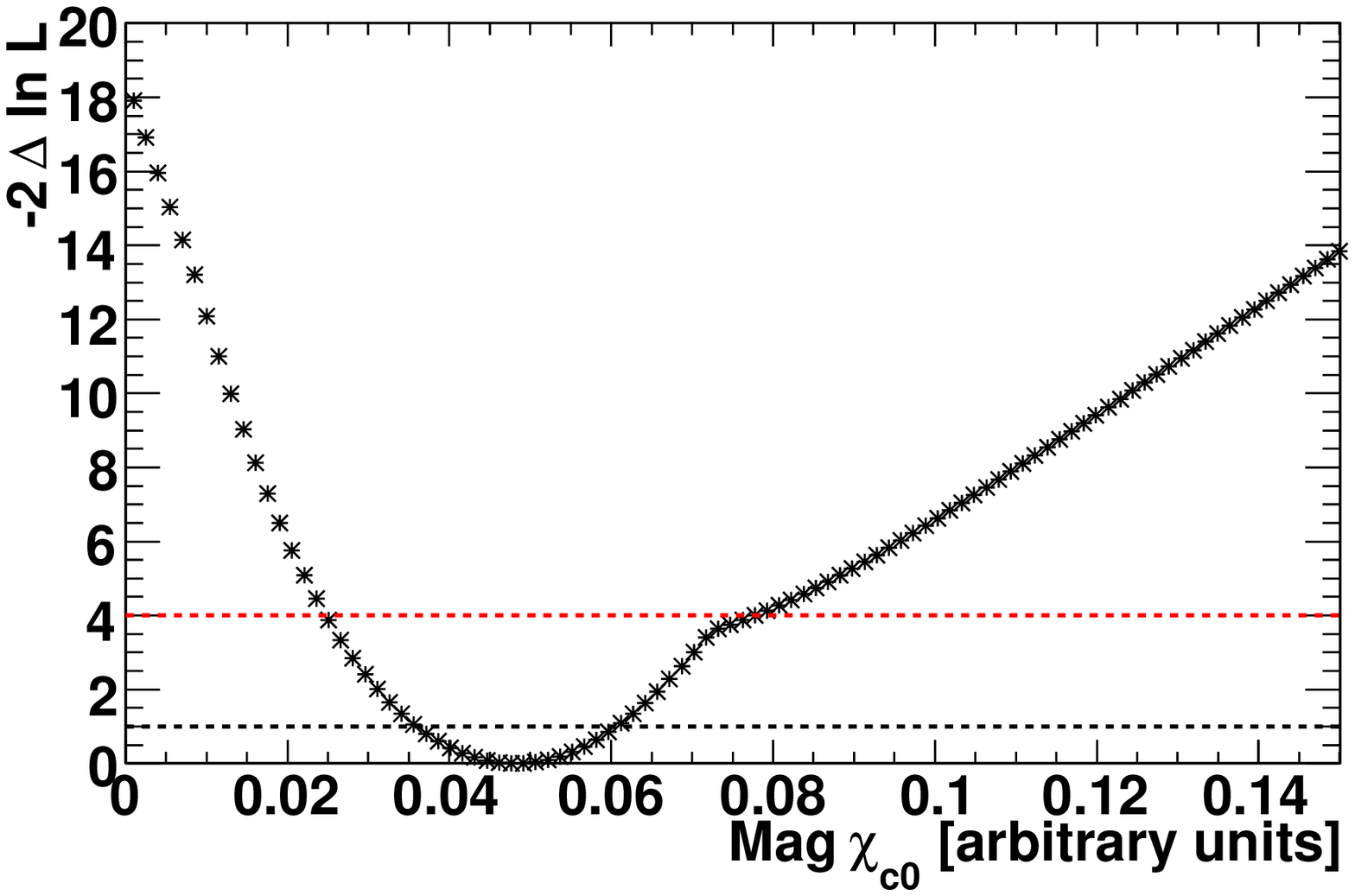}
\includegraphics[width=7.64cm,keepaspectratio]{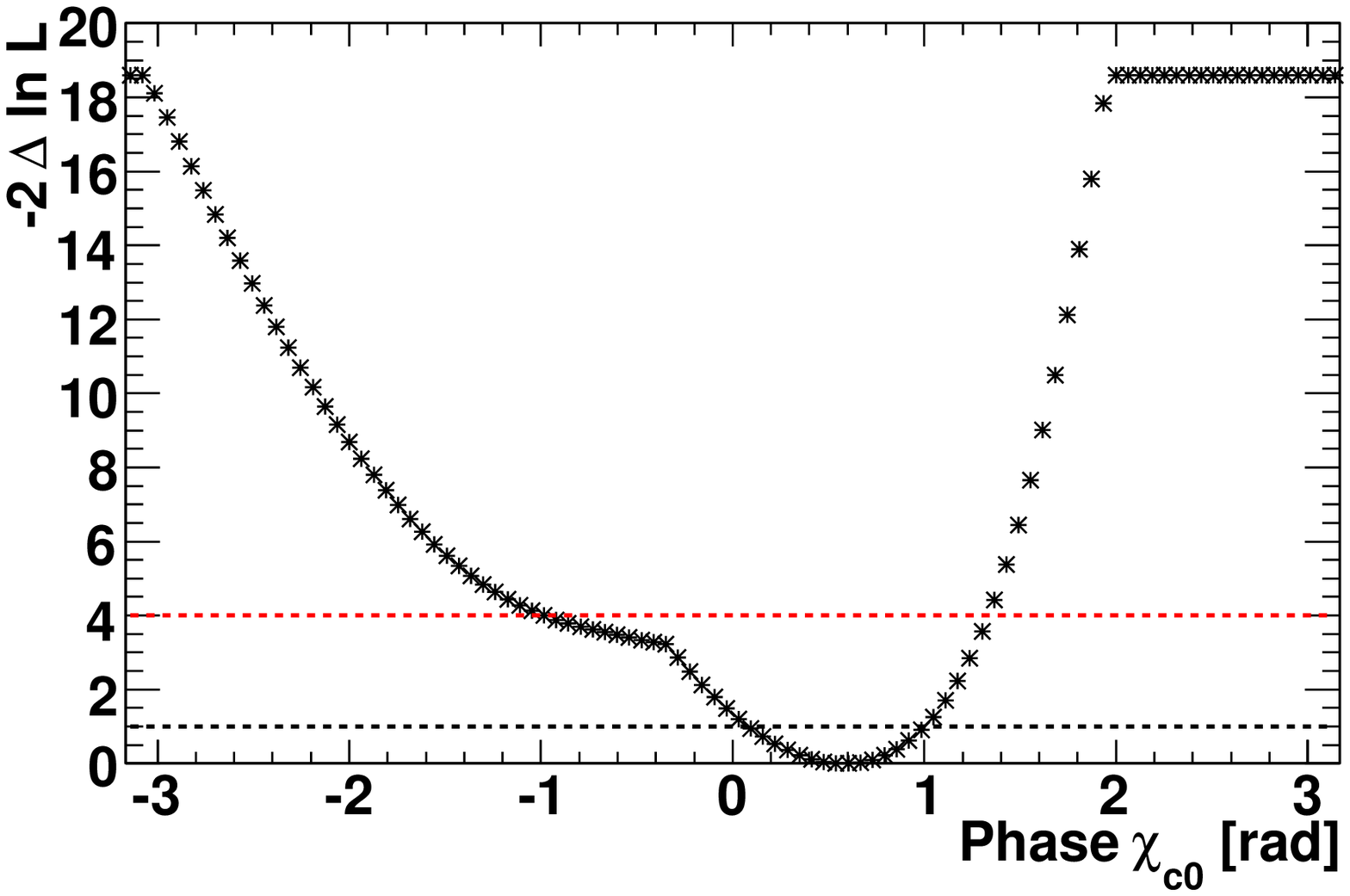}
\caption{
(color online). One-dimensional scans of $-2 \Delta \lnL$ as a function of magnitudes (left) and phases (right) of the resonances $\fI$, $\fV$, $\fVII$, and $\chiczero$ (top to bottom). The horizontal dashed lines mark the one- and two-standard deviation levels.}
\label{fig:ScanMagf0}
\end{figure*}

In the fit, we measure directly the relative magnitudes and phases of the different components of the signal model. The magnitude and phase of the NR amplitude are fixed to $1$ and $0$, respectively, as a reference.
In Fig.~\ref{fig:ScanMagf0} we show likelihood scans of the isobar magnitudes and phases of all the resonances, where both solutions can be noticed. Each of these scans is obtained by fixing the corresponding isobar parameter at several consecutive values, for each of which
the fit to the data is repeated.
The measured relative amplitudes $c_{\mu}$
are used to extract the fit fraction defined as
\begin{equation}
\label{eq:ff}
FF(k) = \frac{\sum_{\mu=3k -2}^{3k}\sum_{\nu=3k -2}^{3k}{c_{\mu}c^*_{\nu}\langle F_{\mu}F^*_{\nu}
\rangle}}{\sum_{\mu \nu}{c_{\mu}c^*_{\nu}\langle F_{\mu}F^*_{\nu}
\rangle}},
\end{equation}
where $k$, which varies from 1 to 5, represents an intermediate state.
Each fit fraction is a sum of three identical contributions, one for each pair of \KS. The indices $\mu$ and $\nu$ run from 1 to 15, as each of the five resonances contributes to three pairs of \KS , which correspond to the three terms ($3k-2$, $3k-1$, and $3k$) in each sum in the numerator of Eq.~\eqref{eq:ff}.
The dynamical amplitudes $F$ are defined in Sec.~\ref{sec:kinematics} and the terms
\begin{equation}
\langle F_{\mu}F^*_{\nu} \rangle = \int\int{F_{\mu}F^*_{\nu}d \smin d \smax}
\end{equation}
are obtained by integration over the DP.
The total fit fraction is defined as the algebraic sum of all fit fractions. This quantity is not
necessarily unity due to the potential presence of net constructive or destructive interference.

In order to estimate the statistical significance of each resonance, we
evaluate the difference $\Delta \ln {\cal L}$ between the log-likelihood of the nominal fit and that of a fit where the magnitude of the amplitude  of the resonance is set to $0$ (this difference can be directly read from the likelihood scans as a function of magnitudes in Fig.~\ref{fig:ScanMagf0}). In this case the phase of the resonance becomes meaningless, and we therefore account for two degrees of freedom removed from the fit.
The value $2 \Delta \ln {\cal L}$ is used to evaluate the $p$-value for $2$ degrees of freedom;
we determine the equivalent one-dimensional significance from this $p$-value.

The results for the phase and the fit fraction are given in Table~\ref{tab:Q2B} for the
two solutions; the change in likelihood when the amplitude of the resonance is set to $0$ and the resulting statistical significance of each resonance is given for Solution 1.

As the fit fractions are not parameters of the PDF itself, their statistical errors are obtained from the $68.3\%$ coverage intervals of the fit-fraction distributions obtained from a large number of pseudoexperiments generated with the corresponding solution (1 or 2). As observed in other three-kaon modes~\cite{babarKKK,belleKKK,Aubert:2007sd,babarKKKs,belleKKKs}, the total FF significantly exceeds unity.
\begin{table}
\caption{
Summary of measurements of the quasi-two-body parameters. The quoted uncertainties are statistical only. 
The change in the log-likelihood ($-2 \Delta \lnL$) corresponds to the case where the magnitude of the amplitude of the resonance is set to 0. This number is used for the estimation of the statistical significance of each resonance.
}
\renewcommand{\arraystretch}{1.5}
\begin{tabular}{llcc}
\hline \hline
Mode          & Parameter               & Solution 1                & Solution 2\\ \hline
$f_0(980)\KS$ & FF                      & $0.44\,^{+0.20}_{-0.19}$  & $1.03\,^{+0.22}_{-0.17}$  \\
              & Phase [rad]             & $0.09  \pm  0.16$         & $1.26  \pm  0.17$      \\
              & $-2 \Delta \lnL$        & $11.7$                    &    -                   \\
              & Significance [$\sigma$] & $3.0$                     &    -                   \\ \hline
$f_0(1710)\KS$& FF                      & $0.07\,^{+0.07}_{-0.03} $ & $0.09\,^{+0.05}_{-0.02}$  \\
              & Phase [rad]             & $1.11  \pm  0.23$         & $0.36 \pm 0.20$        \\
              & $-2 \Delta \lnL$        & $14.2$                    &    -                   \\
              & Significance [$\sigma$] & $3.3$                     &    -                   \\ \hline
$f_2(2010)\KS$& FF                      & $0.09\,^{+0.03}_{-0.03}$  & $0.10 \pm 0.02$        \\
              & Phase [rad]             & $2.50 \pm 0.20$           & $1.58 \pm 0.22$        \\
              & $-2 \Delta \lnL$        & $14.0$                    &    -                   \\
              & Significance [$\sigma$] & $3.3$                     &    -                   \\ \hline
NR            & FF                      & $2.16\,^{+0.36}_{-0.37}$  & $1.37\,^{+0.26}_{-0.21}$  \\
              & Phase [rad]             & $0.0$                     & $0.0$                  \\
              & $-2 \Delta \lnL$        & $68.1$                    &    -                   \\
              & Significance [$\sigma$] & $8.0$                     &    -                   \\ \hline
$\chiczero\KS$& FF                      & $0.07\,^{+0.04}_{-0.02}$  & $ 0.07 \pm 0.02$       \\
              & Phase [rad]             & $0.63 \pm 0.47$           & $-0.24 \pm 0.52$       \\
              & $-2 \Delta \lnL$        & $18.5$                    &    -                   \\
              & Significance [$\sigma$] & $3.9$                     &    -                   \\ \hline
              & Total FF                & $2.84\,^{+0.71}_{-0.66}$  & $2.66\,^{+0.35}_{-0.27}$ \\ \hline \hline
\end {tabular}
\renewcommand{\arraystretch}{1.0}
\label{tab:Q2B}
\end{table}

In Table~\ref{tab:Q2B} it can be seen that the two solutions differ mostly in the fraction assigned to the NR and the $\fI$ components.
Solution 1 corresponds to a small FF of the $\fI$ and a large value for the NR, and Solution 2 has a large $\fI$ fraction and a smaller NR fraction. Other three-kaon modes~\cite{babarKKK,belleKKK,Aubert:2007sd,babarKKKs,belleKKKs} favor the behavior of Solution 1.

Generalizing Eq.~\eqref{eq:ff}, we obtain the interference fractions among the intermediate decay modes $k$ and $j$
\begin{equation}
FF(k,j) = \frac{\sum_{\mu=3k -2}^{3k}\sum_{\nu=3j-2}^{3j}{c_{\mu}c^*_{\nu}\langle F_{\mu}F^*_{\nu}
\rangle}}{\sum_{\mu \nu}{c_{\mu}c^*_{\nu}\langle F_{\mu}F^*_{\nu}
\rangle}},
\end{equation}
which are given in Table~\ref{tab:ff_interference} for Solution 1. Unlike the total FF defined above, the elements of this matrix sum to unity. The large destructive interference between the $\fI\KS$ and the NR components appears clearly in the table.
This is possible due to the large overlap in phase space between the exponential NR term and the broad tail of the $\fI$ resonance above the $KK$ threshold.
\begin{table*}[htbp]
	\centering
	\caption{The interference fractions $FF(k,j)$ among the intermediate decay amplitudes for Solution 1. Note that the diagonal elements are those defined in Eq.~\eqref{eq:ff} and detailed in Table~\ref{tab:Q2B}. The lower diagonal elements are omitted since the matrix is symmetric.}
		\begin{tabular}{l|ccccc}
\hline\hline
           & $\fI\KS$ & $\fV\KS$ & $\fVII\KS$ &  NR       & $\chiczero\KS$ \\ \hline
$\fI\KS$   & $0.44$   & $0.07$   &  $-0.02$   & $-0.80$   &$\ \ 0.01$    \\
$\fV\KS$   &          & $0.07$   &  $-0.01$   & $-0.17$   & $-0.0003$    \\
$\fVII\KS$ &          &          & $\ \ 0.09$ & $\ \ 0.02$&$\ \ 0.0002$  \\
NR         &          &          &            & $\ \ 2.16$& $-0.02$      \\
$\chiczero\KS$ &      &          &            &           &$\ \ 0.07$    \\
\hline\hline
		\end{tabular}
	\label{tab:ff_interference}
\end{table*}

Using the relative fit fractions, we calculate the branching fraction $\mathcal B$ for the intermediate mode $k$ as
        \begin{equation}
        \label{eq:BF_definition}
        FF(k)\times \mathcal{B}(\Bz\to\KS\KS\KS ),
        \end{equation}
where $\mathcal{B}(\Bz\to\KS\KS\KS )$ is the total inclusive branching fraction:
        \begin{equation}
        \label{total_inclusive_BF}
        \mathcal{B}(\Bz\to\KS\KS\KS)=\frac{N_{\rm sig}}{\bar{\varepsilon} N_{B\bar{B}}}.
        \end{equation}
We estimate the average efficiency $\bar{\varepsilon}=6.6\%$ using a fully reconstructed DP-model MC sample generated with the parameters found in data.
The results of the branching fraction measurements are shown in Table~\ref{tab:BR}. As a cross check we attempt to compare our measured branching fractions to results from other measurements; however, many of the branching fractions for the decay into kaons of the resonances included in our model are not (or are only poorly) measured (marked as ``seen" in Ref.~\cite{Amsler:2008zz}). An exception is the charmonium state $\chiczero$, for which the measured value is ${\cal B}(\chiczero\to\KS\KS)=(3.16 \pm 0.18)\times 10^{-3}$~\cite{Amsler:2008zz}. We can then use the \babar\ measurement of ${\cal B}(\Bz\to\chiczero\Kz)=(142\,^{+55}_{-44} \pm 8 \pm 16 \pm 12)\times 10^{-6}$~\cite{Aubert:2009me} to calculate ${\cal B}[\Bz\to\chiczero(\to\KS\KS)\KS]=\frac{1}{2}\,{\cal B}(\Bz\to\chiczero\Kz)\times{\cal B}(\chiczero\to\KS\KS)=(0.224 \pm 0.078)\times 10^{-6}$, which is consistent with our measured branching fraction, given in Table~\ref{tab:BR}.

An interesting conclusion from this first amplitude analysis of the $\Bz \to \KS\KS\KS$ decay mode is that we do not need to include a broad scalar $f_X(1500)$ resonance, as has been done in other measurements~\cite{babarKKK,belleKKK,Aubert:2007sd,babarKKKs,belleKKKs}, to describe the data. The peak in the invariant mass between $1.5$ and $1.6\gevcc$ can be described by the interference between the $f_0(1710)$ resonance and the nonresonant component. However minor contributions from the $f_{2}^{'}(1525)$ and $f_{0}(1500)$ resonances to this structure cannot be excluded.
\begin{table}[htbp]
\caption{Summary of measurements of branching fractions (${\cal B}$). The quoted numbers are obtained by multiplying the corresponding fit fraction from Solution 1 by the measured inclusive $\Bz\to\KS\KS\KS$ branching fraction. The first uncertainty is statistical, the second is systematic, and the third represents the signal DP-model dependence.}
\renewcommand{\arraystretch}{1.5}
\begin{tabular}{lc}
\hline \hline
Mode & ${\cal B}$ [$\times 10^{-6}$] \\ \hline
Inclusive $\Bz\to\KS\KS\KS$ & $6.19 \pm 0.48 \pm 0.15 \pm 0.12$ \\ \hline
$f_0(980)\KS$, $~~f_0(980)\to\KS\KS$& $2.7\,^{+1.3}_{-1.2} \pm 0.4 \pm 1.2$ \\
$f_0(1710)\KS$, $~~f_0(1710)\to\KS\KS$& $0.50\,^{+0.46}_{-0.24} \pm 0.04 \pm 0.10$ \\
$f_2(2010)\KS$, $~~f_2(2010)\to\KS\KS$& $0.54\,^{+0.21}_{-0.20} \pm 0.03 \pm 0.52$ \\
%${\rm NR}$, $~~\KS\KS\KS$& $13.32\,^{+2.23}_{-2.30} \pm 0.55 \pm 2.13$ \\
${\rm NR}$, $~~\KS\KS\KS$& $13.3\,^{+2.2}_{-2.3} \pm 0.6 \pm 2.1$ \\
$\chiczero\KS$, $~\chiczero\to\KS\KS$& $0.46\,^{+0.25}_{-0.17} \pm 0.02 \pm 0.21$ \\ \hline \hline
\end {tabular}
\renewcommand{\arraystretch}{1.0}
\label{tab:BR}
\end{table}
\subsection{Systematic uncertainties}
\label{sec:dp_systematics}
Systematic effects are divided into model and experimental uncertainties. Details on how they have been estimated are given below and the associated numerical values are summarized in Table~\ref{tab:systematics}.

\subsubsection{Model uncertainties}
We vary the mass, width, and any other parameter of all isobar fit components within their errors, as quoted in Table~\ref{tab:model}, and assign the observed differences in our observables as the first part of the model uncertainty (``Model'' in Table~\ref{tab:systematics}).
To estimate the contribution to $\Bz\to\KS\KS\KS$ from resonances that are not included in our signal model but cannot be excluded statistically, namely the $f_{0}(1370)$, $f_{2}(1270)$, $f_{2}^{'}(1525)$, $a_{0}(1450)$, and $f_{0}(1500)$ resonances, we perform fits to pseudoexperiments that include these resonances. The masses and the widths are taken from~\cite{Amsler:2008zz}, except for the $f_{0}(1370)$ for which we take the values from~\cite{pappagallo}. We generate pseudoexperiments with the additional resonances, where the isobar magnitudes and phases have been determined in fits to data, and fit these datasets with the nominal model. We assign the induced shift in the observables as a second part of the model uncertainty. 

\subsubsection{Experimental systematic uncertainties}
To validate the analysis procedure, we perform fits on a large number of pseudoexperiments generated with the measured yields of signal events and continuum background. The signal events are taken from fully reconstructed MC that has been generated with the fit result to data. We observe small biases in the isobar magnitudes and phases. We correct for these biases by shifting the values of the parameters and assign to this procedure a systematic uncertainty, which corresponds to half the correction combined in quadrature with its error. This uncertainty accounts also for correlations between the signal variables, wrongly reconstructed events, and effects due to the limited sample size (``Fit Bias'' in Table~\ref{tab:systematics}).

From MC we estimate that there are six \B\ background events in our data sample. To determine the bias introduced by these events, we add \B\ background events from MC to our data sample, and fit it with the nominal model. We then assign the observed differences in the observables as a systematic uncertainty (``\B-bkg'' in Table~\ref{tab:systematics}).
We assign a systematic uncertainty for all fixed PDF parameters by varying them within their uncertainties according to the covariance matrix.
%where we take into account the correlations between these parameters.

We vary the histogram PDFs, i.e., the SDP PDF for continuum and the NN PDF for signal (``Discr.~Vars'' in Table~\ref{tab:systematics}).
The \mes dependence of the SDP PDF for continuum was found to be negligible.
We account for differences between simulation and data observed in the control sample $\Bz\to\jpsi\KS$ (``MC-Data'' in Table~\ref{tab:systematics}).
These differences were estimated by propagating the differences, in the control sample, between background-subtracted data and signal MC, into the fit PDFs.

For the branching fraction measurement, we assign a systematic uncertainty due to the error on the calculation of $N_{B\bar B}$ (``$N_{B \bar B}$'' in Table~\ref{tab:systematics}) and to the $\KS$ reconstruction efficiency. We correct the $\KS$ reconstruction efficiency by the difference between the efficiency found in a dedicated \KS\ data sample and that found in simulation.
We assign the uncertainty on the correction as a systematic error (``\KS\ reco'' in Table~\ref{tab:systematics}). 
   
\begin{table*}[htbp]
\caption{Summary of systematic uncertainties. The model uncertainty is dominated by the variation of the line shapes due to the contribution of the poorly measured $f_{2}(2010)$.}
\begin{tabular}{lccccccccc}
\hline \hline
Parameter                   &Fit bias&\B-bkg & Discr.~Vars &MC-Data&$N_{B \bar B}$&$\KS$ reco& Sum   & Model \\
${\cal B}(\Bz\to\KS\KS\KS)\,[10^{-6}]$ & $0.011$  & $0.030$ & $0.053$      & $0.015$ & $0.067$      & $0.111$  & $0.145$ & $0.120$ \\ \hline
FF \fI                      & $0.013$  & $0.056$ & $0.006$      & $0.001$ &  -           &  -       & $0.058$ & $0.190$ \\
FF \fV                      & $0.007$  & $0.001$ & $0.001$      & $0.001$ &  -           &  -       & $0.007$ & $0.016$ \\
FF $\fVII$                  & $0.005$  & $0.001$ & $0.003$      & $0.001$ &  -           &  -       & $0.006$ & $0.084$ \\
FF NR                       & $0.024$  & $0.083$ & $0.023$      & $0.001$ &  -           &  -       & $0.090$ & $0.344$ \\
FF $\chiczero$              & $0.002$  & $0.000$ & $0.001$      & $0.000$ &  -           &  -       & $0.002$ & $0.034$ \\
Ph [rad] \fI                & $0.008$  & $0.018$ & $0.014$      & $0.000$ &  -           &  -       & $0.024$ & $0.177$ \\
Ph [rad] \fV                & $0.011$  & $0.020$ & $0.001$      & $0.003$ &  -           &  -       & $0.023$ & $0.185$ \\
Ph [rad] \fVII              & $0.044$  & $0.014$ & $0.004$      & $0.002$ &  -           &  -       & $0.046$ & $0.684$ \\
Ph [rad] $\chiczero$        & $0.039$  & $0.011$ & $0.010$      & $0.007$ &  -           &  -       & $0.042$ & $0.498$ \\ \hline\hline
\end {tabular}   
\label{tab:systematics}    
\end{table*}

\section{Time-dependent analysis}
\label{sec:TD_analysis}

In Sec.~\ref{sec:dt} we describe the proper-time distribution used to extract the time-dependent \CP\ asymmetries.
In Sec.~\ref{sec:td_selection} we explain the selection
requirements used to obtain the signal candidates and
suppress backgrounds. In Sec.~\ref{sec:td_likelihood} we describe the fit method
and the approach used to account for experimental effects.
In Sec.~\ref{sec:TD_fitresults} we present the results of the fit and finally, in Sec.~\ref{sec:td_systematics} we discuss systematic uncertainties in the results.

\subsection{Proper-time distribution}
\label{sec:dt}

The time-dependent \CP\ asymmetries are functions of the proper-time difference $\dt =t_{\CP} - t_{\rm tag}$ between a fully reconstructed \Bz\to\KS\KS\KS\ decay ($B_{\CP}$) and the other $B$ meson decay in the event ($B_{\rm tag}$), which is partially reconstructed. The observed decay rate is the physical decay rate modified to include tagging imperfections, namely ${\langle D \rangle}_c$ and $\Delta D_c$; the former
is the rate of correctly assigning the flavor of the \B\ meson, averaged over \Bz\ and \Bzb,
and the latter is the difference between $D_c$ for \Bz\ and \Bzb.
The index $c$ denotes different quality categories of the tag-flavor assignment.
Furthermore the decay rate is convolved with the per-event $\dt$ resolution ${\cal R}_{\rm sig}(\dt,\sigma_{\dt})$, which is described by the sum of three Gaussians and depends on $\dt$ and its error $\rm \sigma_{\dt}$.
For an event $i$ with tag flavor $q_{\rm tag}$, one has
        \begin{eqnarray}
        \label{eq:dtmeas}
      \lefteqn{{\cal P}^i_{\rm sig}(\Delta t,\sigma_{\dt};q_{\rm tag},c) =}\\
\nonumber & & \frac{e^{-\left|\dt\right|/\tau_{\Bz}}}{4\tau_{\Bz}} \biggl\{ 1 + q_{\rm tag}\frac{\Delta D_{c}}{2} \\
\nonumber & & + q_{\rm tag}\langle D\rangle_{c} \Bigl[{\cal S}\sin(\Delta m_d\dt)-\;{\cal C}\cos(\Delta m_{d}\dt)\Bigr] \biggr\} \\
\nonumber & & \;\otimes\; {\cal R}_{\rm sig}(\dt,\sigma_{\dt}), \nonumber
        \end{eqnarray}
where $q_{\rm tag}$ is defined to be $+1$ ($-1$) for $B_{\rm tag}=\Bz$ ($B_{\rm tag}=\Bzb$), $\tau_{\Bz}$ is the mean \Bz\ lifetime, and $\Delta m_{d}$ is the mixing frequency~\cite{Amsler:2008zz}. The widths of the $\Bz$ and the $\Bzb$ are assumed to be the same.

\subsection{Event selection and backgrounds}
\label{sec:td_selection}

We reconstruct $\Bz\to\KS\KS\KS$ candidates either from three $\KS\to\pip\pim$ candidates, or from two $\KS\to\pip\pim$ and one $\KS\to\piz\piz$, where the $\piz$ candidates are formed from pairs of photons. The vertex fit requirements are the same as in the amplitude analysis, and also the requirement that the charged pions of at least one of the $\KS$ have hits in the two inner layers of the vertex tracker.
The \KS\ candidates in the \pippim\ submode must have mass within $12\mevcc$ of the nominal $\Kz$ mass~\cite{Amsler:2008zz} and decay length with respect to the $B$ vertex between $0.2$ and $40\cm$. 
In addition, combinatorial background is suppressed in both submodes by imposing that the angle between
the momentum vector of each $\KS(\pip\pim)$ candidate
and the vector connecting the beamspot and the $\KS(\pip\pim)$ vertex is smaller than 0.2~radians.
Each \KS\ decaying to charged pions in the \pizpiz\ submode
is required to have decay length  between $0.15$ and $60$~cm and $\pip\pim$ invariant mass less than $11\mev$ from the world average \KS\ mass~\cite{Amsler:2008zz}. The \KS\ decaying to neutral pions in the \pizpiz\ submode must have $\piz\piz$ invariant mass between $0.48$ and $0.52\gevcc$. Additionally, the neutral pions are selected if they have $\gamma\gamma$ invariant mass between $0.100$ and $0.141\gevcc$ and if the photons have energies greater than $50\mev$ in the laboratory frame and a lateral energy deposition profile in the electromagnetic calorimeter consistent with that expected for an electromagnetic shower (lateral moment~\cite{Drescher:1984rt} less than $0.55$). 
The fact that we do not model any PDF using sideband data allows a loose requirement on $\mes$ and $\de$ in the time-dependent analysis, namely, $5.22 < \mes <5.29\gevcc$ and $-0.18<\de<0.12\gev$.
In case of multiple candidates passing the selection, we proceed in the same way as in the amplitude analysis.
We use the same NN as in the amplitude analysis to suppress continuum background.  

With the above selection criteria, we  obtain signal reconstruction efficiencies of $6.7\%$ and $3.1\%$ for the \pippim\ and \pizpiz\ submodes, respectively. These efficiencies are determined from a DP-model MC sample generated using the results of the amplitude analysis.
We estimate from MC that $2.1\%$ of the selected signal events are misreconstructed for \pippim, while the figure is $2.4\%$ in \pizpiz,
and we do not treat these events differently from correctly reconstructed events.
Because of the looser requirements, there are more background events from \B decays than in the amplitude analysis, in particular in the \pizpiz\ submode. These backgrounds are included in the fit model and are summarized in Table~\ref{tab:bbackground}.
As the analysis is phase-space integrated, we cannot model the $\chiczero$ resonance separately, and its contribution to the \CP\ asymmetries could cloud deviations in the charmless contributions.
We therefore apply a veto around the invariant mass of this charmonium state.
\begin{table*}[htbp]
\begin{center}
\caption{ \label{tab:bbackground} Summary of \B background modes included in the fit model of the time-dependent analysis. The expected number of events takes into account the branching fractions (${\cal B}$) and efficiencies. In case there is no measurement, the branching fraction of an isospin-related channel is used. All the fixed yields are varied by $\pm 100\%$ for systematic uncertainties.}
\setlength{\tabcolsep}{0.0pc}
\begin{tabular*}{\textwidth}{@{\extracolsep{\fill}}lcccc}
\hline\hline
Submode      & Background mode           & Varied & ${\cal B}$ [$\times 10^{-6}$]  & Number of events \\ \hline\\[-9pt]
\pippim\     & $\KS\KS\KL$               & no     & $2.4$                   & $ 0.71  $ \\
             & $\KS\KS\Kstarz$           & no     & $27.5$                  & $ 9.55  $ \\
             & $\KS\KS\Kp$               & no     & $11.5$                  & $ 4.27  $ \\ 
             & $\Bz \rar \{\text{neutral generic decays}\}$    & yes  & not applicable             & $21.7$  \\
             & $\Bp \rar \{\text{charged generic decays}\}$    & yes  & not applicable             & $15.5$ \\
\hline\\[-9pt]
\pizpiz\     & $\KS\KS\KL$               & no     & $2.4 $ & $0.67  $ \\
             & $\KS\KS\Kstarz$           & no     & $27.5 $ & $5.3  $ \\
             & $\KS\KL\Kstarz$           & no     & $27.5 $ & $0.3  $ \\
             & $\KS\KS\Kp$               & no     & $11.5 $ & $2.9  $ \\
             & $\KS\KS\Kstarp$           & no     & $27.5 $ & $7.2  $ \\
             & $\Bz \rar \{\text{neutral generic decays}\}$    & yes  & not applicable             & $73.6$  \\
             & $\Bp \rar \{\text{charged generic decays}\}$    & yes  & not applicable             & $73.8$ \\
\hline\hline
\end{tabular*}

\end{center}
\end{table*}

\subsection{The maximum-likelihood fit}
\label{sec:td_likelihood}
We perform an unbinned extended maximum-likelihood fit to extract the $\Bz\to\KS\KS\KS$ event yields along with
the ${\cal S}$ and ${\cal C}$ parameters of the time-dependent analysis. 

The fit uses as variables $\mes$, $\de$, the NN output, $\dt$, and $\sigma_{\dt}$. The selected on-resonance data sample is assumed to consist of signal, continuum background, and backgrounds from $B$ decays. Wrongly reconstructed signal events are not considered separately. The likelihood function ${\cal L}_i$ for event $i$ is the sum 

\begin{equation}
{\cal L}_i = \sum_j{N_j {\cal P}^i_j(\mes,\Delta E,\Delta t,\sigma_{\dt},{\rm NN};q_{\rm tag},c,p)},
\end{equation}
where $j$ stands for the species (signal, continuum background, one for each $\B$ background category) and $N_j$ is the corresponding yield; $q_{\rm tag}$, $c$, and $p$ are the tag flavor, the tagging category, and the physics category, respectively.

To determine $q_{\rm tag}$ and $c$ we use the \B\ flavor-tagging
algorithm of Ref.~\cite{BabarS2b}. This algorithm combines several
different signatures, such as charges, momenta, and decay
angles of charged particles in the event to achieve optimal
separation between the two \B\ flavors. This produces six
mutually exclusive tagging categories.
We also retain untagged events in a seventh category;
although these events do not contribute to the measurement
of the time-dependent CP asymmetry they do provide
additional sensitivity for the measurement of direct \CP\
violation~\cite{Gardner:2003su}.

The two physics categories correspond to \pippim\ and \pizpiz\ decays.
The PDF for species $j$ evaluated for event $i$ is given by the product of individual PDFs:
\begin{align}
&{\cal P}^i_j(\mes,\Delta E,\Delta t,\sigma_{\dt},{\rm NN};q_{\rm tag},c,p) = \\
\nonumber & {\cal P}^i_j(\mes;p)\:{\cal P}^i_j(\Delta E;p)\:{\cal P}^i_j({\rm NN};c,p)\:{\cal P}^i_j(\Delta t,\sigma_{\dt};q_{\rm tag},c,p).
\end{align}

To take into account the different reconstruction of the two submodes,
we use separate PDFs for the two physics categories. Separate NN and
$\dt$ PDFs are included for each tagging category within each physics
category. The separate $\dt$ PDFs for the two physics categories allow
us to fit the
${\cal S}$ and ${\cal C}$ parameters either separately for the two submodes, or together.
The total likelihood is given by
\begin{equation}
\label{eq:td_Likelihood}
 {\cal L}=\exp(-\sum_j N_j)\prod_i {\cal L}_i.
\end{equation}

\subsubsection{$\Delta t$ PDFs}
\label{sec:td_likeDalitz}
The signal PDF for $\Delta t$ is given in Eq.~\eqref{eq:dtmeas}.
Parameters that depend solely on the tag side of the events (namely, ${\langle D \rangle}_c$ and $\Delta D_c$) are taken from the analysis of $\B \to c \bar c K^{(*)}$ decays~\cite{LastSin2betaBabar}. On the other hand, parameters that depend on the signal-side reconstruction, due to the absence of direct tracks from the \B decay, cannot be taken from modes that include such direct tracks. This is the case for the parameters that describe the resolution function, which are found in a fit to simulated events. A systematic uncertainty for data-MC differences is assigned using the control sample $\Bz\to\jpsi\KS$, as explained in Sec.~\ref{sec:td_systematics}.

For continuum events we use a zero-lifetime component. This parametrization is convolved with the
same resolution function as for signal, with different parameters that are varied in the fit to data.
The parameters of this PDF are not separated in the tagging categories.
The small contribution from $\epem \to \ccbar$ events is well described by the tails of the resolution function. 
For \B\ background events we use the signal PDF, with resolution parameters
from the \babar\ $\B \to c\bar c K^{(*)}$ analysis~\cite{LastSin2betaBabar}.
The parameters ${\cal S}$ and ${\cal C}$ are set to zero and varied
to assign a systematic uncertainty.
 
\subsubsection{Description of the other variables}
\label{sec:td_likeDiscrim}

The \mes\  and \DeltaE\ distributions of signal events are parametrized by an asymmetric Gaussian with power-law tails, as given in Eq.~\eqref{eq:cruijff}, and, for \mes, a small additional component, parameterized by an ARGUS shape function~\cite{argus}, to correctly describe misreconstructed events. The means in these two PDFs for \pippim\ events are allowed to vary in the fit to data, and the other parameters are taken from MC simulation.
For the NN distributions of signal we use histogram PDFs taken from MC simulation for each physics and tagging category.
The \mes, $\DeltaE$, and NN PDFs for continuum events are parametrized by an ARGUS shape function, a straight line and the sum of power functions from Eq.~\eqref{eq:spexp}, respectively. All continuum parameters, except for
$c_2$ and $c_3$
of the NN PDF, are allowed to vary in the fit. All the fixed parameters are varied, within the uncertainties found in a fit to sideband data, to estimate systematic errors.

All the $B$ background PDFs are described by fixed histograms taken from MC simulation. 
\subsection{Results}
\label{sec:TD_fitresults}

The maximum-likelihood fit of $3261$ candidates in the $\pippim$ submode and $7209$ candidates in the $\pizpiz$ submode results in the event yields detailed in Table~\ref{tab:td_yields}.
\begin{table}[htb]
\begin{center}
\caption{ \label{tab:td_yields} Event yields for the different event species, resulting from the maximum-likelihood fit for the time-dependent analysis.  ``$\Bp\Bm$ ($\Bz\Bzb$) bkg'' represents background from charged (neutral) \B decays. Quoted uncertainties are statistical only.}
\begin{tabular}{ lcc}
\hline \hline \noalign{\smallskip}
Species       & $\ 3\KS(\pip\pim)\ $     & $\ 2\KS(\pip\pim)\KS(\piz\piz)\ $\\ 
\noalign{\smallskip}\hline\noalign{\smallskip}
Signal        & $\rm 201\,^{+16}_{-15}$    & $\rm 62\,^{+13}_{-12}$ \\[0.5mm]
Continuum     & $\rm 3086\,^{+56}_{-54} $  & $\rm 7086\,^{+85}_{-83} $ \\[0.5mm]
$\Bp\Bm$ bkg  & $\rm -54\,^{+29}_{-24}$ &  $\rm 45\,^{+34}_{-30}$ \\[0.5mm]
$\Bz\Bzb$ bkg & $\rm 9\,^{+31}_{-30}$ & $\rm 4\,^{+38}_{-29}$ \\[0.7mm] \hline \hline
\end{tabular}             
\end{center}
\end{table}

\begin{figure*}[htbp]
\begin{center}
\includegraphics[width=7.0cm,keepaspectratio]{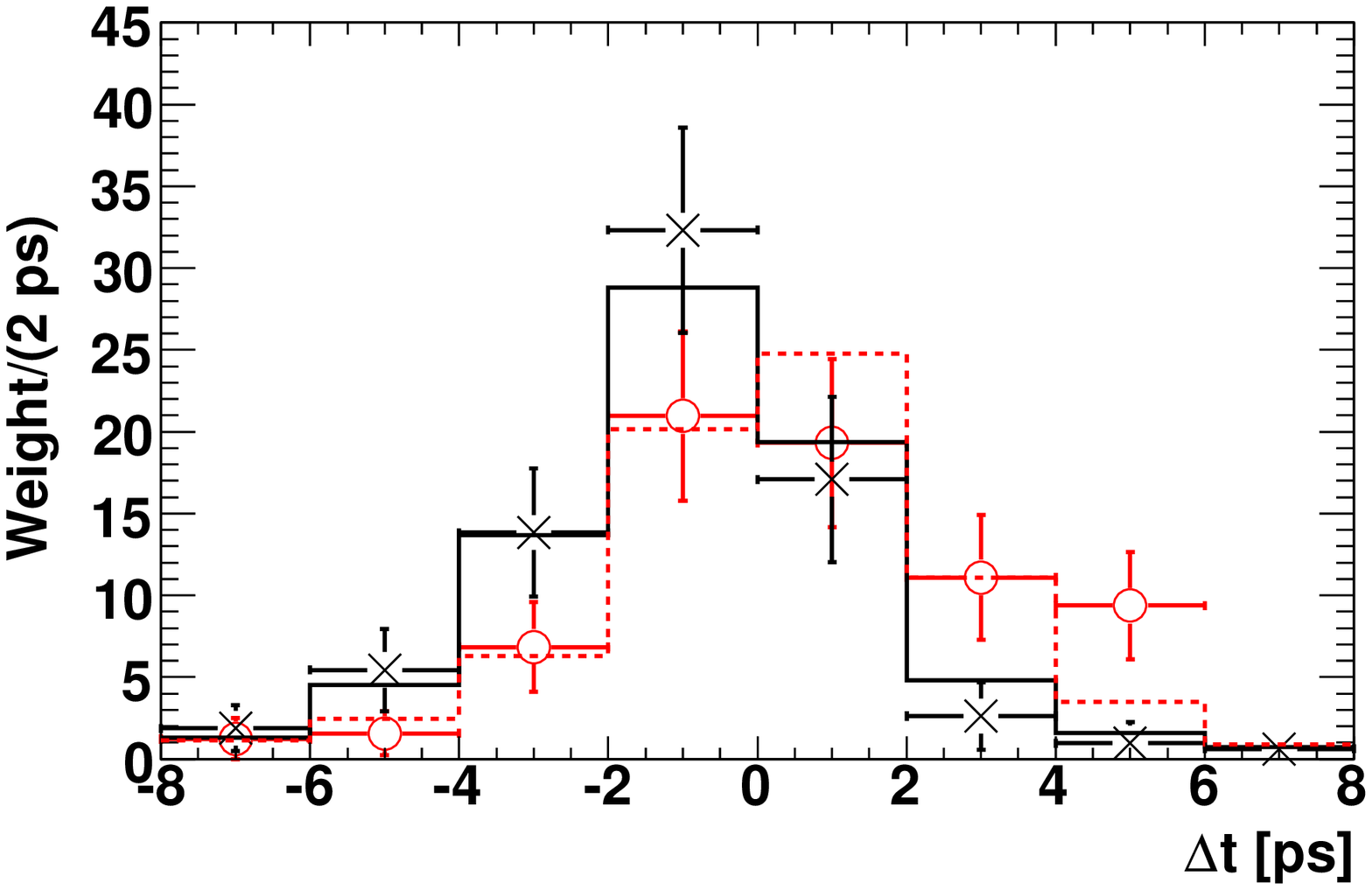}
\includegraphics[width=7.0cm,keepaspectratio]{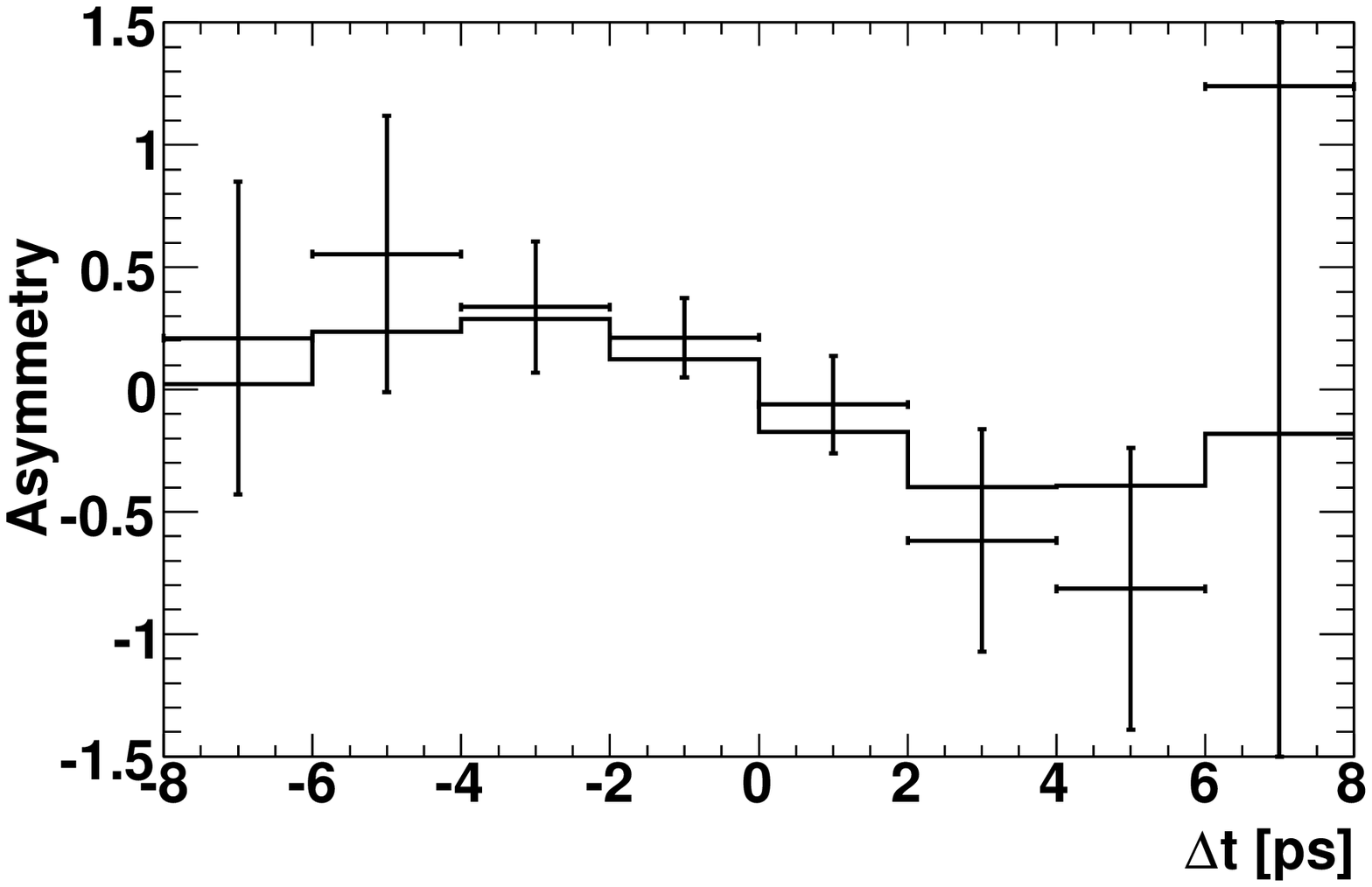}
\includegraphics[width=7.0cm,keepaspectratio]{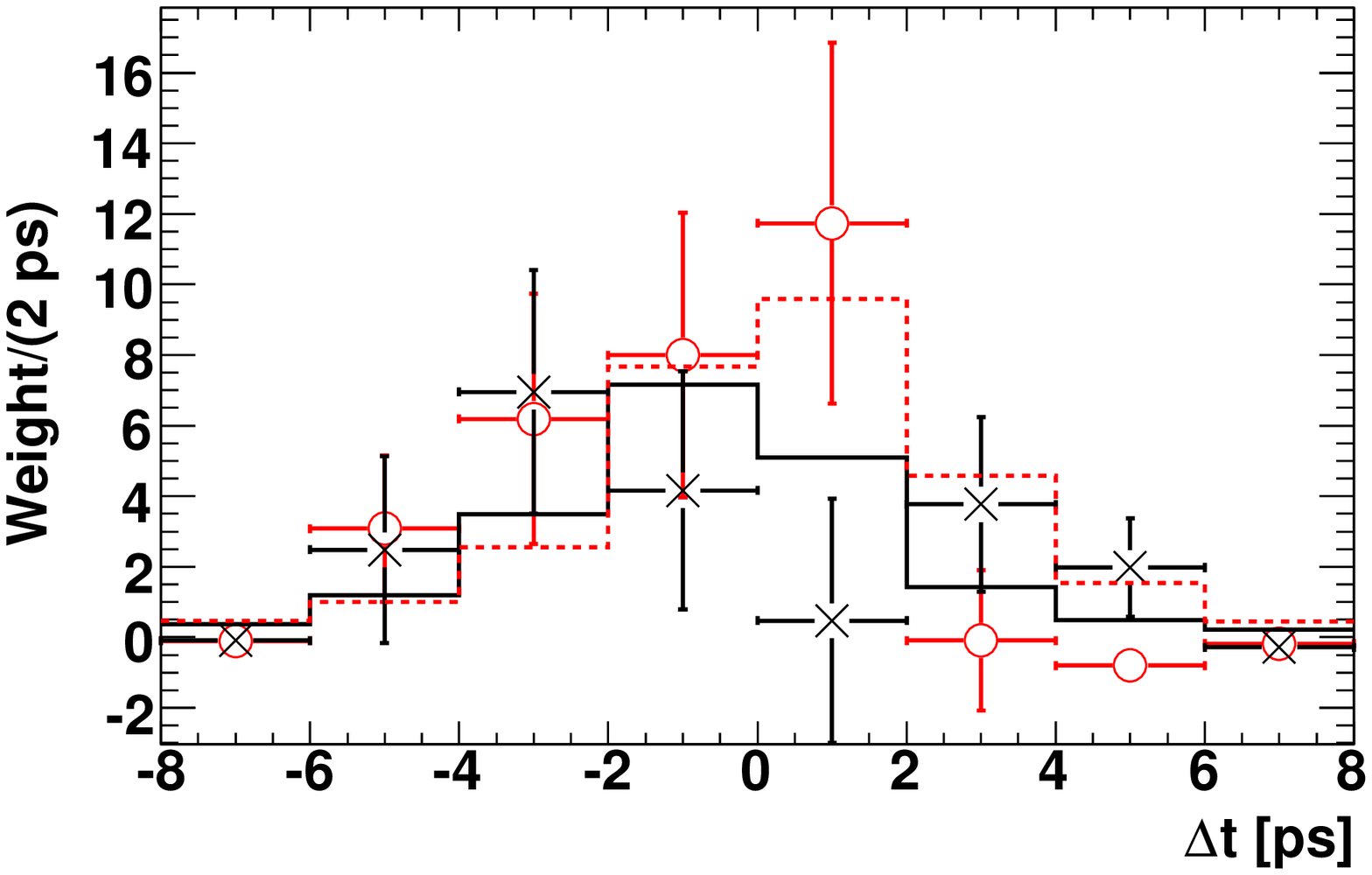}
\includegraphics[width=7.0cm,keepaspectratio]{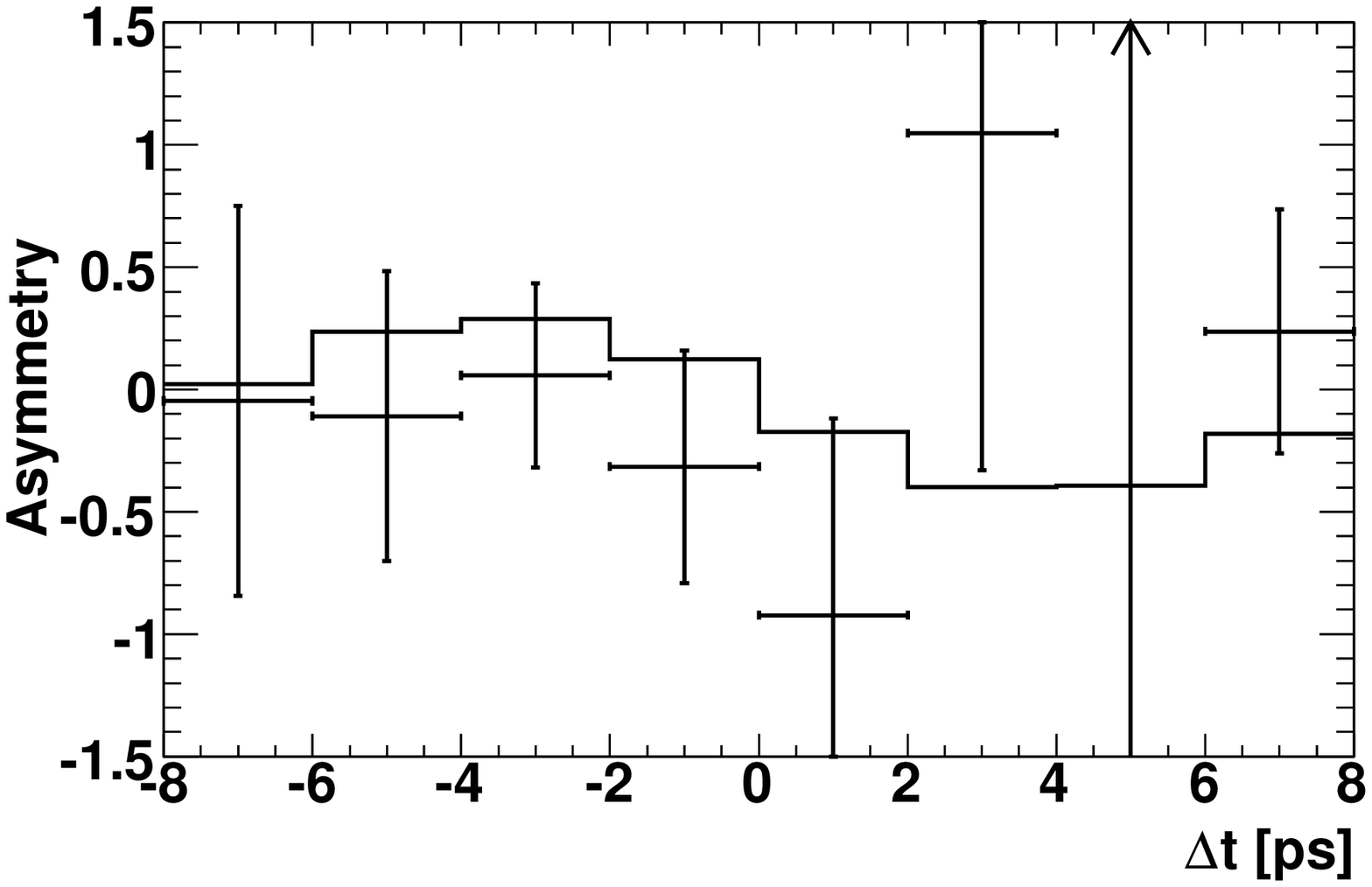}
\includegraphics[width=7.0cm,keepaspectratio]{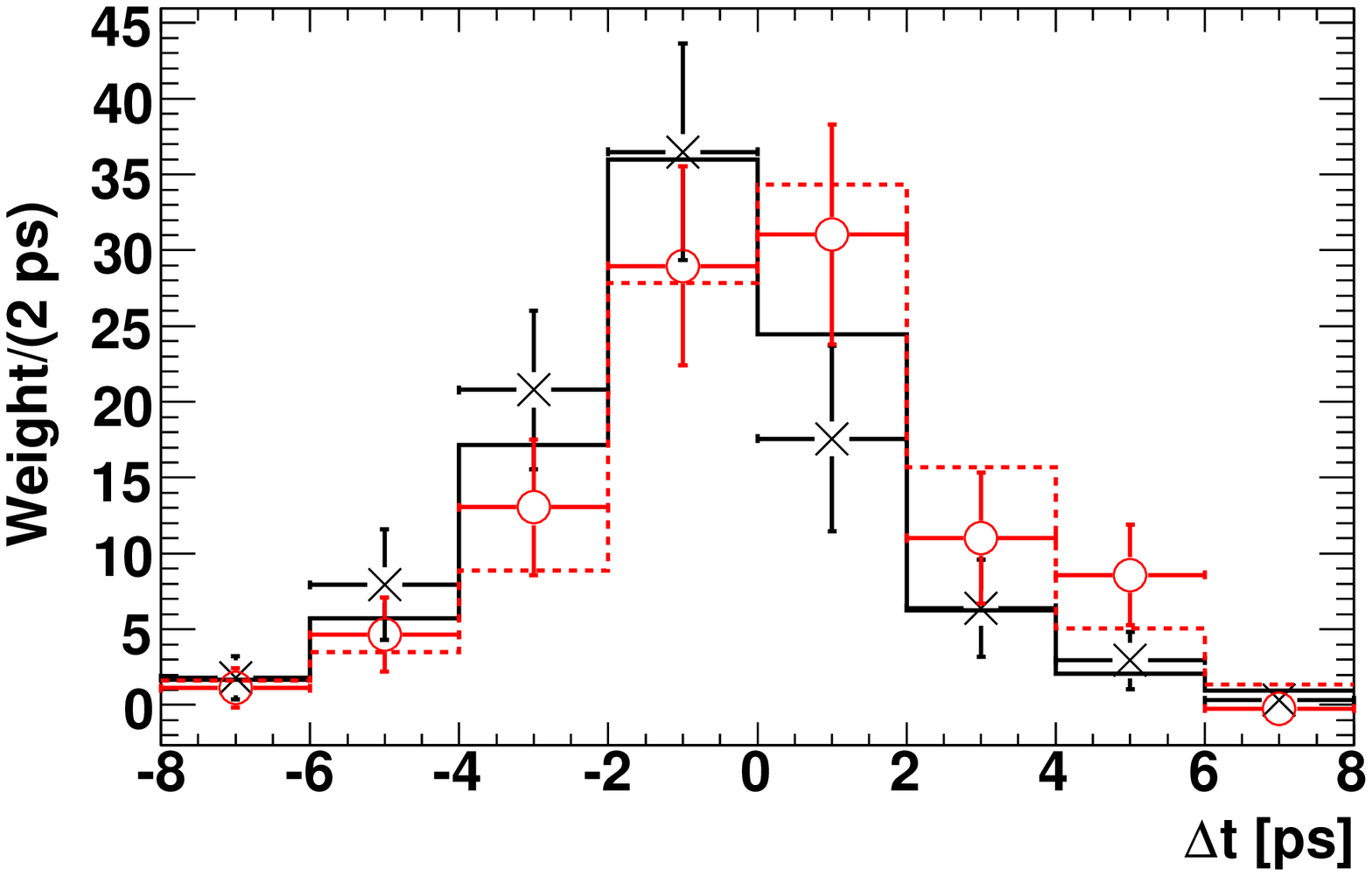}
\includegraphics[width=7.0cm,keepaspectratio]{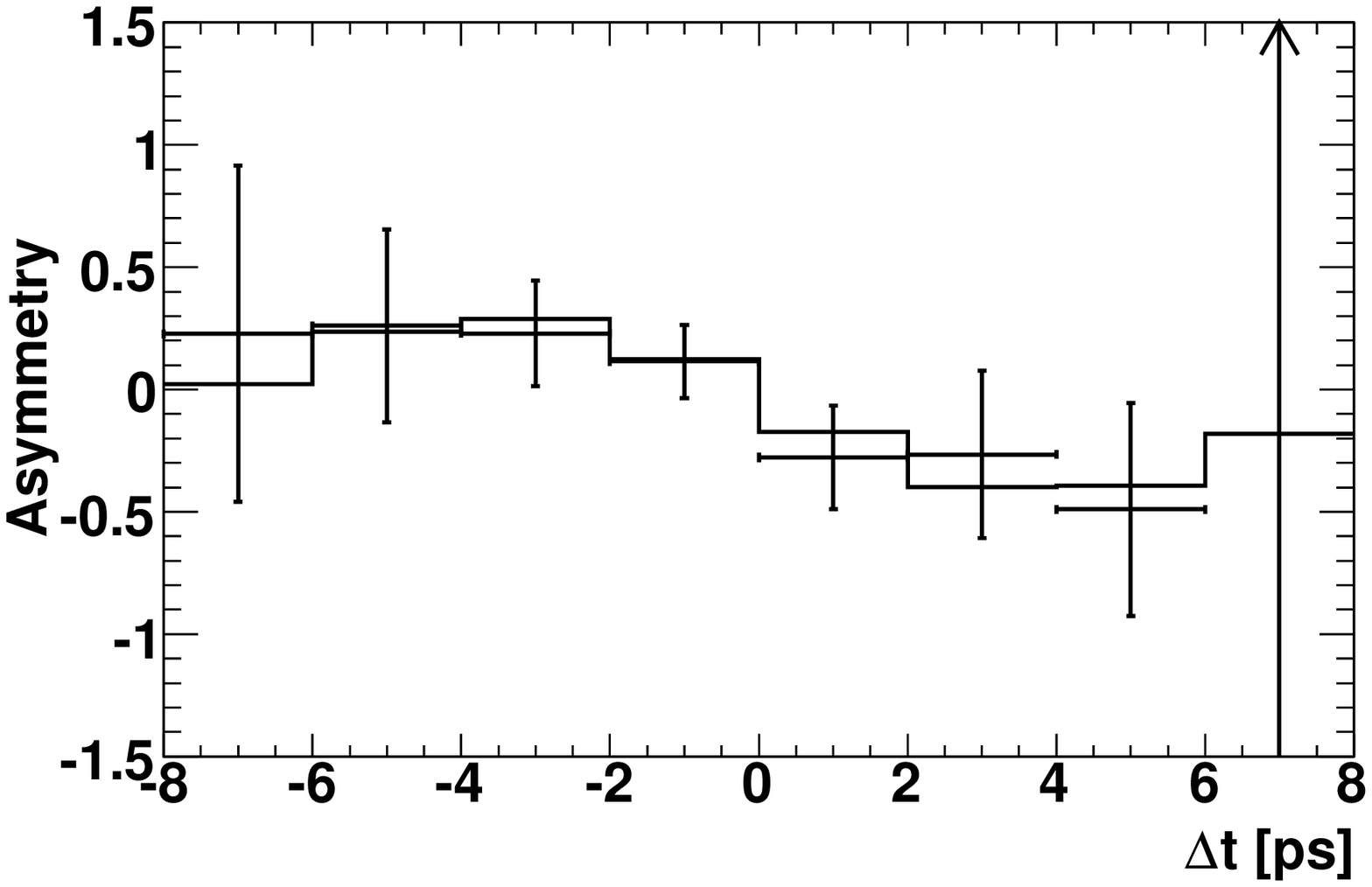}
\end{center}
\caption{(color online). Signal {\ensuremath{\hbox{$_s$}{\cal P}lots}} (points with error bars) and PDFs (histograms) of $\dt$ (left) and the derived asymmetry (right) for the $\Bz\to 3\KS(\pip\pim)$ submode (top), the $\Bz\to 2\KS(\pip\pim)\KS(\piz\piz)$ submode (middle), and the combined fit (bottom). In the $\dt$ distributions on the left-hand side, points marked with $\times$ and solid lines correspond to decays where $\B_{\rm tag}$ is a $\Bz$ meson; points marked with $\circ$ and dashed lines correspond to decays where $\B_{\rm tag}$ is a $\Bzb$ meson.
Points of the asymmetry {\ensuremath{\hbox{$_s$}{\cal P}lots}} that are outside the range of a figure are marked by arrows.}
\label{fig:Deltat_sPlots_TD}
\end{figure*}

\begin{figure*}[htbp]
\begin{center}
\includegraphics[width=7.0cm,keepaspectratio]{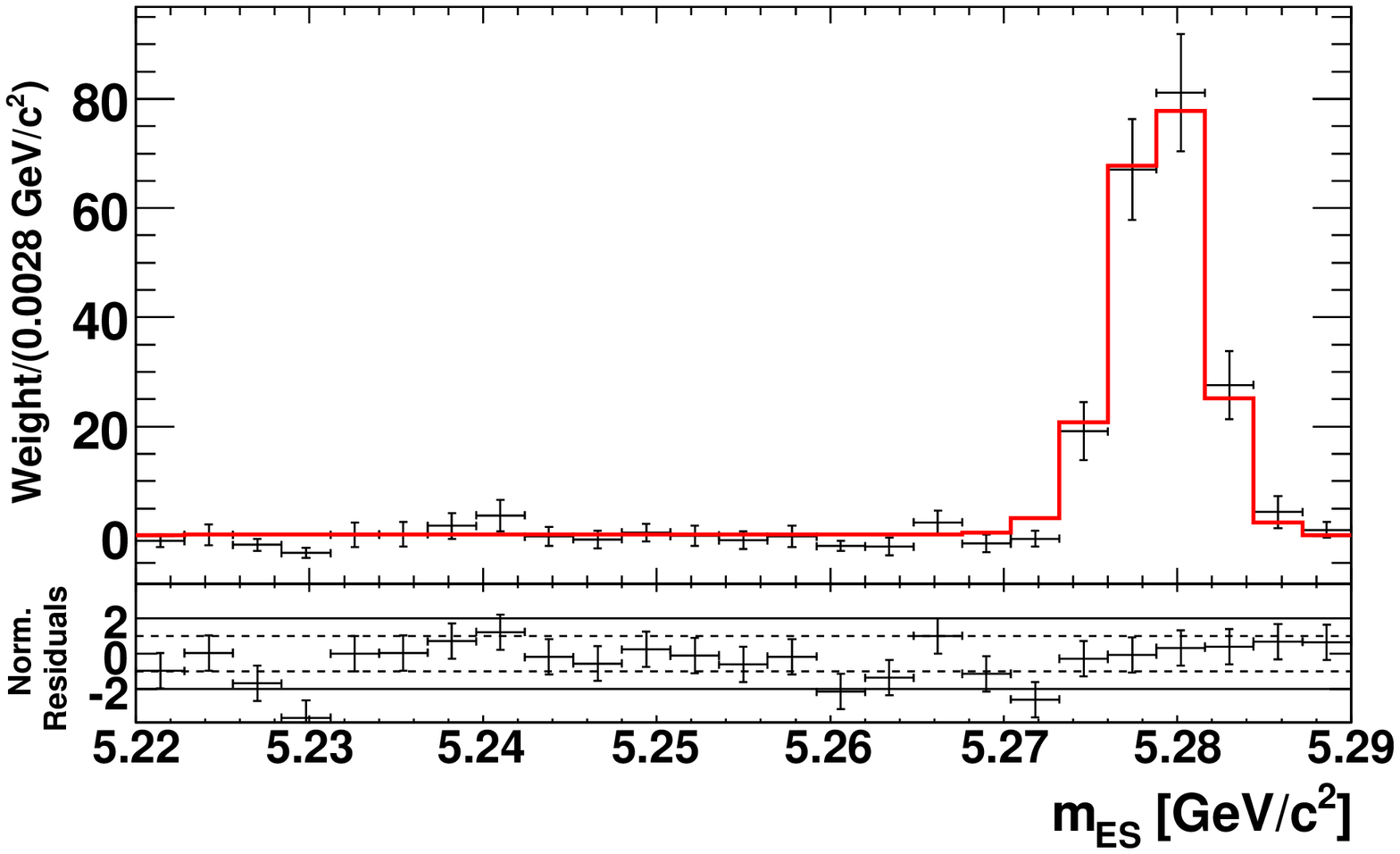}
\includegraphics[width=7.0cm,keepaspectratio]{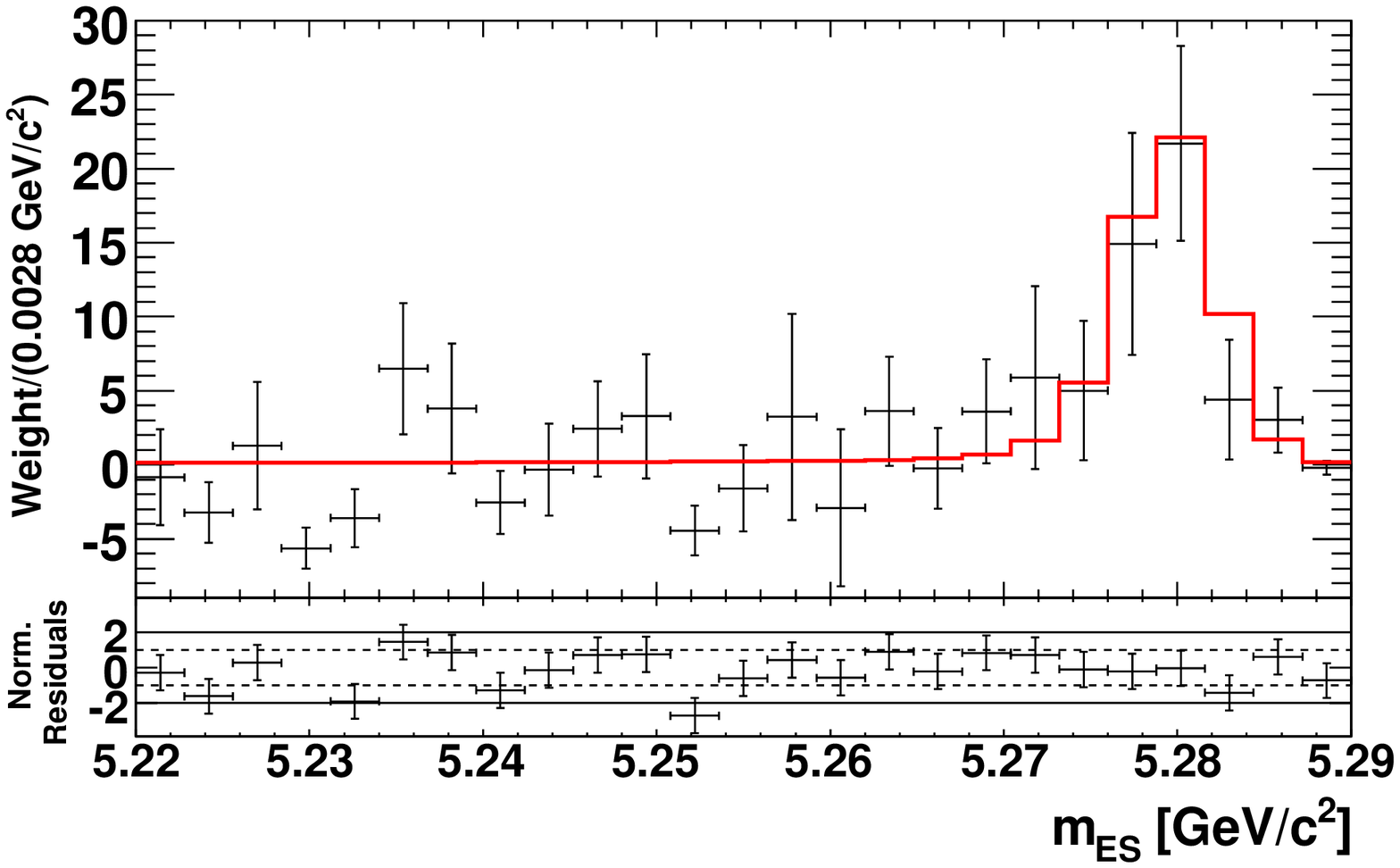}
\includegraphics[width=7.0cm,keepaspectratio]{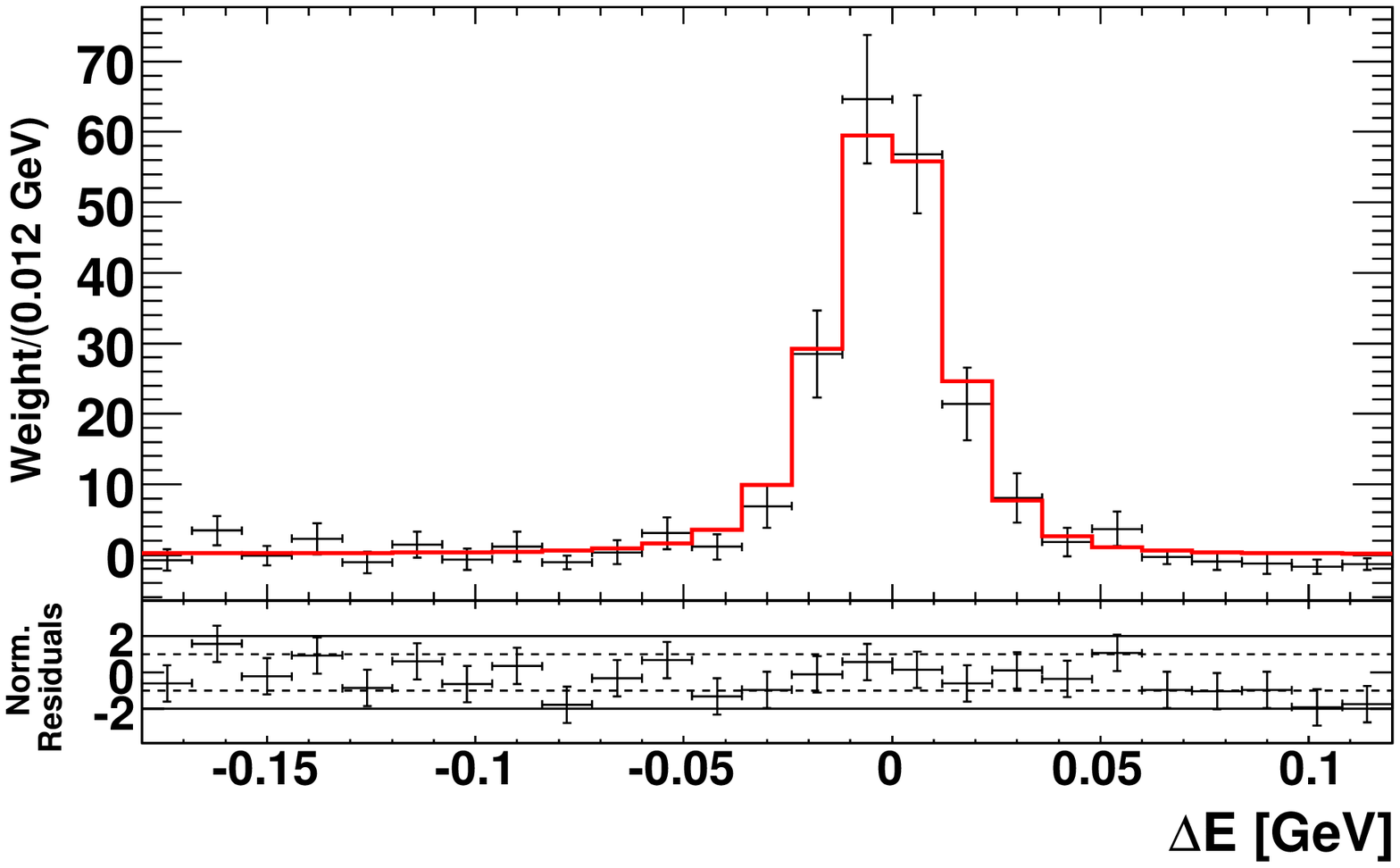}
\includegraphics[width=7.0cm,keepaspectratio]{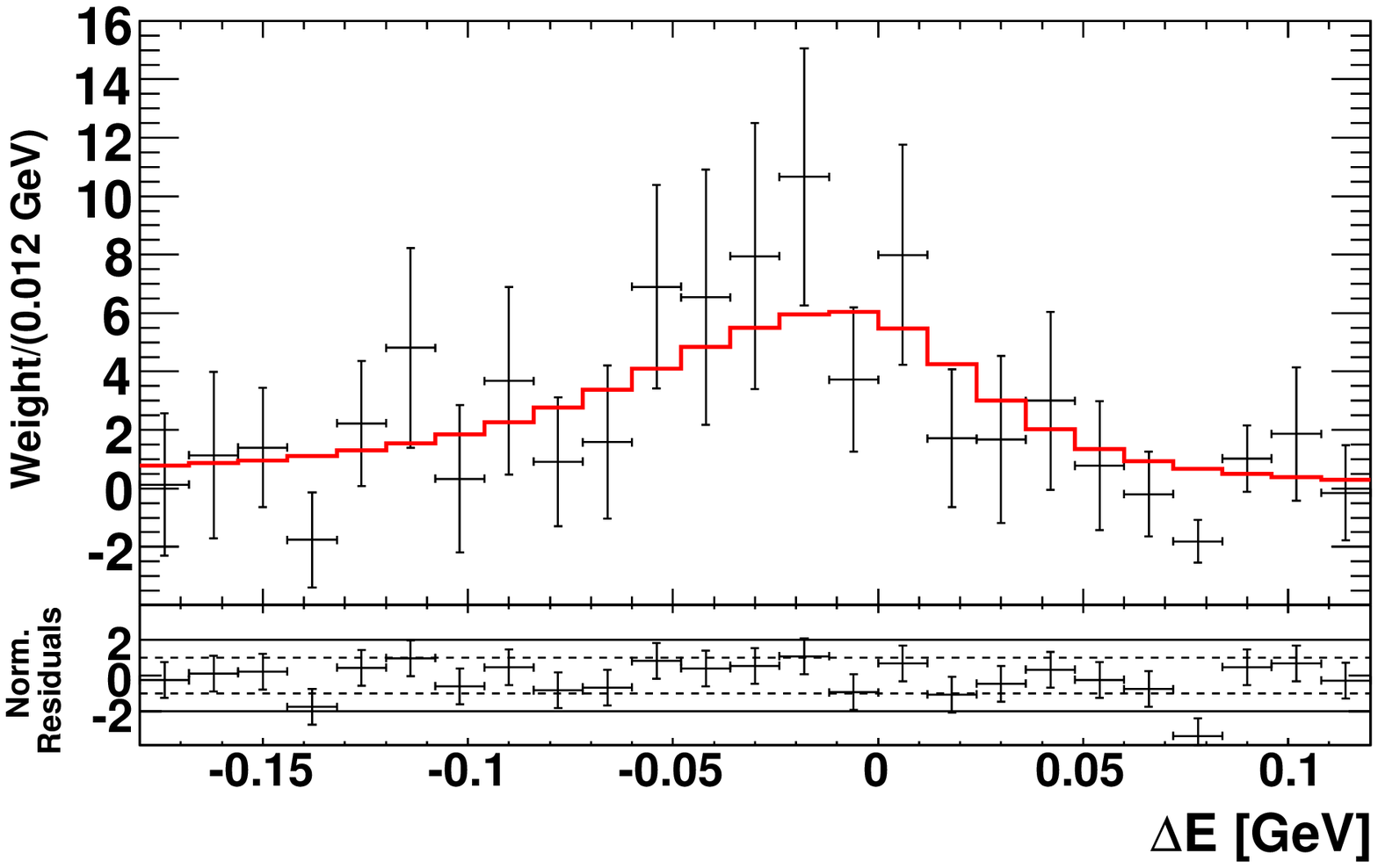}
\includegraphics[width=7.0cm,keepaspectratio]{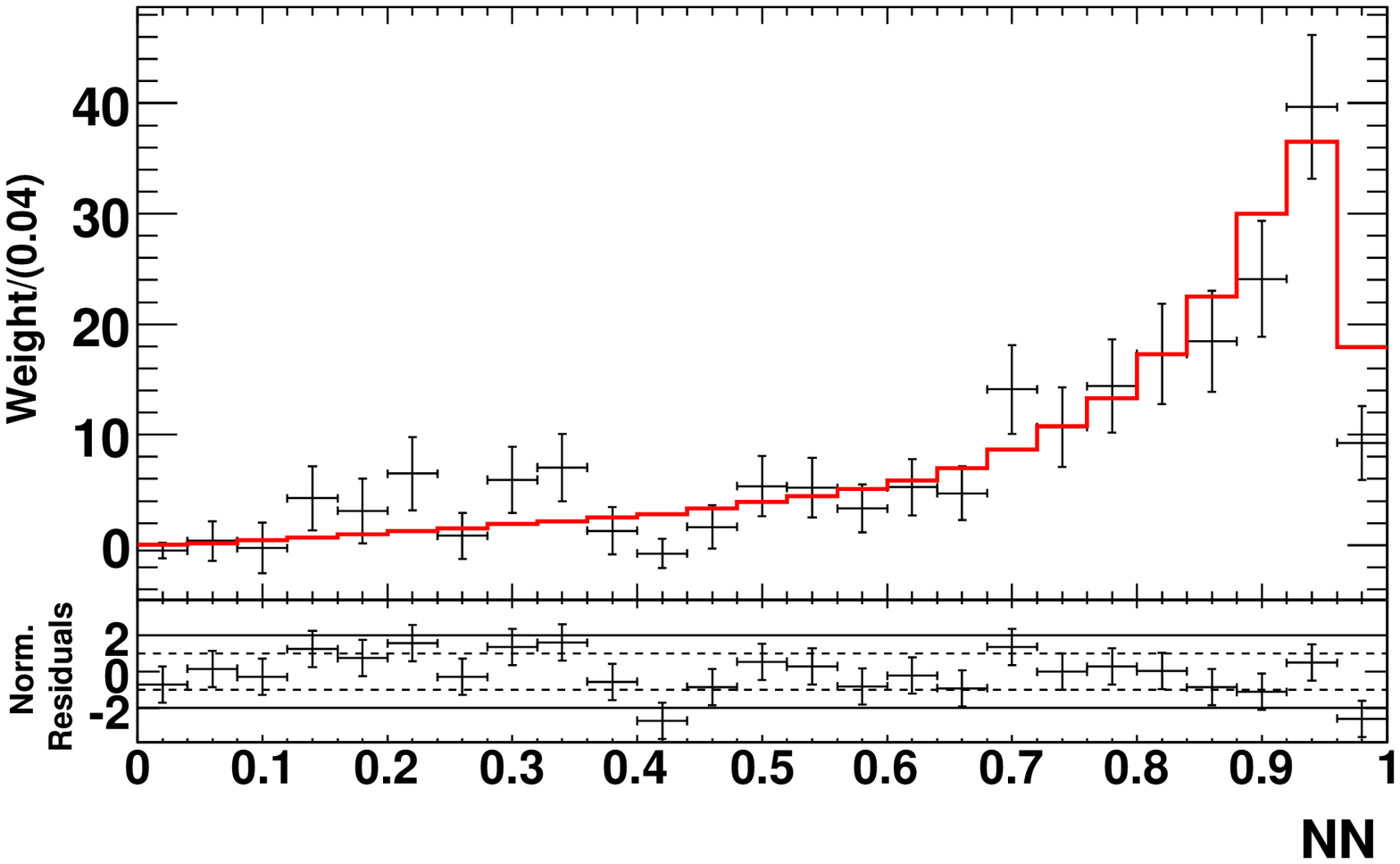}
\includegraphics[width=7.0cm,keepaspectratio]{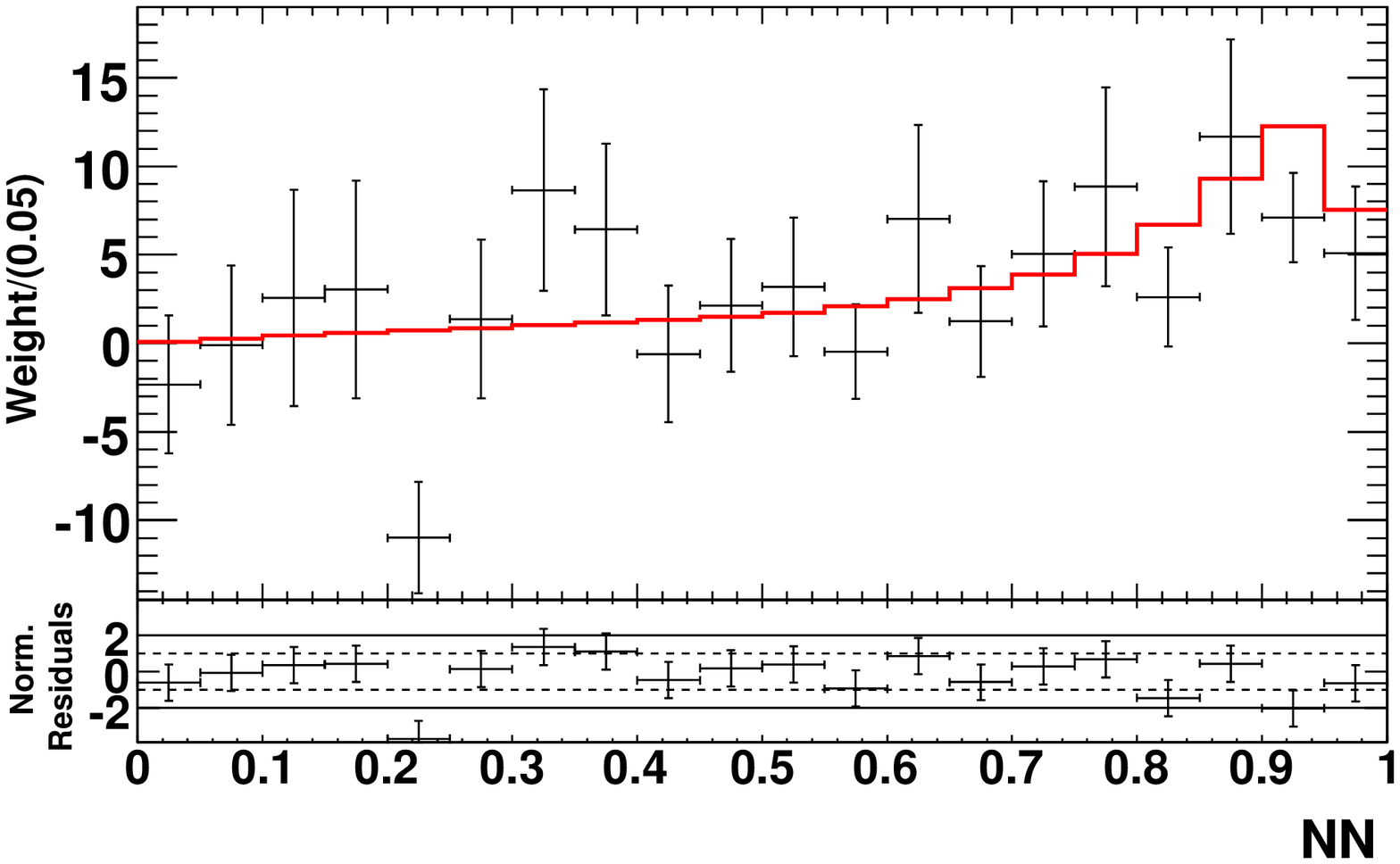}
\end{center}
\caption{
(color online). Signal {\ensuremath{\hbox{$_s$}{\cal P}lots}} (points with error bars) and PDFs (histograms) of the discriminating variables: $\mes$ (top), $\de$ (middle), and the NN output (bottom) for the $\Bz\to 3\KS(\pip\pim)$ submode (left) and for the $\Bz\to 2\KS(\pip\pim)\KS(\piz\piz)$ submode (right). Below each bin are shown the residuals, normalized in error units. The horizontal dotted and full lines mark the one- and two-standard deviation levels, respectively.}
\label{fig:Discrimvars_sPlots_TD}
\end{figure*}

\begin{figure*}[htbp]
\begin{center}
\includegraphics[width=7.0cm,keepaspectratio]{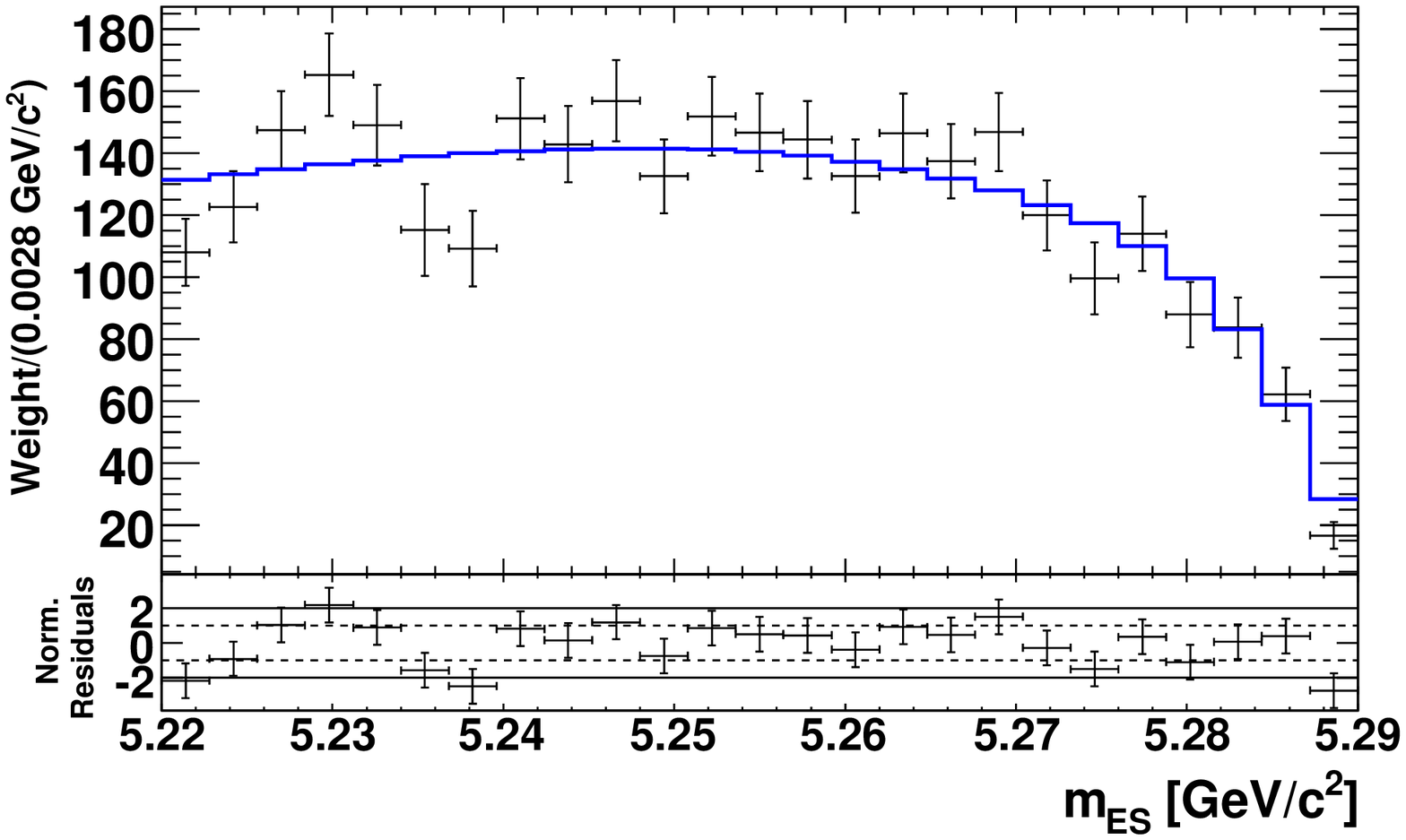}
\includegraphics[width=7.0cm,keepaspectratio]{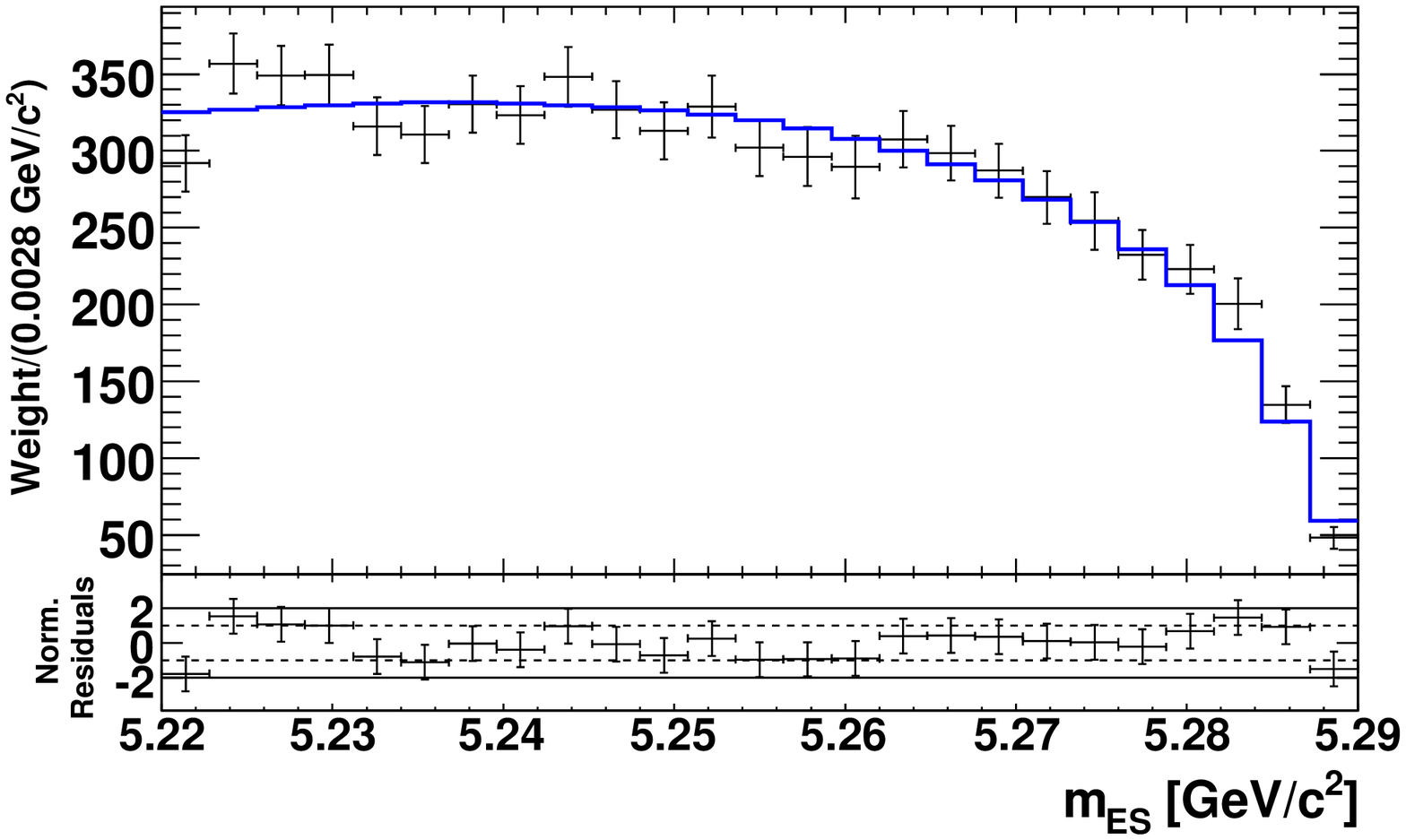}
\includegraphics[width=7.0cm,keepaspectratio]{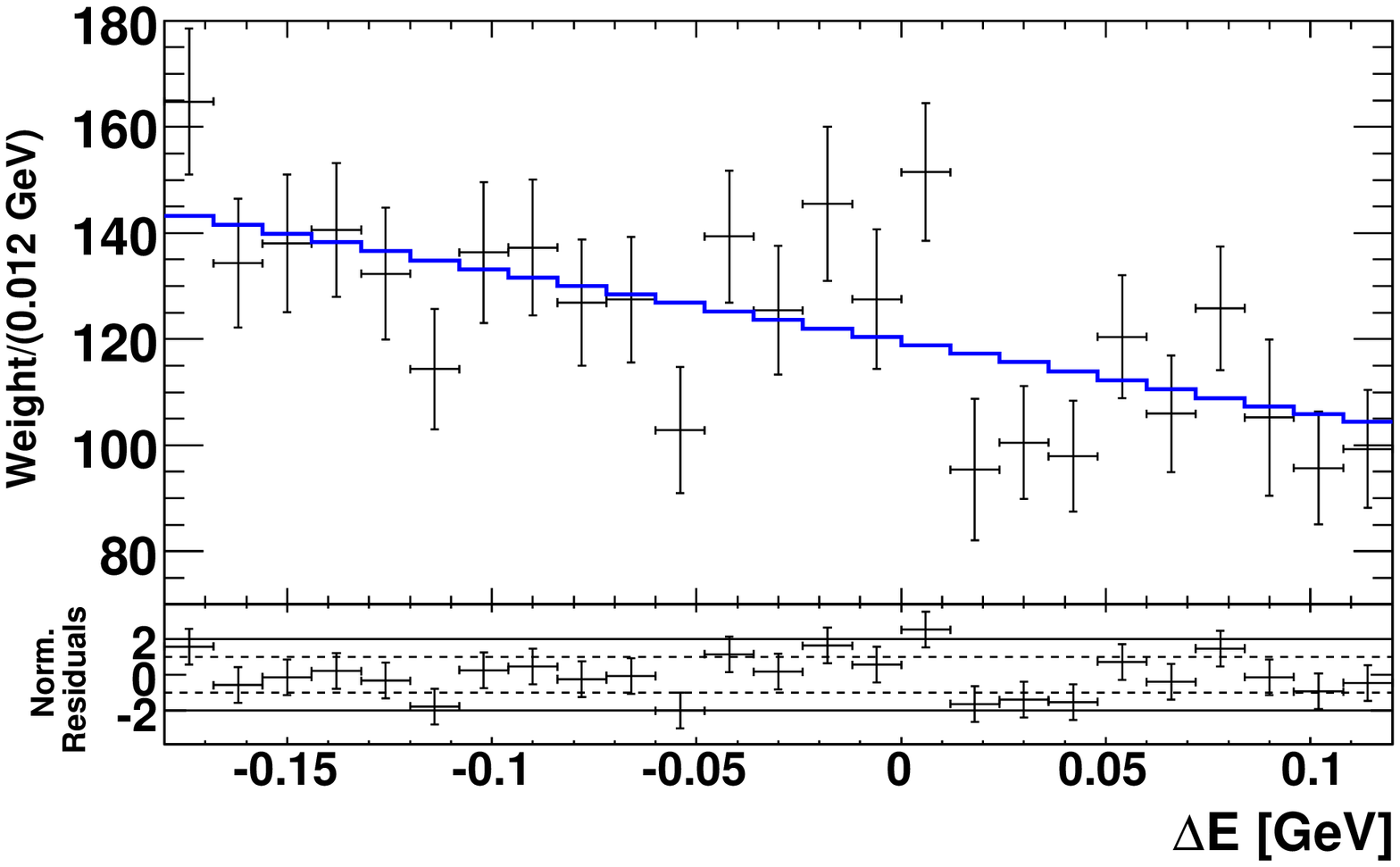}
\includegraphics[width=7.0cm,keepaspectratio]{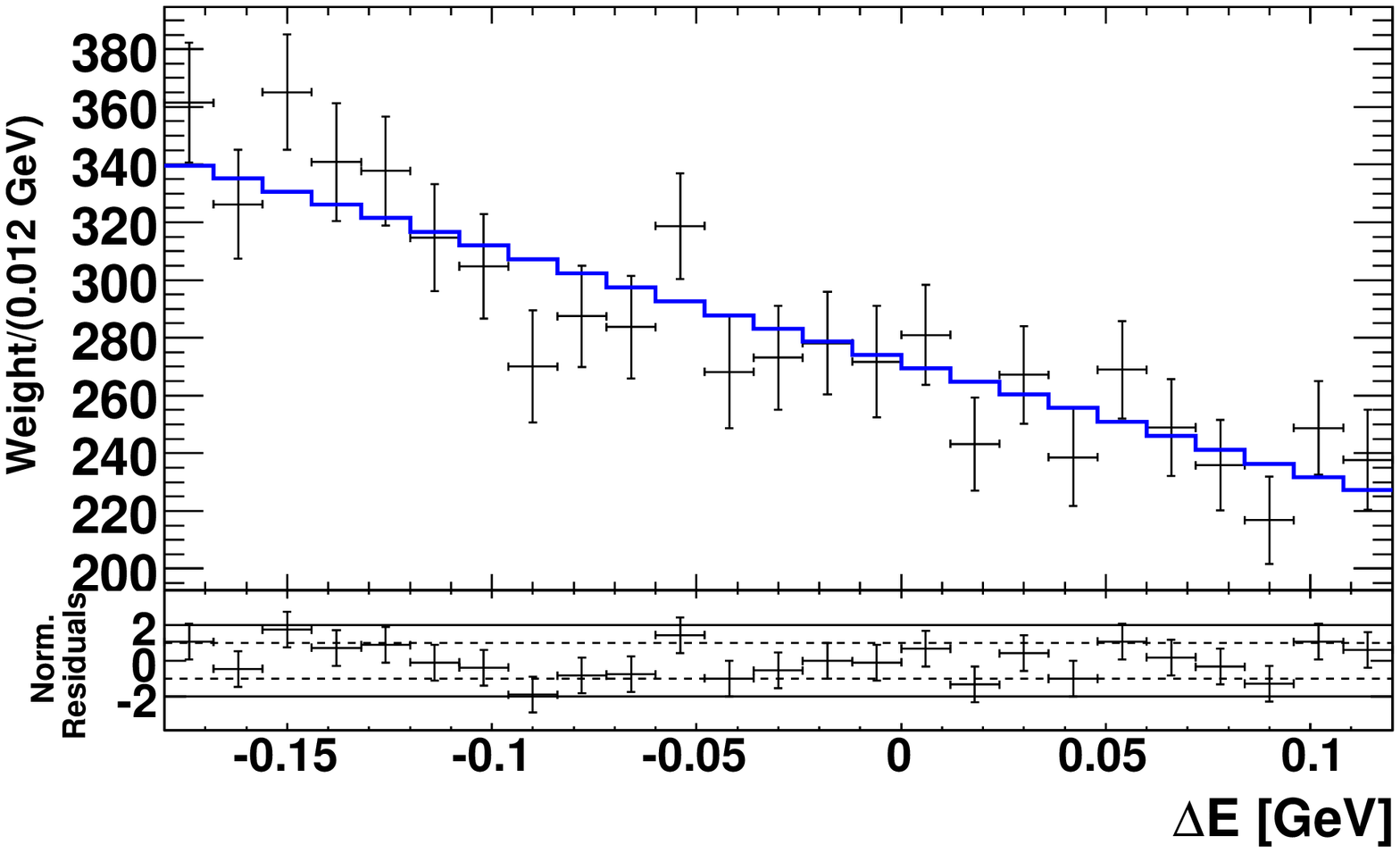}
\includegraphics[width=7.0cm,keepaspectratio]{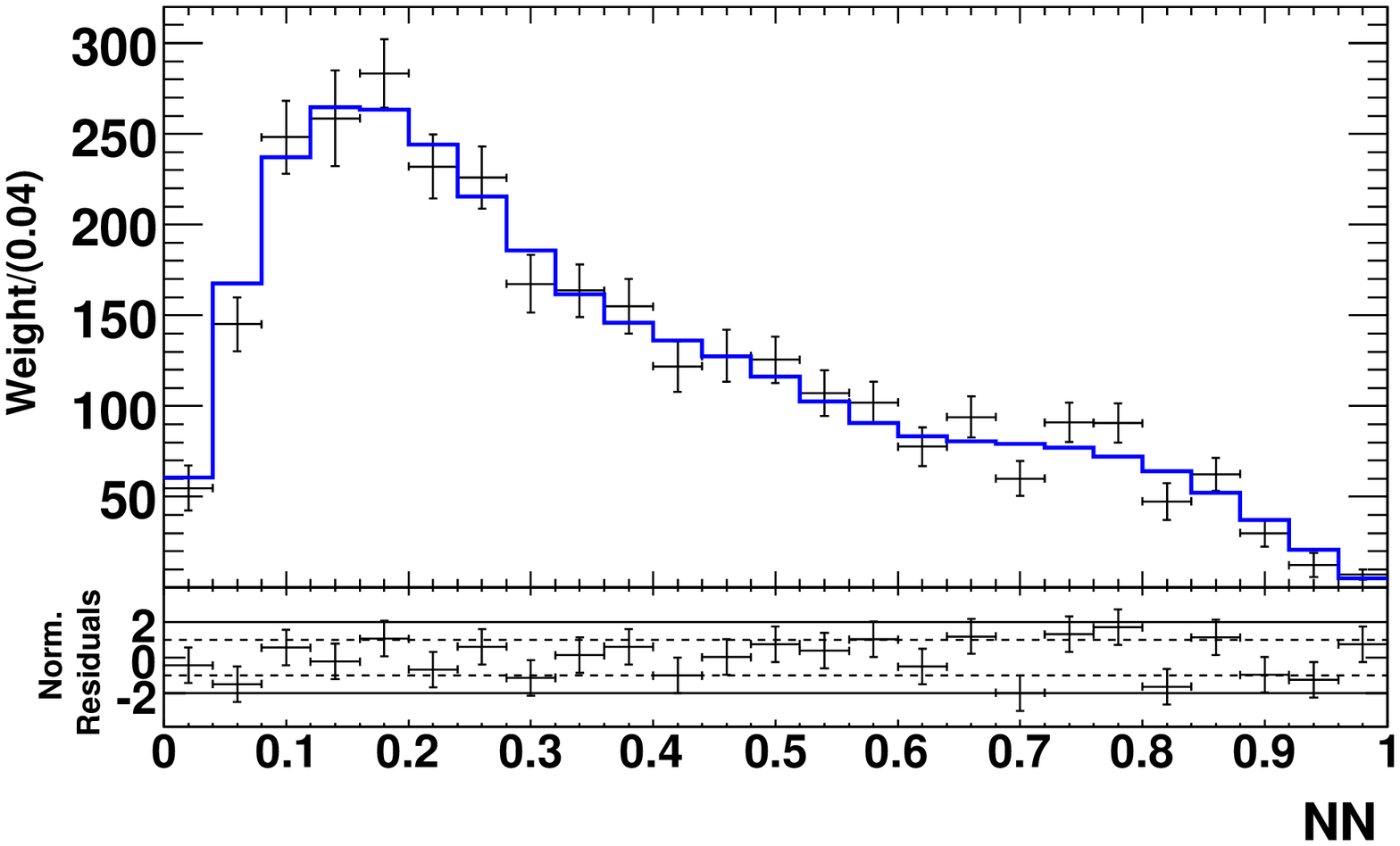}
\includegraphics[width=7.0cm,keepaspectratio]{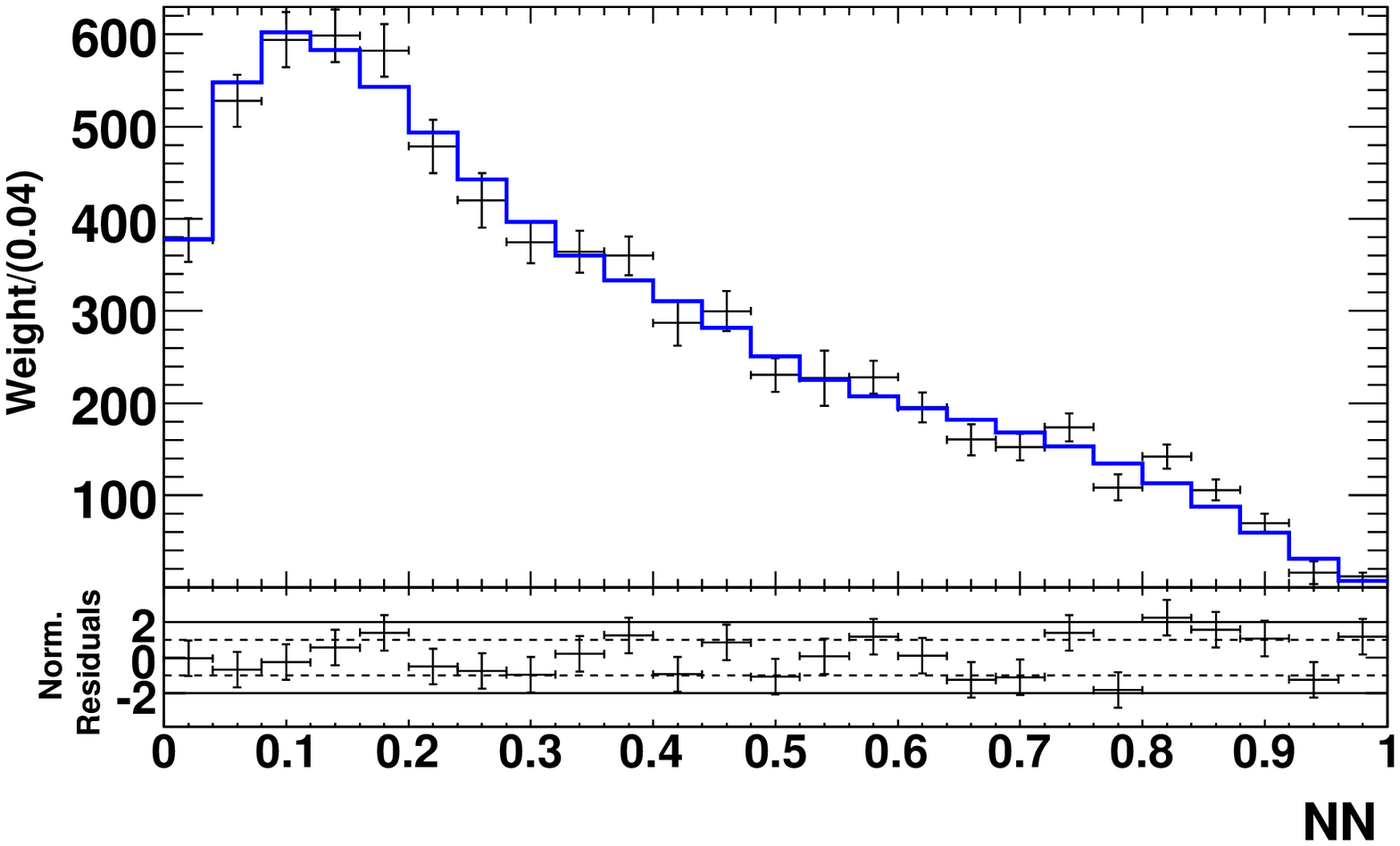}
\includegraphics[width=7.0cm,keepaspectratio]{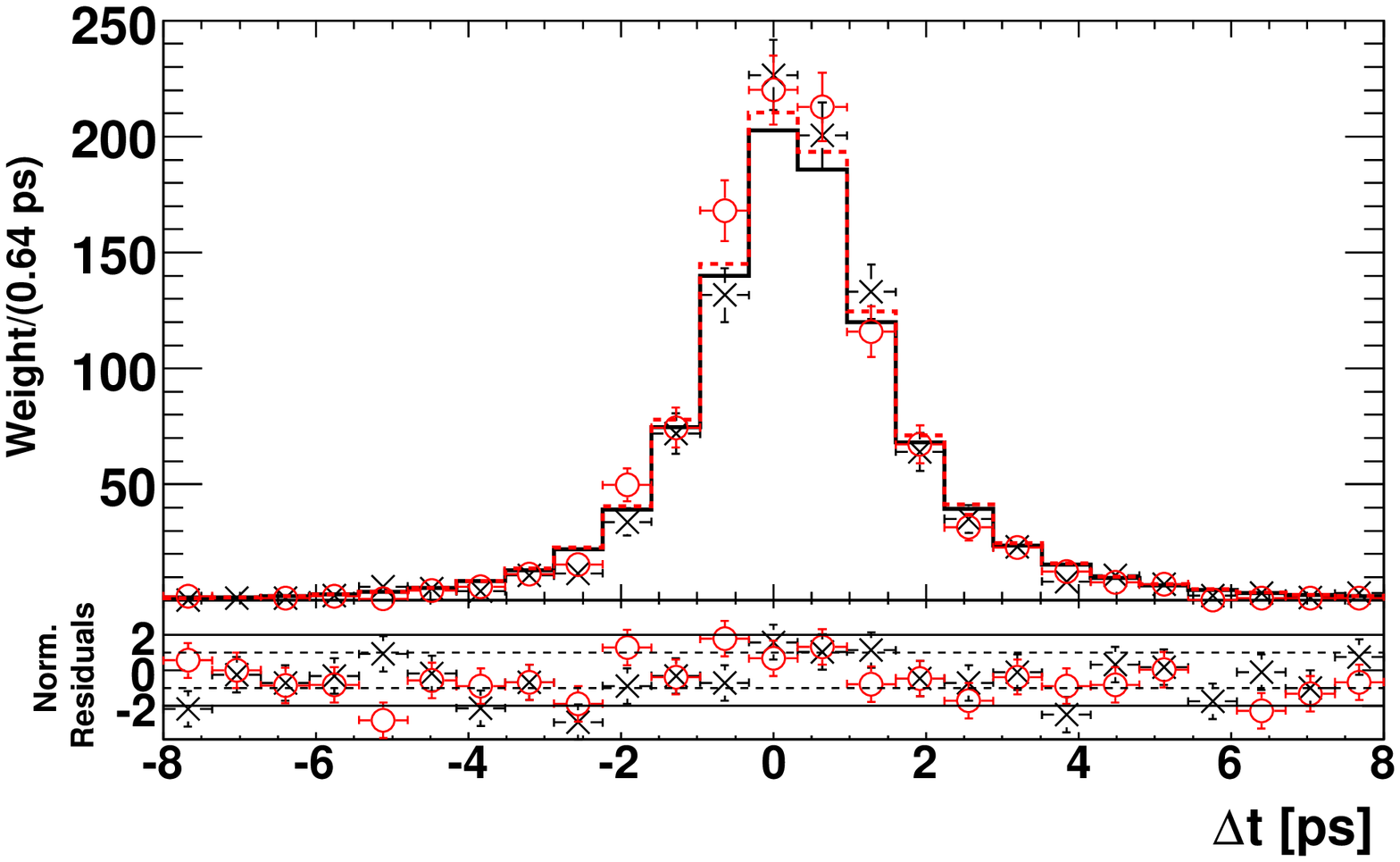}
\includegraphics[width=7.0cm,keepaspectratio]{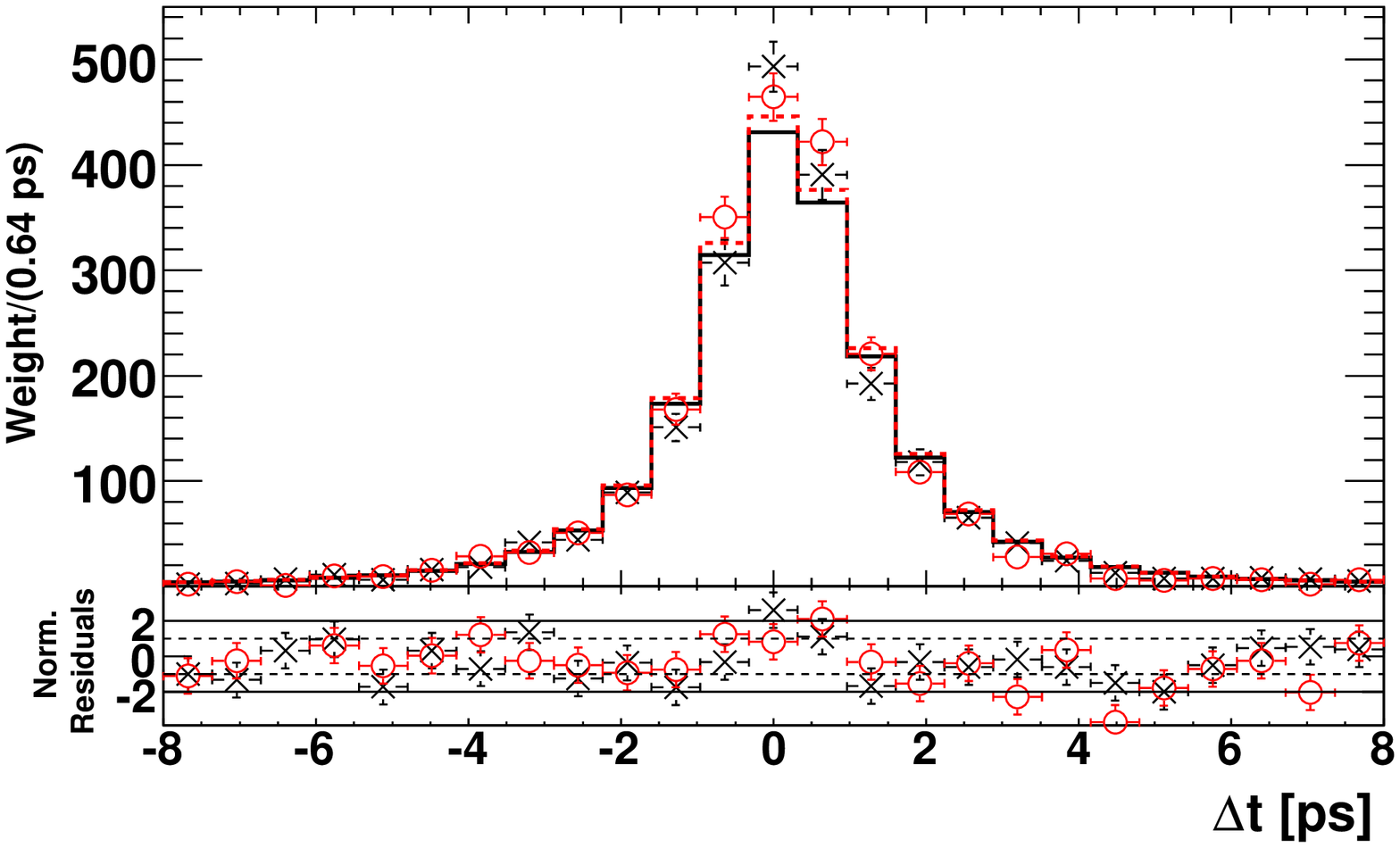}
\end{center}
\caption{(color online). Continuum {\ensuremath{\hbox{$_s$}{\cal P}lots}} (points with error bars) and PDFs (histograms) of $\mes$, $\de$, the NN output, and $\dt$ (top to bottom). Plots on the left-hand side correspond to the $\Bz\to 3\KS(\pip\pim)$ submode, and on the right-hand side to the $\Bz\to 2\KS(\pip\pim)\KS(\piz\piz)$ submode. In the $\dt$ distributions, points marked with $\times$ and solid lines correspond to decays where $\B_{\rm tag}$ is a $\Bz$ meson; points marked with $\circ$ and dashed lines correspond to decays where $\B_{\rm tag}$ is a $\Bzb$ meson. Below each bin are shown the residuals, normalized in error units. The horizontal dotted and full lines mark the one- and two-standard deviation levels, respectively.}
\label{fig:Discrimvars_Cont_sPlots_TD}
\end{figure*}

The fit result for the time-dependent \CP-violation parameters ${\cal S}$ and ${\cal C}$ is
\begin{eqnarray}
\nonumber {\cal S} &=& -0.94\,^{+0.24}_{-0.21}, \\
\nonumber {\cal C} &=& -0.17 \pm 0.18, 
\end{eqnarray}
where the uncertainties are statistical only. The correlation between ${\cal S}$ and ${\cal C}$ is $-0.16$.
We use the fit result to create {\ensuremath{\hbox{$_s$}{\cal P}lots}} of the signal distributions of $\dt$, the time-dependent asymmetry, and the discriminating variables. Figure~\ref{fig:Deltat_sPlots_TD} shows the $\dt$ {\ensuremath{\hbox{$_s$}{\cal P}lots}} for the combined fit result and for the individual submodes.
Figure~\ref{fig:Discrimvars_sPlots_TD} shows the signal distributions and Fig.~\ref{fig:Discrimvars_Cont_sPlots_TD} the continuum background distributions of the discriminating variables.
The distributions shown in these three figures illustrate the good agreement between the data and the fit model.

\begin{figure*}[htbp]
\begin{center}
\includegraphics[width=7.0cm,keepaspectratio]{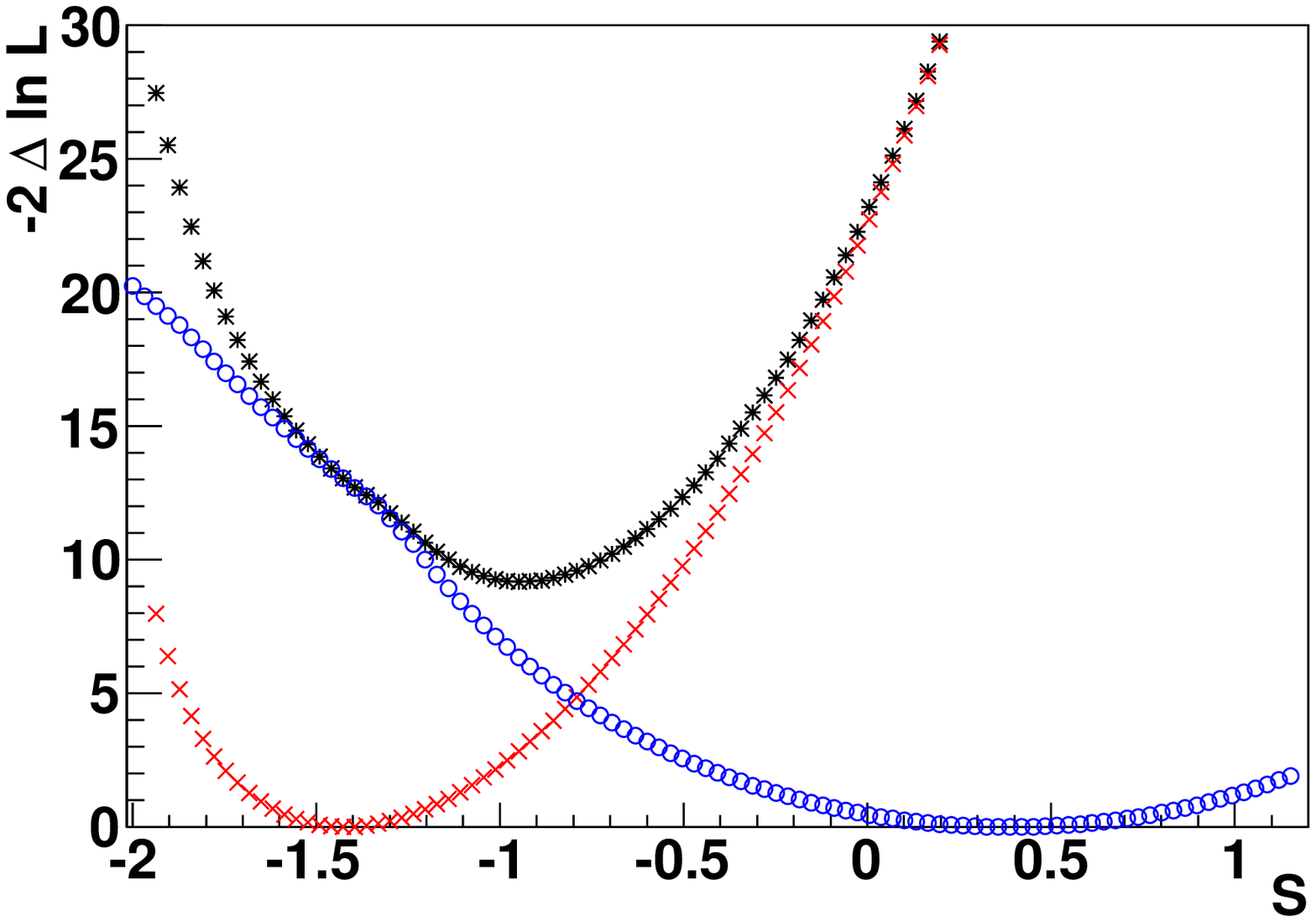}
\includegraphics[width=7.0cm,keepaspectratio]{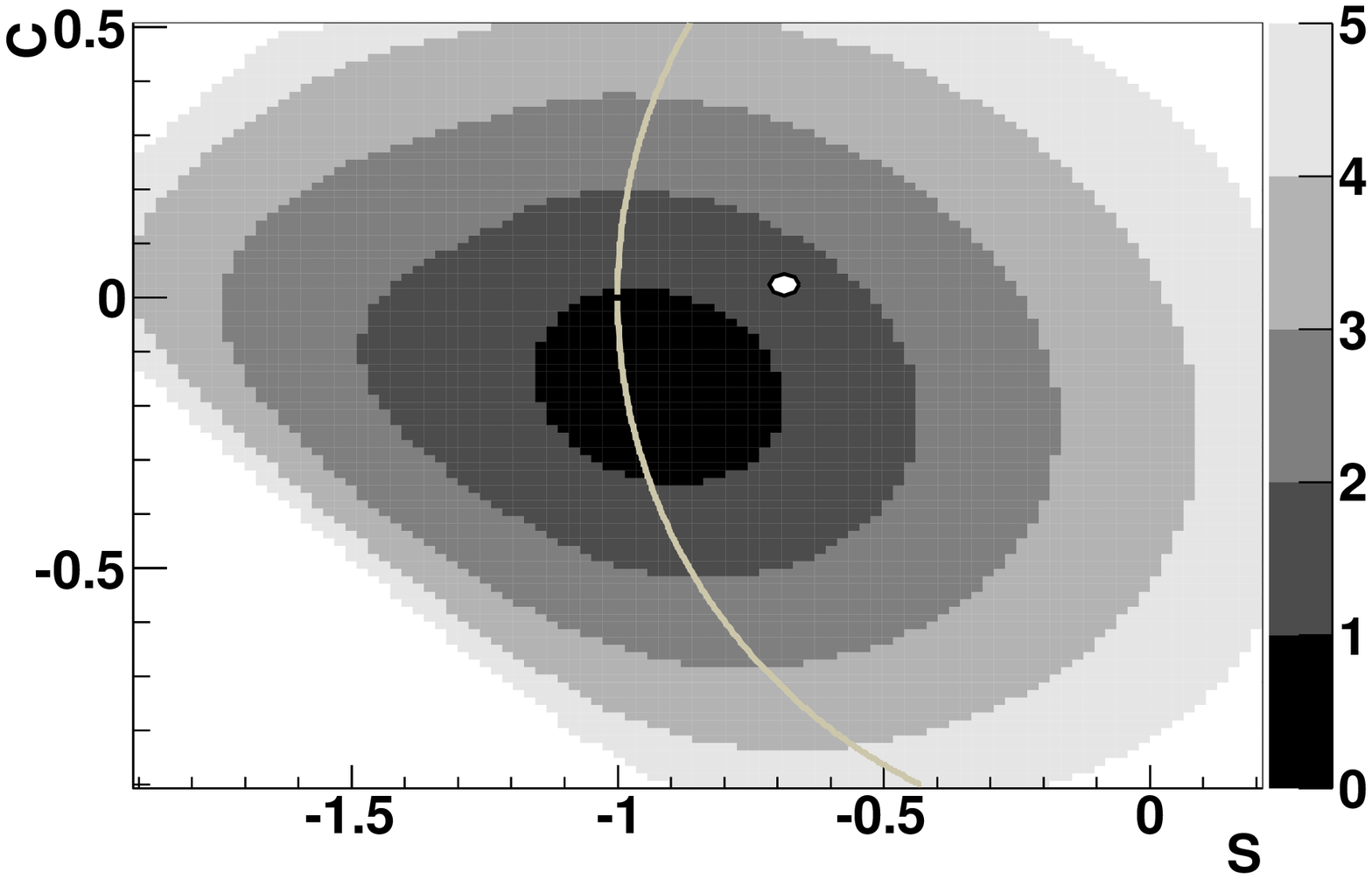}
\end{center}
\caption{(color online). One-dimensional statistical scan of $-2 \Delta \lnL$ as a function of ${\cal S}$ (left) and the two-dimensional scan, including systematic uncertainty, as a function of ${\cal S}$ and ${\cal C}$ (right). In the left-hand plot, red points marked with $\times$ correspond to the $\Bz\to 3\KS(\pip\pim)$ submode, blue points marked with $\circ$ to the $\Bz\to 2\KS(\pip\pim)\KS(\piz\piz)$ submode, and  black points marked with $\ast$ to the combined fit. In the right-hand plot,
the gray scale is given in units of $\sqrt{-2 \Delta \lnL}$.
The result of the \babar\ analyses of $\Bz\to c \bar c K^{(*)}$ decays~\cite{LastSin2betaBabar} is indicated as a white ellipse and the physical boundary (${\cal S}^2+{\cal C}^2 \leq 1$) is marked as a gray line. The scan appears to be trimmed on the lower left since the PDF becomes negative outside the physical region (i.e., the white region does not indicate that the scan flattens out at 5$\sigma$).}
\label{fig:likelihoodScans_TD}
\end{figure*}

We scan the statistical-only likelihood of the ${\cal S}$ parameter for both submodes and for the combined fit. The result, on the left-hand side of Fig.~\ref{fig:likelihoodScans_TD}, shows a sizable difference between the ${\cal S}$ values for the two submodes;
the level of consistency, conservatively estimated from the sum of the two individual likelihood scans, is approximately $2.6\sigma$ (a $p$-value of $1.0\%$ with $2$ degrees of freedom). 
This value is obtained including only the dominant statistical uncertainty and neglecting the small correlation between the \CP-violation parameters.
The results obtained when ${\cal S}$ and ${\cal C}$ are allowed to vary individually for each of the submodes are ${\cal S}=-1.42\,^{+0.27}_{-0.24}$, ${\cal C}=-0.14\,^{+0.17}_{-0.17}$ for $\Bz\to 3\KS(\pip\pim)$ and ${\cal S}=0.40\,^{+0.56}_{-0.57}$, ${\cal C}=0.19\,^{+0.42}_{-0.43}$ for $\Bz\to 2\KS(\pip\pim)\KS(\piz\piz)$. In both cases the quoted uncertainties are statistical only.

As there is some correlation between the ${\cal S}$ and ${\cal C}$ parameters, we perform a two-dimensional statistical likelihood scan of the combined likelihood, which is then convolved by the systematic uncertainties on ${\cal S}$ and ${\cal C}$ (systematic uncertainties are discussed in Sec.~\ref{sec:td_systematics}.)
The result is shown on the right-hand side of Fig.~\ref{fig:likelihoodScans_TD}.
We find that \CP\ conservation is excluded at $3.8$ standard deviations, and thus, for the first time, we measure an evidence of \CP\ violation in $\Bz\to\KS\KS\KS$ decays.
The difference between our result and that from $\Bz\to c \bar c K^{(*)}$ is less than $2$ standard deviations.
The scan also shows that the result is close to the physical boundary, given by the constraint ${\cal S}^2+{\cal C}^2 \leq 1$.

\subsection{Systematic uncertainties}
\label{sec:td_systematics}
The systematic uncertainties are summarized in Table~\ref{tab:sys_TD}. The ``${\rm MC_{stat}}$'' uncertainty accounts for the limited size of the simulated data samples used to create the PDFs. The ``$\B_{\rm reco}$'' uncertainty propagates the experimental uncertainty in the measurement of tag-side-related quantities taken from~\cite{LastSin2betaBabar} to our measurement. The ``\B-bkg'' contribution results from the uncertainty in the \CP content and the branching fractions of fixed yields in the model of background from \B decays. The dominant ``MC-Data: \dt'' systematic uncertainty is due to possible differences between data and simulation
concerning the procedure used to obtain the signal \B decay vertex from tracks originating from $\KS$ decays.
We quantify this uncertainty using the control sample $\Bz\to\jpsi\KS$ by comparing the difference between $\Delta t$ values obtained with and without the $\jpsi$ in data and simulation. We then propagate the observed differences and their uncertainties to the resolution function. We use this new resolution function to refit the data and obtain an estimate of the effect on ${\cal S}$ and ${\cal C}$. We also use the samples $\Bz\to\jpsi\KS(\pip\pim)$ and $\Bz\to\jpsi\KS(\piz\piz)$ to estimate simulation-data differences for the other variables in the submodes \pippim\ and \pizpiz, respectively. This contribution is referred to as ``MC-Data: Discr.~Vars'' in Table~\ref{tab:sys_TD}. The ``Fit Bias'' uncertainty is evaluated using fits to fully reconstructed MC samples. It accounts for effects from wrongly reconstructed events and correlations between fit variables. The ``Vetoes'' uncertainty is related to the veto on the invariant mass. It is evaluated using events that pass the veto in pseudo-experiments studies. Finally, the ``Misc'' uncertainty includes contributions from doubly-Cabibbo-suppressed decays, silicon vertex tracker alignment, and the uncertainties in the boost of the $\FourS$. These contributions are taken from the \babar\ $\B \to c\bar c K^{(*)}$ analysis~\cite{LastSin2betaBabar}.  
\begin{table*}[htbp]
\begin{center}
\caption{Summary of systematic uncertainties on the ${\cal S}$ and ${\cal C}$ parameters.
\label{tab:sys_TD}}
\begin{tabular}{lcc}
\hline \hline
Source               & ${\cal S}$ &  ${\cal C}$ \\ \hline
$\rm MC_{\rm stat}$  & $0.002$    &  $0.001$    \\
$\B_{\rm reco}$      & $0.004$    &  $0.003$    \\
$\B$-bkg             & $0.032$    &  $0.012$    \\
MC-Data: \dt         & $0.045$    &  $0.027$    \\
MC-Data: Discr.~Vars & $0.021$    &  $0.004$    \\
Fit Bias             & $0.022$    &  $0.018$    \\
Vetoes               & $0.006$    &  $0.004$    \\
Misc                 & $0.004$    &  $0.015$    \\ \hline
Sum                  & $0.064$    &  $0.038$    \\ \hline \hline
\end{tabular}
\end{center}
\end{table*}

\section{SUMMARY}
\label{sec:Summary}

We have performed the first amplitude analysis of $\Bz \rightarrow \KS\KS\KS$ events and measured the total inclusive branching fraction to be $\rm (6.19 \pm 0.48 \pm 0.15 \pm 0.12) \times 10^{-6}$, where the first uncertainty is statistical, the second is systematic, and the third represents the signal DP-model dependence. We have identified the dominant contributions to the DP to be from $\fI$, $\fV$, $\fVII$, and a nonresonant component, and measured the individual fit fractions and phases of each component.  We do not observe any significant contribution from the so-called $f_X(1500)$ resonance seen in, for example, $\Bp\rightarrow \Kp\Km\Kp$~\cite{babarKKK}. The peak in the invariant mass between $1.5$ and $1.6\gevcc$ can be described by the interference between the $f_0(1710)$ resonance and the nonresonant component. We see some hints from the $f_{2}^{'}(1525)$ and $f_{0}(1500)$ resonances that could also contribute to this structure, but due to limited sample size we cannot make a significant statement. Future investigations of the $KK$ system could shed more light on the situation.
Furthermore we have performed an update of the phase-space-integrated time-dependent analysis of the same decay mode, using \pippim\ and \pizpiz\ decays, with the final \babar\ dataset. We measure the \CP-violation parameters to be ${\cal S} = -0.94\,^{+0.24}_{-0.21} \pm 0.06 $ and ${\cal C} = -0.17 \pm 0.18 \pm 0.04 $, where the first quoted uncertainty is statistical and the second is systematic.
These measured values are consistent with and supersede those reported in Ref.~\cite{KsKsKsBabar}. They are compatible within two standard deviations with those measured in tree-dominated modes such as $\Bz\to \jpsi\KS$, as expected in the SM. For the first time, we report evidence of \CP\ violation in $\Bz\to\KS\KS\KS$ decays; \CP\ conservation is excluded at $3.8$ standard deviations including systematic uncertainties.

\section*{ACKNOWLEDGMENTS}
\label{sec:acknowledgments}
We are grateful for the 
extraordinary contributions of our \pep2\ colleagues in
achieving the excellent luminosity and machine conditions
that have made this work possible.
The success of this project also relies critically on the 
expertise and dedication of the computing organizations that 
support \babar.
The collaborating institutions wish to thank 
SLAC for its support and the kind hospitality extended to them. 
This work is supported by the
US Department of Energy
and National Science Foundation, the
Natural Sciences and Engineering Research Council (Canada),
the Commissariat \`a l'Energie Atomique and
Institut National de Physique Nucl\'eaire et de Physique des Particules
(France), the
Bundesministerium f\"ur Bildung und Forschung and
Deutsche Forschungsgemeinschaft
(Germany), the
Istituto Nazionale di Fisica Nucleare (Italy),
the Foundation for Fundamental Research on Matter (The Netherlands),
the Research Council of Norway, the
Ministry of Education and Science of the Russian Federation, 
Ministerio de Ciencia e Innovaci\'on (Spain), and the
Science and Technology Facilities Council (United Kingdom).
Individuals have received support from 
the Marie-Curie IEF program (European Union), the A. P. Sloan Foundation (USA) 
and the Binational Science Foundation (USA-Israel).

%\clearpage
\bibliography{paper_KsKsKs}

\begin{thebibliography}{27}
\expandafter\ifx\csname natexlab\endcsname\relax\def\natexlab#1{#1}\fi
\expandafter\ifx\csname bibnamefont\endcsname\relax
  \def\bibnamefont#1{#1}\fi
\expandafter\ifx\csname bibfnamefont\endcsname\relax
  \def\bibfnamefont#1{#1}\fi
\expandafter\ifx\csname citenamefont\endcsname\relax
  \def\citenamefont#1{#1}\fi
\expandafter\ifx\csname url\endcsname\relax
  \def\url#1{\texttt{#1}}\fi
\expandafter\ifx\csname urlprefix\endcsname\relax\def\urlprefix{URL }\fi
\providecommand{\bibinfo}[2]{#2}
\providecommand{\eprint}[2][]{\url{#2}}

\bibitem[{\citenamefont{Grossman and Worah}(1997)}]{Grossman:1996ke}
\bibinfo{author}{\bibfnamefont{Y.}~\bibnamefont{Grossman}} \bibnamefont{and}
  \bibinfo{author}{\bibfnamefont{M.~P.} \bibnamefont{Worah}},
  \bibinfo{journal}{Phys. Lett.} \textbf{\bibinfo{volume}{B 395}},
  \bibinfo{pages}{241} (\bibinfo{year}{1997}).

\bibitem[{\citenamefont{Cheng et~al.}(2005)\citenamefont{Cheng, Chua, and
  Soni}}]{Cheng:2005ug}
\bibinfo{author}{\bibfnamefont{H.-Y.} \bibnamefont{Cheng}},
  \bibinfo{author}{\bibfnamefont{C.-K.} \bibnamefont{Chua}}, \bibnamefont{and}
  \bibinfo{author}{\bibfnamefont{A.}~\bibnamefont{Soni}},
  \bibinfo{journal}{Phys. Rev.} \textbf{\bibinfo{volume}{D 72}},
  \bibinfo{pages}{094003} (\bibinfo{year}{2005}).

\bibitem[{\citenamefont{Aubert et~al.}(2007{\natexlab{a}})}]{KsKsKsBabar}
\bibinfo{author}{\bibfnamefont{B.}~\bibnamefont{Aubert}} \bibnamefont{et~al.}
  (\bibinfo{collaboration}{\babar}), \bibinfo{journal}{Phys. Rev.}
  \textbf{\bibinfo{volume}{D 76}}, \bibinfo{pages}{091101}
  (\bibinfo{year}{2007}{\natexlab{a}}).

\bibitem[{\citenamefont{Chen et~al.}(2007)}]{belleCP}
\bibinfo{author}{\bibfnamefont{K.-F.} \bibnamefont{Chen}} \bibnamefont{et~al.}
  (\bibinfo{collaboration}{Belle}), \bibinfo{journal}{Phys. Rev. Lett.}
  \textbf{\bibinfo{volume}{98}}, \bibinfo{pages}{031802}
  (\bibinfo{year}{2007}).

\bibitem[{\citenamefont{Gershon and Hazumi}(2004)}]{Gershon:2004tk}
\bibinfo{author}{\bibfnamefont{T.}~\bibnamefont{Gershon}} \bibnamefont{and}
  \bibinfo{author}{\bibfnamefont{M.}~\bibnamefont{Hazumi}},
  \bibinfo{journal}{Phys. Lett.} \textbf{\bibinfo{volume}{B 596}},
  \bibinfo{pages}{163} (\bibinfo{year}{2004}).

\bibitem[{\citenamefont{Aubert et~al.}(2006)}]{babarKKK}
\bibinfo{author}{\bibfnamefont{B.}~\bibnamefont{Aubert}} \bibnamefont{et~al.}
  (\bibinfo{collaboration}{\babar}), \bibinfo{journal}{Phys. Rev.}
  \textbf{\bibinfo{volume}{D 74}}, \bibinfo{pages}{032003}
  (\bibinfo{year}{2006}).

\bibitem[{\citenamefont{Aubert et~al.}(2007{\natexlab{b}})}]{Aubert:2007sd}
\bibinfo{author}{\bibfnamefont{B.}~\bibnamefont{Aubert}} \bibnamefont{et~al.}
  (\bibinfo{collaboration}{\babar}), \bibinfo{journal}{Phys. Rev. Lett.}
  \textbf{\bibinfo{volume}{99}}, \bibinfo{pages}{161802}
  (\bibinfo{year}{2007}{\natexlab{b}}).

\bibitem[{\citenamefont{Aubert et~al.}(2008)}]{babarKKKs}
\bibinfo{author}{\bibfnamefont{B.}~\bibnamefont{Aubert}} \bibnamefont{et~al.}
  (\bibinfo{collaboration}{\babar}) (\bibinfo{year}{2008}),
  \bibinfo{note}{presented at ICHEP 2008}, \eprint{arXiv:0808.0700}.

\bibitem[{\citenamefont{Garmash et~al.}(2005)}]{belleKKK}
\bibinfo{author}{\bibfnamefont{A.}~\bibnamefont{Garmash}} \bibnamefont{et~al.}
  (\bibinfo{collaboration}{Belle}), \bibinfo{journal}{Phys. Rev.}
  \textbf{\bibinfo{volume}{D 71}}, \bibinfo{pages}{092003}
  (\bibinfo{year}{2005}).

\bibitem[{\citenamefont{Nakahama et~al.}(2010)}]{belleKKKs}
\bibinfo{author}{\bibfnamefont{Y.}~\bibnamefont{Nakahama}} \bibnamefont{et~al.}
  (\bibinfo{collaboration}{Belle}), \bibinfo{journal}{Phys. Rev.}
  \textbf{\bibinfo{volume}{D 82}}, \bibinfo{pages}{073011}
  (\bibinfo{year}{2010}).

\bibitem[{\citenamefont{Aubert et~al.}(2007{\natexlab{c}})}]{babarKKpi}
\bibinfo{author}{\bibfnamefont{B.}~\bibnamefont{Aubert}} \bibnamefont{et~al.}
  (\bibinfo{collaboration}{\babar}), \bibinfo{journal}{Phys. Rev. Lett.}
  \textbf{\bibinfo{volume}{99}}, \bibinfo{pages}{221801}
  (\bibinfo{year}{2007}{\natexlab{c}}).

\bibitem[{\citenamefont{Aubert et~al.}(2009{\natexlab{a}})}]{KsKspi}
\bibinfo{author}{\bibfnamefont{B.}~\bibnamefont{Aubert}} \bibnamefont{et~al.}
  (\bibinfo{collaboration}{\babar}), \bibinfo{journal}{Phys. Rev. D. RC}
  \textbf{\bibinfo{volume}{79}}, \bibinfo{pages}{051101}
  (\bibinfo{year}{2009}{\natexlab{a}}).

\bibitem[{\citenamefont{Rey-Le~Lorier and London}(2012)}]{London_Lorier:2011}
\bibinfo{author}{\bibfnamefont{N.}~\bibnamefont{Rey-Le~Lorier}}
  \bibnamefont{and} \bibinfo{author}{\bibfnamefont{D.}~\bibnamefont{London}},
  \bibinfo{journal}{Phys.Rev.} \textbf{\bibinfo{volume}{D 85}},
  \bibinfo{pages}{016010} (\bibinfo{year}{2012}).

\bibitem[{\citenamefont{Aubert et~al.}(2002)}]{babar}
\bibinfo{author}{\bibfnamefont{B.}~\bibnamefont{Aubert}} \bibnamefont{et~al.}
  (\bibinfo{collaboration}{\babar}), \bibinfo{journal}{Nucl. Instrum. Methods
  Phys. Res.} \textbf{\bibinfo{volume}{A 479}}, \bibinfo{pages}{1}
  (\bibinfo{year}{2002}).

\bibitem[{\citenamefont{Blatt and Weisskopf}(1952)}]{blatt-weisskopf}
\bibinfo{author}{\bibfnamefont{J.}~\bibnamefont{Blatt}} \bibnamefont{and}
  \bibinfo{author}{\bibfnamefont{V.~E.} \bibnamefont{Weisskopf}},
  \emph{\bibinfo{title}{Theoretical Nuclear Physics}} (\bibinfo{publisher}{J.
  Wiley, {\it New York}}, \bibinfo{year}{1952}).

\bibitem[{\citenamefont{Amsler et~al.}(2008)}]{Amsler:2008zz}
\bibinfo{author}{\bibfnamefont{C.}~\bibnamefont{Amsler}} \bibnamefont{et~al.}
  (\bibinfo{collaboration}{Particle Data Group}), \bibinfo{journal}{Phys.
  Lett.} \textbf{\bibinfo{volume}{B 667}}, \bibinfo{pages}{1}
  (\bibinfo{year}{2008}).

\bibitem[{\citenamefont{Flatte}(1976)}]{Flatte}
\bibinfo{author}{\bibfnamefont{S.~M.} \bibnamefont{Flatte}},
  \bibinfo{journal}{Phys. Lett.} \textbf{\bibinfo{volume}{B 63}},
  \bibinfo{pages}{224} (\bibinfo{year}{1976}).

\bibitem[{\citenamefont{Ablikim et~al.}(2005)}]{valFlatte}
\bibinfo{author}{\bibfnamefont{M.}~\bibnamefont{Ablikim}} \bibnamefont{et~al.}
  (\bibinfo{collaboration}{BES}), \bibinfo{journal}{Phys. Lett.}
  \textbf{\bibinfo{volume}{B 607}}, \bibinfo{pages}{243}
  (\bibinfo{year}{2005}).

\bibitem[{\citenamefont{Gay et~al.}(1995)\citenamefont{Gay, Michel, Proriol,
  and Deschamps}}]{Gay:1995sm}
\bibinfo{author}{\bibfnamefont{P.}~\bibnamefont{Gay}},
  \bibinfo{author}{\bibfnamefont{B.}~\bibnamefont{Michel}},
  \bibinfo{author}{\bibfnamefont{J.}~\bibnamefont{Proriol}}, \bibnamefont{and}
  \bibinfo{author}{\bibfnamefont{O.}~\bibnamefont{Deschamps}}
  (\bibinfo{year}{1995}), \bibinfo{note}{prepared for 4th International
  Workshop on Software Engineering and Artificial Intelligence for High-energy
  and Nuclear Physics (AIHENP 95), Pisa, Italy, 1995}.

\bibitem[{\citenamefont{Albrecht et~al.}(1990)}]{argus}
\bibinfo{author}{\bibfnamefont{H.}~\bibnamefont{Albrecht}} \bibnamefont{et~al.}
  (\bibinfo{collaboration}{ARGUS}), \bibinfo{journal}{Z. Phys.}
  \textbf{\bibinfo{volume}{C 48}}, \bibinfo{pages}{543} (\bibinfo{year}{1990}).

\bibitem[{\citenamefont{Pivk and Le~Diberder}(2005)}]{Pivk:2004ty}
\bibinfo{author}{\bibfnamefont{M.}~\bibnamefont{Pivk}} \bibnamefont{and}
  \bibinfo{author}{\bibfnamefont{F.~R.} \bibnamefont{Le~Diberder}},
  \bibinfo{journal}{Nucl. Instrum. Methods Phys. Res.}
  \textbf{\bibinfo{volume}{A 555}}, \bibinfo{pages}{356}
  (\bibinfo{year}{2005}).

\bibitem[{\citenamefont{Aubert et~al.}(2009{\natexlab{b}})}]{Aubert:2009me}
\bibinfo{author}{\bibfnamefont{B.}~\bibnamefont{Aubert}} \bibnamefont{et~al.}
  (\bibinfo{collaboration}{\babar}), \bibinfo{journal}{Phys. Rev.}
  \textbf{\bibinfo{volume}{D 80}}, \bibinfo{pages}{112001}
  (\bibinfo{year}{2009}{\natexlab{b}}).

\bibitem[{\citenamefont{del Amo~Sanchez et~al.}(2011)}]{pappagallo}
\bibinfo{author}{\bibfnamefont{P.}~\bibnamefont{del Amo~Sanchez}}
  \bibnamefont{et~al.} (\bibinfo{collaboration}{\babar}),
  \bibinfo{journal}{Phys. Rev.} \textbf{\bibinfo{volume}{D 83}},
  \bibinfo{pages}{052001} (\bibinfo{year}{2011}).

\bibitem[{\citenamefont{Drescher et~al.}(1985)}]{Drescher:1984rt}
\bibinfo{author}{\bibfnamefont{A.}~\bibnamefont{Drescher}}
  \bibnamefont{et~al.}, \bibinfo{journal}{Nucl. Instrum. Methods Phys. Res.}
  \textbf{\bibinfo{volume}{A 237}}, \bibinfo{pages}{464}
  (\bibinfo{year}{1985}).

\bibitem[{\citenamefont{Aubert et~al.}(2005)}]{BabarS2b}
\bibinfo{author}{\bibfnamefont{B.}~\bibnamefont{Aubert}} \bibnamefont{et~al.}
  (\bibinfo{collaboration}{\babar}), \bibinfo{journal}{Phys. Rev. Lett.}
  \textbf{\bibinfo{volume}{94}}, \bibinfo{pages}{161803}
  (\bibinfo{year}{2005}).

\bibitem[{\citenamefont{Gardner and Tandean}(2004)}]{Gardner:2003su}
\bibinfo{author}{\bibfnamefont{S.}~\bibnamefont{Gardner}} \bibnamefont{and}
  \bibinfo{author}{\bibfnamefont{J.}~\bibnamefont{Tandean}},
  \bibinfo{journal}{Phys. Rev.} \textbf{\bibinfo{volume}{D69}},
  \bibinfo{pages}{034011} (\bibinfo{year}{2004}).

\bibitem[{\citenamefont{Aubert et~al.}(2009{\natexlab{c}})}]{LastSin2betaBabar}
\bibinfo{author}{\bibfnamefont{B.}~\bibnamefont{Aubert}} \bibnamefont{et~al.}
  (\bibinfo{collaboration}{\babar}), \bibinfo{journal}{Phys. Rev.}
  \textbf{\bibinfo{volume}{D 79}}, \bibinfo{pages}{072009}
  (\bibinfo{year}{2009}{\natexlab{c}}).

\end{thebibliography}
\bibliographystyle{apsrev}

\end{document}